\documentclass[12pt, preprint]{aastex}
\usepackage{graphicx}

\newcommand{\mn}{\ion{Mn}{2}}
\newcommand{\ca}{\ion{Ca}{2}}
\newcommand{\mg}{\ion{Mg}{2}}
\newcommand{\si}{\ion{Si}{2}}
\newcommand{\co}{\ion{Co}{2}}
\newcommand{\fe}{\ion{Fe}{2}}
\newcommand{\oi}{\ion{O}{1}}
\newcommand{\ci}{\ion{C}{1}}
\newcommand{\kms}{km s$^{-1}$}
\newcommand{\gcm}{g cm$^{-3}$}
\newcommand{\mum}{$\mu$m}

\begin{document}

\title{A Catalog of Near Infrared Spectra from Type Ia Supernovae}

\author{G. H. Marion}
\affil{Astronomy Department, University of Texas at Austin, 
       Austin, TX 78712, USA}

\author{P. H\"oflich and C. L. Gerardy}
\affil{Physics Department, Florida State University,
       Tallahassee, FL, 32306}

\author{W. D. Vacca}
\affil{SOFIA-USRA,
       NASA Ames Research Center,   
       MS N211-3,   
       Moffett Field, CA 94035-1000}

\author{J.C. Wheeler and E. L. Robinson}
\affil{Astronomy Department, University of Texas at Austin, 
       Austin, TX 78712, USA}

\begin{abstract}
We present forty-one near infrared (NIR, 0.7--2.5 \mum) spectra from normal Type Ia supernovae (SNe Ia) obtained at epochs ranging from fourteen days before to seventy-five days with respect to the maximum light date in the V-band.  All data were obtained at the IRTF using the SpeX instrument.  We identify many spectral features, measure the Doppler velocities, and discuss the chemical distribution of explosion products in SNe Ia.  We describe procedures for smoothing data, fitting continua, and measuring absorption features to insure consistency for measurement and analysis.  

This sample provides the first opportunity to examine and compare a large number of SNe Ia in this wavelength region. NIR data are a rich source of information about explosion products whose signatures are blended or obscured in other spectral regions and NIR observations probe a greater radial depth than optical wavelengths.  We analyze similarities and differences in the spectra and we show that the progressive development of spectral features for normal SNe Ia in the NIR is consistent with time.  We confirm the presence of \oi, \mg, \ca, \si, \ion{Fe}{2}, and \ion{Co}{2} in these SNe.  Possible identifications are made for \ion{S}{1}, \ion{Si}{3}, \mn, and \ion{Fe}{3}.  There is no evidence in these data for \ion{H}{1}, \ion{He}{1},  \ci, or \ion{C}{2}.  

As the explosion products expand and cool, progressively deeper layers are revealed.  Thus a time sequence of spectra examines the chemical structure and provides direct evidence of the physical properties of SNe Ia from the outer layers to deep inside the SN.  Measured Doppler velocities indicate that burning products in SNe Ia are distributed in distinct layers with no large scale mixing.  Carbon is not detected in these data, in agreement with previous results with NIR data establishing very low limits on carbon abundance in SNe Ia.  Carbon burning products, O and Mg, are plentiful in the outer layers suggesting that the entire progenitor is burned in the explosion.  The data provide a resource for investigations of cross-correlations with other data libraries that may further constrain SN Ia physics and improve the effectiveness of SNe Ia as cosmological distance indicators.  
\end{abstract}
\maketitle


\section{Introduction}
This sample of forty-one near infrared (NIR, 0.7--2.5 \mum) spectra from Type Ia supernovae (SNe Ia) provides the first opportunity to study NIR spectral evolution in a large number of SNe Ia.  

SNe Ia are intrinsically interesting because they are very powerful explosions ($\approx 10^{51}$ ergs) involving extreme and exotic physics. SNe Ia are also key contributors to the chemical evolution of the Universe.  In recent years, the relative uniformity and high luminosity of SNe Ia have made them important as ``standard candles'' used to make distance estimates at large redshifts.  SNe Ia data are a valuable tool for measuring many cosmological parameters.   The fact that distant SNe Ia appear slightly dimmer than expected for a flat, coasting universe can be used to show that the expansion rate of the universe is accelerating \citep{gar98, Riess98a, schmidt98, Perl_99}.  This important result is combined with information from studies of the cosmic microwave background and large scale structure to suggest the existence of a ``dark energy'' in the universe.

SNe Ia are not precisely standard candles, and variability in SN Ia observables introduces uncertainties that limit their effectiveness as distance indicators.  The intrinsic dispersion in the peak brightness of SNe Ia events can be constrained to about 0.2 magnitudes by calibrating the measured luminosity using corrections determined by the shape of the light curve \citep{phillips93, Riess95, hamuy96, jha07, guy07, conley08}.  This accuracy is sufficient for some cosmological analysis, but in order to achieve the level of high precision cosmology required for dark energy measurements the dispersion must be reduced by an order of magnitude \citep{WA2001, kowalski08}.  

The NIR is a productive source of information about light and intermediate mass elements such as He, C, O, Mg, and Mn that are undetectable or obscured by line blending at other wavelengths.  The progenitor of a SN Ia is expected to be a carbon/oxygen white dwarf star (C/O WD), so that after the explosion a region where C and O are detected together is composed of primordial material that has not experienced nuclear burning.  The extent of nuclear burning in the SN will consequently be defined by a boundary between a region in which C or O or both have been consumed and a region that contains both C and O.  The abundance of Mn is temperature dependent so Mn can be used to probe the burning temperature in regions of incomplete silicon burning \citep{hwt98, pah02}.  Si, S, Ca, Fe, and Co are burning products that also produce strong lines in the NIR, making data from this wavelength region a rich source of information about the physical characteristics of SNe Ia. 

Many of these important explosion products are revealed by spectral observations of SNe Ia made near or before the date of maximum brightness in the V-band ($V_{max}$). Our sample is an excellent resource for investigation of SNe Ia because it includes eleven spectra obtained before $V_{max}$, and 20 spectra obtained within one week of maximum brightness. 

Supernovae expand and cool during the first weeks after the explosion which is coincident with the time period covered by these observations.  As the envelope thins and the dominance of Thomson scattering in the continuum diminishes, progressively deeper regions of the supernova are revealed.  The data show that NIR spectral features evolve consistently for normal SNe Ia, so a time sequence of spectra can be used to trace the chemical structure from the outer layers toward the center. NIR observations are particularly effective for this purpose because the optical depth is less in the NIR than at shorter wavelengths so that a greater radial depth can be probed with each spectrum. 

With a sample of this size, NIR data begin to realize their potential to further constrain estimates for the intrinsic brightness of individual SNe Ia by providing direct evidence of their physical properties.  This is an essential part of the interaction between theoretical modeling of supernova physics and observational analysis.  NIR spectra help define the chemical structure and constrain abundances of some burning products in SNe Ia.  

Relatively few NIR spectra are found in the literature and they were obtained from even fewer individual SNe Ia.  Six spectra from -8 to +8 days were obtained from SN1994D in narrow wavelength bands \citep{Meikle_96}. An excellent time sequence for SN 1999ee from -9 to +42 days was obtained by \citet{Hamuy02}, but without comparison with other objects, the data provide no information about the typical behavior for SNe Ia.  A fine set of NIR spectra were obtained by Gerardy from SNe Ia SN 1999by between -4 to +28 days \citep{pah02}, but this object was significantly sub-luminous and does not contribute to the discussion of normal behavior in SNe Ia.  Our program previously published 13 spectra from 12 objects (spectra that are included in the current sample) that provided clues for many of the conclusions derived from the larger sample \citep{m03}.

 The spectra in this sample also provide an opportunity to search for secondary relationships between NIR observables and other data libraries that may further constrain SN Ia physics and improve the effectiveness of SNe Ia as cosmological distance indicators.  For example: this sample may provide a link to accurate ``K corrections'' for photometric observations in the J, H, and K bands.

Details of the data acquisition and reduction procedures are presented in \S \ref{obs}, including the use of the Fourier transform to smooth the data. The spectra are presented in \S \ref{spectra}.  We identify many spectral features, measure the Doppler velocities for individual ions, and discuss the radial distribution of explosion products.  In \S \ref{results} we discuss some of the physical implications of our results. We examine the evolution of spectral features and compare spectra obtained at similar ages noting the similarities and differences.  Generally accepted characteristics of the physical properties and behaviors of SNe Ia are reviewed in \S \ref{phys} to provide a basis for analysis of the spectra.  \S \ref{sum} summarizes the results.  

Appendix \ref{features} provides a detailed discussion of individual features in the spectra.  Included with the appendix are Tables \ref{5K} and \ref{10K} that have estimated lines strengths of the strongest lines for many of the ions expected to be present in NIR spectra from SNe Ia.  The line strength estimates are separately computed at two temperatures: 5,000K, which is expected to be reasonable for extended line-forming regions and 10,000K which is expected where lines are formed closer to the photosphere.

\section{Acquisition and Reduction of Data}
\label{obs}
We obtained low and medium resolution NIR spectra from SNe Ia using the 3.0 meter telescope at the NASA Infrared Telescope Facility (IRTF) with the SpeX medium-resolution spectrograph \citep{Rayner03}.  The SpeX instrument provides single exposure coverage of the wavelength region from $0.8-2.5$ \mum.  For SpeX observations using a grating and prism cross-dispersers (SXD mode), the average instrumental spectral resolution ($R_{\lambda}=\lambda/\Delta\lambda$) is 750--2000 and with a single prism (LRS mode) the resolution 120--300.  Most of our data were obtained using one of three settings: the 0.5'' slit in LRS ($R_{\lambda}\approx200$, which makes the velocity resolution $R_{vel}=(\Delta\lambda \times c)/\lambda\approx750$\kms), the 0.8'' slit in SXD ($R_{\lambda}\approx750$, $R_{vel}\approx400$\kms), and the 0.5'' slit in SXD ($R_{\lambda}\approx1200$, $R_{vel}\approx250$\kms).  The velocity resolution estimates are calculated for 1.0 \mum\ and assume that for the LRS data we can accurately interpolate midway between data points (see Tables \ref{snelist_epoch} and \ref{snelist_disc}).  This resolution in velocity space is sufficient because the features we want to investigate have widths of several thousand \kms.  The LRS observing mode can extend coverage at the blue end of the NIR to about 0.65 \mum. SpeX also contains an infrared slit-viewer/guider covering a 60x60 arcsec field-of-view at 0.12 arcsec/pixel.  The detectors are a Raytheon 1024x1024 InSb array in the spectrograph and a Raytheon 512x512 InSb array in the infrared slit-viewer.

For most observations the slit was aligned to the parallactic angle at the beginning of the observing sequence.  In a few cases, the proximity of the SN to the host galaxy made it impossible to achieve the optimal parallactic alignment.  However, most of our observations were made at less than 1.5 air-masses and the features in which we are interested are near the guiding wavelength so the errors due to atmospheric refraction are not significant.  We do not attempt to compare line-strengths and continuum levels with a precision that would be affected by light losses due to mis-alignment of the slit.

For bright targets, the SpeX guider effectively maintained the centroid of the target in the slit by guiding on the spill-over flux from the object in the slit.  When the target did not clearly appear outside the slit, we used the SpeX guider in offset mode with another object in the field of the SpeX imager.  Another guiding option was the IRTF optical guider and this was used in a few instances.

Saturation was not a concern due to the faintness of our objects, but OH lines are numerous and highly variable in the NIR.  To avoid an increase in background noise due to poor OH removal, individual exposure times are capped at 150s.  A typical set is limited to ten exposures for a total of 25 minutes integration time in order to remain within time scales of atmospheric variability.  The source is nodded along the slit using an A\_B\_B\_A\_A\_B\_B\_A\_A\_B pattern.  Calibration images are obtained from A0V standard stars. The standards were selected to be as near as possible to the time and airmass of the SN. Each observation set also includes calibration images from internal lamps for flat fielding and wavelength calibrations.

The data were reduced using a package of IDL routines specifically designed for the reduction of SpeX data (Spextool v. 3.2; \citeauthor{Cushing04} 2004). These routines perform pair subtraction, flat-fielding, aperture definition, spectral tracing and extraction, residual sky subtraction, host galaxy subtraction, and wavelength calibration for data acquired in both the prism mode and the cross-dispersed mode.  Corrections for telluric absorption were performed using the extracted spectrum of an A0V star and a specially designed IDL package developed by \citet{Vacca03}. These routines generate a telluric correction spectrum by comparing the spectrum of an A0V star to a model A0V spectrum that has been scaled to the observed magnitude, smoothed to the observed resolution and shifted to the observed radial velocity. The telluric correction spectrum is then shifted to align the telluric absorption features seen in the SN spectrum and divided into the target spectrum.  

Spextool removes problems with background irregularities by defining the aperture width and background level for each subtracted image. The flux level along the slit includes a positive image of the SN at the A position and a negative image of the SN at the B position.  The background level in the slit is defined by Spextool using a fit through four regions, in two pairs, that bracket each aperture. Typically this is a linear fit, but for noisy spectra a second or third order fit is more effective.

Noise levels are calculated and recorded by Spextool for each wavelength bin.  Because the data are subject to bad pixels which can skew the combined mean value and the error on the mean, we substitute the median and median absolute deviation as robust estimates of these statistics.  The median absolute deviation (MAD) is defined as:

\begin{displaymath}
MAD = 1.4826 \times median(|(f_{\lambda}-\langle f_{\lambda}\rangle_{1/2})|)
\end{displaymath} 

\noindent{where $f=$ flux, $\langle f_{\lambda}\rangle_{1/2}$ is the median of the combined values, and 1.4826 is a constant introduced on the assumption of a Gaussian distribution of initial values.}

The {\it error on the median} at each wavelength is equal to the one-sigma noise level and equal to the median absolute deviation divided by $\sqrt{N}$, where $N$ is the number of spectra that have been combined.

The data have also been cleaned by replacing spikes greater than twice the local noise level (Figure \ref{spikes}).  The replacement value from removed spikes is the mean of 40 adjacent data points for SXD and 8 adjacent points for LRS spectra.

Observing time scheduled months in advance presents unusual problems when working with transient objects such as SNe.  We recognized that the number of potential targets would be improved by more frequent but shorter observing runs.  Working with IRTF personnel we employed a remote observing program to obtain data on systematically scheduled observing dates at a frequency of 8--12 times per year for five years.  We obtained high quality NIR spectra from whatever SNe Ia were available at the scheduled time.  In most cases, each target was accessible only once during our observing dates.  As a result, we obtained a single epoch or ``snapshot' from most of our targets. 

\subsection {Using the Fourier Transform technique to smooth spectra}
\label{ftx3}

Using a Fourier transform (FT) permits us to smooth the data by removing the highest frequency portion of the signal, leaving only the lower frequency components which are the actual information from the target.  Since the features we are studying are very broad compared to the spectral resolution, this is an excellent way of reducing the noise to allow more consistent measurements and more effective comparisons between spectra.  

The energy flux from SNe Ia drops off rapidly with increasing wavelength, and as a result the signal-to-noise in our data diminishes at longer wavelengths.   Due to this different quality of the data, we use different parameters for the FT smoothing in different parts of the spectra.  SXD spectra contain $\approx$ 3,900 data points covering the wavelength range 0.8-2.5 \mum.  This large number of bins allows us to separate SXD spectra into three regions: 0.7--1.35 \mum, 1.35--1.8 \mum, and  1.8--2.5 \mum.  LRS spectra span the same wavelength band with $\approx540$ data points.  We separate LRS spectra into two sections with the division at about 1.35 \mum.

Each region of a spectrum is individually fit with a straight line between the endpoints and normalized to a flat continuum with the continuum equal to zero.  Setting both ends of the spectrum to zero eliminates most of the ringing that can be introduced by the transform procedure.  After normalization, a fast Fourier transform routine is used to change the spectrum to frequency space.  

The power spectrum (signal power by frequency) is obtained by taking the square of the absolute value of the frequency spectrum.  A log plot of the power spectrum reveals the noise level forming a horizontal line through the central region of the spectrum.  The signal power as a function of frequency descends rapidly from a maximum power in the lowest frequencies located at the edge of the spectrum and it crosses the noise level usually within the first 50 frequency bins.  Figure \ref{power}a shows a log plot of the 50 data points (out of 851) in the power spectrum derived from the region 0.8--1.35 \mum\ in the SXD spectrum obtained from SN 2005am at -4 days.  The noise level (N) is taken to be constant with a value equal to the mean of the 100 data points immediately to the right of the region displayed (50--150).  N is marked on the figure by a horizontal dashed line.  

We estimate that the slope of the signal for this data intersects the noise level near frequency bin 27.  We represent the signal level (S) for each frequency bin using a straight line in log space from the y-intercept ($\approx -4.0$) through the intersection point: (27, -12.5).  To avoid an abrupt transition from the signal line to the noise at the intersection, we generate a filter using S/(S+N) and multiply the frequency spectrum at each bin by the filter at that bin (Figure \ref{power}b).  We explored using a Wiener filter ($S^2/(N^2+S^2)$) and found no measurable difference in the resulting smoothed spectrum.  The result of applying the filter to the FT data is shown in Figure \ref{ft1} with the original frequency spectrum plotted in black and the filtered spectrum is superimposed in red.

As a test of the consistency of our method, we compare the effects of different smoothing parameters on differently shaped absorption features from the same data (SN 2005am at -4d).  The smoothing parameter is defined as the number of frequency bins to be included in the signal (S).   Figure \ref{ft3} shows raw data plotted in black with the results of various smoothing parameters superimposed in colors.  The top panel displays the feature from \mg\ at 0.9227 \mum\ which is asymmetric due to impingement on the red side from emission due to the \ca\ IR-triplet.  The bottom panel shows the feature from the \mg\ 1.0927 \mum\ line which is more nearly symmetrical.  

The FT smoothed spectrum, derived according to our standard procedure described above and using 27 bins, is plotted in black.  We reduce the number of bins until the smoothing obviously compromises the positions of the absorption minima. Different colors are used in the figure to represent the spectrum smoothed using different parameters.  Table \ref{fttable} gives the measured velocities with the different smoothing parameters.  The table shows that a very large change in the smoothing parameter, beyond any reasonable choice made by the user, is required to change the measured velocity by more than a 200 \kms.  

When the smoothing parameter is increased, the change in measured velocity is very small. The primary effect of too large a value of the parameter is that local noise may be included in the smoothed spectrum.  However the features we are measuring are much larger than any local spikes.  Consequently, when there were ambiguities in the choice of smoothing parameters, we were careful to avoid choosing smaller values.  Interpretations made by individual users may vary the total number of bins selected by 3--5 in the smaller direction and 5-10 in the greater direction but the figure and table clearly show that differences on that scale have no significant effect on the results.  An exaggerated change on the order of 40\% (from 27 to 16 bins) is far beyond the uncertainty in our parameter choices and it only moves the measured velocity by 100 \kms\ for the more symmetrical feature in the bottom of Figure \ref{ft3} and 900 \kms\ for the less symmetrical feature in the top of Figure \ref{ft3}.  

The amount that the spectral resolution is degraded by smoothing depends on the original resolution and the amount of smoothing required.  In the example here with the data from SN 2005am at -4d and a smoothing parameter of 27, the instrumental resolution of $\approx250$ \kms\ is reduced to $\approx1500$ \kms, but the measured location of the absorption minimum only changes by $200-300$ \kms.

The reverse Fourier transform of the filtered frequency spectrum returns a smoothed data spectrum that greatly improves our ability to interpret the features.  The top spectrum in Figure \ref{ft2} shows the raw data from a relatively high quality spectrum obtained from SN 2005am at -4d plotted in green, with the smoothed spectrum plotted in black over the original. The lower spectrum in the figure shows data from a noisier spectrum, SN 2004bl at -2d, and again the smoothed spectrum is plotted in black over the original data.  The figure illustrates the effectiveness of this technique to facilitate inspection and comparison of spectra in the sample.

\subsection{Fitting a local continuum}
\label{flc23}

The energy flux from SNe Ia decreases rapidly with increasing wavelength.  Within this steeply sloping continuum, the wavelength of minimum flux may not correspond to the wavelength of the maximum departure from the continuum.  For absorption features, this effect is not significant where the continuum is relatively flat, as in the optical region of the \ion{Si}{2} line at 6355 \mum, but it can potentially affect features in the NIR by $\approx 0.005$ \mum\ or $\approx$ 1,500 \kms.  To remove this influence, we estimate the location of the continuum in the region of each absorption feature and normalize the spectrum to a flat continuum before identifying the wavelength of absorption minimum.

Line-forming regions in SNe Ia are extended in velocity space (\S \ref{phlfr}).  Thus a full analysis requires measurement of the full range of Doppler velocities between the absorption minimum and the detection limit of the blue wings for each feature.  Due to uncertainties introduced by the sloping continuum, this region can be more accurately measured by normalizing the data to a level continuum.  

Figure \ref{confit1} illustrates the difficulty of fitting a continuum to SNe Ia spectra over the wavelength region 0.8--2.5 \mum.  Four spectra from different epochs are displayed with estimated continua superimposed in red (-3.0 power law) and green (-4.5 power law).   It is clear that a single power law does not fit any spectrum over the entire region.  Although we have been as consistent as possible with our observing and reduction techniques, it is possible that some of the differences in slope and shape between spectra may have been introduced by observation or data processing choices.  There are however, obvious changes in the overall shape of the spectra with time and there appear to be measurable differences in the shapes of spectra from different objects at similar epochs.  

For the purpose of comparing features from the same line in different spectra, we abandon our efforts to fit a continuum over a large wavelength range and concentrate on fitting a local continuum in the vicinity of each feature.  Figure \ref{confit2} shows the same spectra as the top two panels in Figure \ref{confit1} with local continua fit to the spectra in the wavelength regions close to each absorption feature. This technique produces better results than fitting a single continuum to the entire spectrum. The fitted regions are expanded in Figure \ref{confit3}

There remains ambiguity about how to fit the local continua because of differences in the spectra within each region but variations within the limits of our estimates do not significantly affect the results.  For example: we assume that the peaks near 0.85 and 0.93 \mum\ rise above the continuum as emission components of P-Cygni profiles from the very strong \ca\ triplet (mean=0.8579 \mum) and the strong \mg\ line (0.9227 \mum); however, it is not obvious where the true continuum level lies.  Our estimate for the location of the continuum in this region appears as the green line in both panels of Figure \ref{confit2}.  

The absorption features in Figure \ref{confit2} with minima near 1.05 and 1.20 \mum\ display different profiles in the two spectra.   It is not clear how much of the bump on the red side of the absorption is emission and how much is continuum.  The estimated locations of the local continua for these regions are shown as red and blue lines in Figure \ref{confit2}.

Figure \ref{confit3} expands the spectral regions with locally fit continua from Figure \ref{confit2} but with the continuum normalized to zero.  All figures on the left are from the SXD spectrum of SN 2002en at -4 days.  The figures on the right are from a LRS spectrum of SN 2002cr at -5d days.   The top row corresponds to the regions marked in green on Figure \ref{confit2}, the middle row to the red regions and the bottom row to the blue.  The FT smoothed spectrum is superimposed in red on the spectra in all figures.  We tested this method by gradually changing the fitted continua until it was clearly beyond a reasonable fit to the data.  The largest deviations in the wavelengths of the measured minima due to the manual placement of the continua was found to be $\pm 300$ \kms\ and for most features the deviations are smaller.

\section{The Spectra}
\label{spectra}
Our sample consists of forty-one (41) NIR spectra from SNe Ia, obtained between fourteen days before (-14d) and seventy-five days (+75d) after $V_{max}$ (Figure \ref{all}).  All spectra in our sample were obtained at the IRTF using the SpeX instrument. Observation and reduction practices are discussed in \S \ref{obs}.  

The spectra are listed in Table \ref{snelist_epoch} by epoch with respect to $V_{max}$ and Table \ref{snelist_disc} in order of discovery.  The Tables include: SN names, estimated ages at observation (with respect to $V_{max}$), the dates of observation, the average instrumental resolution observing mode, the estimated signal-to-noise ratio for wavelength regions approximately corresponding to Y, J, H, and K bands, and the redshift (z) of the host galaxy.  The epoch of observation with respect to $V_{max}$ is estimated by a variety of means and we have attempted to provide as accurate information as possible.  More complete light curve data for many of the SNe in our sample are in preparation and the observation date for each spectrum is included in the tables in order to facilitate comparison when these data become available.  

In all figures containing multiple spectra, we arrange the spectra in sequence with the earliest spectrum at the top according to the time of observation with respect to $V_{max}$. The abscissa is wavelength, in microns, shifted to the rest frame of the host galaxy and the ordinate is log flux.

Each spectrum has been normalized to 1.0 at 1.0 \mum.  The names of the SNe and their age when observed (in days relative to $V_{max}$) are displayed to the right of each spectrum in the same color as the spectrum.  The colors for each spectrum are consistent throughout the colored figures to facilitate identification.  The black line superimposed on the original data is the smoothed spectrum obtained by Fourier transform as described in \S \ref{obs}. 

Atmospheric opacity depends strongly on wavelength in the NIR.  A transmission spectrum is provided at the top of some figures to help identify regions where opacity reduces the signal.  Complete transparency is at the top axis of the plot and full opacity at one unit down.  The transmission spectrum has been shifted using $z= 0.02$ which is the approximate mean redshift of the SNe Ia in our sample.  The noise levels in the data are very high through regions where the opacity exceeds 20\% ($\approx 1.30-1.42$ \mum\ and $1.78-1.90$ \mum).  We have removed the data in these regions from all spectra.

Order transitions in SXD spectra occur near 0.95, 1.12, and 1.45 \mum.  Small local discontinuities are possible at these locations where the orders are merged.  This problem is more pronounced in the noisier spectra.  SXD spectra also have gaps between $\approx 1.35-1.42$ where there are no data between orders.  The gap locations were selected to correspond to regions of high atmospheric opacity.  

Type Ia spectra are characterized by the steadily diminishing flux at longer wavelengths, so the S/N ratio degrades with increasing wavelength as the level of the signal approaches background noise levels (Tables \ref{snelist_disc} and \ref{snelist_epoch}).

\subsection{Line Identification by Epoch}
\label{lineid}

For spectral analysis we removed seven spectra that have S/N ratio less than 10 in the region $1.10-1.30$ \mum\ from the complete sample (see Table \ref{snelist_epoch}). This wavelength region approximately corresponds to the J-band and contains many of the most important features under discussion.  These seven spectra are displayed using the color yellow in Figure \ref{all}. The thirty-four remaining spectra are presented in Figure \ref{gh}.  

For analysis and line identification, we divide the remaining spectra into three groups selected by epoch and spectral features.  The boundaries between these phases is not always distinct and there is some temporal overlap between the phases.  We define the groups as: \emph{Photospheric Phase} (Outer Layers of the SN; 18 spectra obtained between -14d and +5d), \emph{Extended Photospheric Phase} (Intermediate Layers of the SN; 10 spectra between +4d and +24d), and \emph{Transitional Phase} (Inner Region of the SN; 7 spectra obtained between +37d and +75d). 

We identify many of the prominent features, measure their Doppler velocities, and provide a synopsis of spectral development in each group.  Appendix \ref{features} provides a more detailed analysis of the spectra and features with possible alternative line identifications and a discussion of each ion present and the physical implications.    We display the spectra in log space to assist with the identification of features at longer wavelengths, but care must be taken when comparing the relative strengths of features from widely separated wavelength regions.  

Figure \ref{fk} shows the spectrum from SN 2002fk at -14d, plotted in linear space and also in log space. Spectral features near 1.6 and 2.1 \mum\ are clearly smaller than the features appearing at 1.05 and 1.22 \mum\ when plotted in linear space, but these features appear to be larger when plotted in log space. 

Acceleration of matter in the explosion lasts for only a few seconds but within minutes the envelope expands by several orders of magnitude from the initial WD radius \citep{kho91, hmk93}.  The geometry of the expanding SN is assumed to be approximately spherical because the progenitor WD was nearly spherical, and the ejecta continue to move in homologous expansion with radial velocities proportional to the radial distances from the center \citep{kho97, gam03} and see discussion in \S \ref{phys}.  A continuous range of expansion velocities can therefore be used to represent contiguous physical space in the supernova, and we often refer to comparative radial distances by their measured velocities.  For example, a layer, or shell, of material in the spherical envelope may be described as occupying a region in space between 10,000 and 12,000 \kms.  

Polarization studies indicate that some SNe Ia, at least in the outer layers, may not be completely spherical \citep{wang03, wang08}.  Possible explanations for the asphericity include rotation effects of the progenitor, mergers of two C/O WD stars, high-velocity filaments of explosion products, and influences on the ejecta by the binary companion.  The observed polarization appears to be at maximum in high-velocity \ca\ regions.  The fact that high-velocity \ca\ components are present in most of the early spectra in our sample suggests that NIR data will be valuable for further study of asymmetry in SNe Ia.

After a preliminary identification of an association between a spectral feature and an individual atomic line, it is prudent to examine the spectrum for other lines from the same ion.  Comparison of relative line strengths can establish whether other lines from that ion would be expected to produce detectable features and confirm the initial identification.  To easily make these comparisons we have calculated line strength estimates for several of the strongest lines at two temperatures: 5,000K, which represents the temperatures found in extended line-forming regions, and also at 10,000K which represents temperatures found closer to the photosphere.  These temperatures are sufficiently separated to explore the effect of temperature on the relative strengths of lines from the same ion.  It is not our intention to precisely define temperatures in the line-forming regions.

The relative line strength estimate uses the LTE approximation for the fraction of ions from a population that will be excited to the $n^{th}$ level.  The population of level $n$ is given by the Boltzmann factor: $N_n \propto g_n f_n*10^{-\chi_n/kT}$, where $f_n$ is the oscillator strength, $g_n$ is the statistical weight, $\chi_n$ is the excitation potential, and $T$ is the temperature.

In Table \ref{5K} the estimated line strengths for ions is ordered by strength at 5,000K and in Table \ref{10K}, they are ordered by strength at 10,000K.  Computed values for different lines from each ion are normalized to the strongest line and rest wavelengths are given in air.  In situations where multiple lines from the same ion are close in wavelength, we calculate a mean value, weighted by oscillator strength, for two or more lines.  For these blends, the $\log(gf)$\ value and excitation minima are listed for the strongest line in the blend.  Our generalized treatment of line strength estimates has limitations but it is not model dependent.  The use of NLTE departure coefficients introduces uncertainty in the estimated line strengths by factors of 2--10.  Uncertainties of this magnitude do not change our results.

Transitions from one ionization stage to another are expected to be abrupt in both time and velocity space due to the dominance of radiative transitions over collisional processes.  Fe-group lines dominate the photo-ionization spectrum so ionization boundaries for other elements are influenced by flux levels from the Fe transitions.  Consequently most line-forming regions will contain only one ionization stage for each atom \citep{m06}.

\subsubsection{A Very Early Spectrum (-14d)}
\label{ves}

Figure \ref{fk} shows the spectrum of SN 2002fk, obtained 14 days before $V_{max}$ which is only five or six days after the explosion.  This is a single spectrum obtained in SXD mode.  For a detailed discussion of individual features, please reference \S \ref{ves_featanal}.

\paragraph{Probable Identifications:} \mg, \si, and \ca.  
\paragraph{Possible Identifications:}  \oi,  \ion{Si}{3}, \fe, \ion{Fe}{3}, and \co

\paragraph{Synopsis:}  The measured Doppler velocities in this spectrum are 10,000--12,000 \kms\ which is slower than the measured velocities for spectra from other SNe Ia obtained at later epochs.  There is also possible evidence for multiple ionization stages of silicon.  These unusual characteristics may be caused by lower temperatures within the envelope at this early epoch than will be found a few days later.  The lower temperatures allow the photosphere to temporarily recede more deeply into the ejecta.  Soon after the epoch of this spectrum, energy deposition from radioactive decay of $^{56}$Ni increases the temperature and opacity which pushes the photosphere out to higher velocities (A more detailed discussion of this effect is found in \S \ref{phlfr}).  

\ion{He}{1} and \ci\ are not detected in this spectrum.  \oi\ is possible from the 0.9264 \mum\ line at a Doppler velocity of 11,500 \kms\ in a blend with \mg.  \mg\ is clearly detected from lines at 0.9227 and 1.0927 \mum\ and is probable from lines at 1.0092 and 1.6787 \mum.  Doppler velocities for \mg\ are 10,300--11,600 \kms.  A feature due to \si\ from the 0.9413 \mum\ line is likely to be present at 11,300 \kms.  \ion{Si}{3} is possible but at a low Doppler velocity of 8,300 \kms.  The \ca\ IR-triplet is strong at 12,300 \kms.  \fe\ is unlikely and \ion{Fe}{3} is possible but at a low Doppler velocity of 8,100 \kms.  \co\ is detected at 12,000 \kms.

\subsubsection{The Photospheric Phase (Outer Layers; -14d to +5d)}
\label{mg}

Figure \ref{mg_fig} shows eighteen spectra obtained between -14 and +5 days relative to $V_{max}$.  During this epoch the photosphere is expected to be in the outer layers of the SN and the line-forming regions will be close to the photosphere.  \mg\ is prominent from multiple lines in all spectra in this group.  Six of the spectra were obtained in LRS mode which extends blueward to about 0.7 \mum\ at R=200 (compared with R=1200 for the SXD mode).  Figure \ref{mg_fwl} shows four representative spectra from this group: SNe 2004bw (LRS obtained at -9d), 2005am (SXD obtained at -4d), 2000dm (SXD obtained at +0d), and 2001bf (SXD obtained at +4d). Only the region $0.70-1.32$ \mum\ is displayed in Figures \ref{mg_fig} and \ref{mg_fwl} in order to enhance details in the area where most of the features are under discussion.  Features are labeled in the figure for which we have reasonably confident identifications.   For a detailed discussion of individual features from \emph{Photospheric Phase} group, please reference \S \ref{mg_featanal}.

Velocity references are compiled in Table \ref{vtable}.

\paragraph{Probable Identifications:} \oi, \mg, \si, \ca  
\paragraph{Possible Identifications:}  \mn, \fe, \ion{Fe}{3}, and \co.

\paragraph{Synopsis:}   Doppler velocities for lines identified in this group are between 10,000--14,000 \kms\ with lower velocities from both spectra of SN 2002cr (-7d and -5d) and from SN 2001bf (+4d).  

\ci\ is not detected in any of the spectra (\S \ref{noCI}). Strong features near 0.75 \mum\ due to \oi\ are detected in all spectra in this group that include the wavelength region. \mg\ produces multiple features in all spectra from this group and the \mg\ lines clearly correlate by velocity and strength.  The \ca\ IR triplet is prominent in all spectra from this group and a high-velocity component of the \ca\ feature is present to some extent in all spectra that include the wavelength region shorter than 0.8 \mum. \si\ is detected either alone or in a blend with \mn.  \fe\ is likely in blends with other atoms, as is \co.

A bump appearing near 1.25 \mum\ is strongly peaked in some of the spectra, unlike emission from a P-Cygni profile. This feature does not have a confident identification although it is observed in all spectra throughout our entire sample from -14d to +75d.  The feature near 0.97 \mum\ becomes stronger after maximum light due to contributions from \ion{Fe}{2}.  Another feature of uncertain origin appears at the same time near 1.48 \mum.  Beginning about maximum brightness, the spectra begin to be affected by changes in the apparent continuum due to line blanketing from Fe-group lines. ``Emission'' bumps begin to appear near 1.25, 1.52, and from 1.6 - 2.1 \mum\ and in broader regions (see discussion \S \ref{phlfr}).

\subsubsection{The Extended Photospheric Phase (Intermediate Layers; +4d to +24d)}
\label{fe}

Figure \ref{fe_fig} shows ten spectra from SNe Ia obtained between 4 and 24 days after $V_{max}$.  During the phase represented by this group, the shape of the continuum becomes influenced by line-blanketing from Fe-group lines which extends the effective photospheric radius (see \S \ref{phlfr}).  Spectra obtained in LRS mode extend blueward of 0.80 \mum\ and all others were obtained in SXD mode.  Figure \ref{fe_fwl} shows spectra from SNe 2000dk (+5d), 2001bg (+10d), SN 2004ab (+18d), and SN 2004da (+24d) as representative of this group with features labeled for which we have confident identifications.  For a detailed discussion of individual features from this group, please reference \S \ref{fe_featanal}.

Velocity references are compiled in Table \ref{vtable}.

\paragraph{Probable Identifications:} \oi, \mg, \si, \ion{S}{1}, \ca, \fe, and \co.  
\paragraph{Possible Identifications:} \ion{Si}{1}, \mn

\paragraph{Synopsis:} During the temporal period represented by this group, transitions are made from spectra dominated by \oi\ and \mg\ lines to spectra with lines from \ion{S}{1}, \fe, and possibly \mn\ and \ion{Si}{1}.  Measured Doppler velocities remain above 9,000 \kms\ for most ions with the exception of \fe\ which begins near 9,000 \kms\ at earlier epochs but slows to less than 3,000 \kms\ by +18d.    

\oi\ continues to be detected up to +10d and is possibly identified to +18d.  \mg\ is likely up to +5d but is absent thereafter.  The absorption near 0.89 \mum\ shifts to 0.90 and apparently makes a transition from \mg\ to \ion{S}{1} with possible contributions from \oi.  This feature becomes much weaker after +5d which is consistent with losing the signal from \mg\ and \oi.   

\si\ and \mn\ are possible alone or in a blend producing the feature found near 0.92 \mum\ that is not present in all spectra.  The \ca\ IR triplet remains prominent and the data continue to separately resolve the two strongest components of this feature.  \ion{Fe}{2} is detected from the line at 0.9998 \mum\ in all spectra through +18d, and \co\ is strong in the spectra from this group obtained at +14d and later. 

Individual Fe-group lines are complicated by many potential transitions at the same wavelength and nearly the same strength within each multiplet so the simple formula we have been using to estimate line strength may not be accurate for \fe\ as it was for smaller atoms. In the spectra from this group, the 0.9998 \mum\ line is the only \fe\ line detected (+4d through +18d), although the calculated line strengths suggest that lines at 0.7462, 0.7712, and 1.0500 \mum\ should also be detected (\S \ref{fe_featanal}).

During the \emph{Extended Photospheric Phase}, the shape of the continuum becomes influenced by line-blanketing from Fe-group lines (see \S \ref{phlfr}). Looking at this group in the context of the entire sample (Figure \ref{gh}), we can see how unresolved groups of Fe lines change the effective shape of the continuum during the epoch covered by these spectra. Significant peaks arise near 1.25, 1.55, 1.75 2.0, 2.1, and 2.25 \mum.  Thus many of the apparent ``absorption'' features may be interpreted as gaps between regions of inflating continuum.  For example: minima found near 1.12, 1.21, 1.50, and 1.65 \mum\ are probably formed by local flux increasing on both sides of an apparent minimum rather than by absorption at that location. 

Another example can be seen in Figure \ref{gh}, where the entire region from 0.9 to 1.3 \mum\ displays a distinct change from a concave bend below a straight line in the epoch +0d to +4d, to a nearly straight pattern from +4d to +10d, and finally to a convex bulge above a straight line after +10d that continues in later epochs.  

This group exhibits a reduced continuum level from $1.2-1.5$ \mum\ as compared with the earlier spectra.  Figures \ref{all}, \ref{gh}, \ref{fe_fig}, and \ref{fe_fwl} clearly show this effect.  This depressed continuum level is due to the region having fewer blends of iron group lines compared to adjacent wavelengths \citep{Wheeler98}.  Note that the spectrum of SN 2001bf (+4d) does not exhibit any deficit in this region, even though the epoch is estimated to be only one day from that of SN 2000dk (+5d) in which the effect is pronounced.

We note that the notch found at 1.65 \mum\ between two emission peaks, occurs at approximately the same wavelength as the Mg/Si/Co absorption in earlier spectra, but the features are probably not related. The spectrum of SN 2000dk at +5d is the oldest to show the strong notch indicative of \mg\ absorption at this location and later spectra have a smoother, shallower depression in this wavelength region.  The spectrum from SN 2000dk at +5d is also the oldest spectrum to show \mg\ absorption from lines at 0.9227 and 1.0927 \mum\ (see Figure \ref{mg_fig}).

The steep slope on the blue side of the peak found near 1.55 \mum\ defines the transition from partial to complete silicon burning.  Due to the high optical depth, this feature is formed close to the outer edge of the Fe/Co/Ni core at a rest wavelength of $\sim 1.57$ \mum.  Precise determination of the transition feature is not possible but velocities measured to a point approximately half way up the slope are found between 9,900--12,100 \kms\ for spectra in this group obtained from +4 to +18 days after maximum light.

\subsubsection{Transitional Phase (Inner Region; +37d to +75d)}
\label{transs} 

Figure \ref{transfig} shows seven spectra from our sample obtained between 37 and 75 days after $V_{max}$. This group displays a lot of new structure relative to earlier NIR spectra and the features are remarkably similar from one spectrum to the next.  The spectrum of SN 2004bv at +60d fits well with the others in this group although the spectrum from this SN at -5d was unusual (see Figure \ref{mg_fig}). Spectra obtained in LRS mode extend blueward of 0.80 \mum. All others were obtained in SXD mode.  For a detailed discussion of individual features from this group, please reference \S \ref{transs_featanal}.

\paragraph{Synopsis}  This epoch is a transitional phase between the photospheric era and a true nebular phase.  This is the same time frame in which unreddened SNe Ia have a very low dispersion for B-V colors (Lira, 1995). The common assumption about this era is that expansion has reduced the opacity beyond the point where a distinct photosphere can be defined and individual absorption features can no longer be produced.  Recent work by \citet{Branch08} suggests spectra from this epoch can be modeled by absorption features from permitted lines.  

The apparent emission peaks at 1.60, 1.75, 2.15, 2.25, and 2.35 \mum\ are the result of increased flux due to line blending from iron group elements as explained in \S \ref{phlfr}.  The radius of optical depth unity has increased at these wavelengths producing a larger effective radiation area and an increased flux.

\ion{Fe}{2} lines at 0.7712, 0.9998, and 1.0863 \mum\ appear in absorption beginning about 30 days after maximum light. In spectra from our sample obtained before +37d, the 0.9998 \mum\ line is the only \fe\ line detected, although the calculated line strengths suggest that lines at 0.7462, 0.7712, 1.0500, and 1.0863 \mum\ lines should also be detected.

\co\ lines are clearly detected from lines at 1.7772, 1.7462, 2.2205, 2.4596, and 2.3613 \mum\ at velocities near 9,000 \kms.  \co\ is found in wavelength regions where a pseudo photosphere is created at extended radii (1.5--1.85 and 2.05--2.45 \mum).  Strong absorptions are detected from several other \co\ lines.  

Figure \ref{trans_coii} shows evidence for \co\ and \fe\ lines in 13 spectra from our sample obtained between +14d and +75d with respect to $V_{max}$.  The figure includes two additional spectra from SNe 1998aq and 1998bu (plotted in yellow) that were obtained by \citeauthor{Rudy} (private communication) and have not been previously published.  These spectra are added to verify the consistency of our results.  The positions of strong \co\ and \fe\ lines drawn on the figure to indicate 11,000 \kms\ at the top spectrum and 6,000 \kms\ for the bottom spectrum.  This corresponds to the reduction in measured velocities during this epoch. 

The steep increase in flux found near 1.55 \mum\ is used by some models to define the transition from partial to complete silicon burning.  This feature is well defined up to about +60d.  As shown in Figure \ref{transfig}, the spectrum from SN 2004bv at +60 displays this feature but it is smeared out in the spectrum of SN 2001fe, also obtained at +60d.

\section{Results}
\label{results}
\subsection{Uniformity of Spectral Evolution}
\label{use}
 
Figures \ref{all}, \ref{gh}, \ref{mg_fig}, \ref{fe_fig}, and \ref{transfig} reveal that the development of spectral features with time is very consistent throughout the sample.  The data show that spectra obtained from normal SNe Ia at close to the same age have similar but not identical characteristics.  There are many small variations in features between adjacent spectra and a few instances where one spectrum displays a prominent feature that is absent or weak in other spectra with similar relative ages.   Nevertheless, within uncertainties introduced by variations in data quality and dating methods, the progress of feature development and overall shape is remarkably coherent from spectrum to spectrum in our sample.  That means that NIR spectra from different objects can be directly compared and arranged sequentially to mimic a time series.  Even spectroscopically unusual SNe Ia such as 2004bv and 2004da, which are missing some of the features found in most other SNe Ia, fit well with the general profile of other spectra obtained at similar epochs.   The agreement in spectral features from so many different SNe Ia eliminates the possibility that observations of finite line-forming regions (\S \ref{vtalk}) are due to clumps of material.

We test this hypothesis by comparing our ``snapshot'' spectra from separate SNe Ia to an excellent time series of NIR spectra obtained by \citet{Hamuy02} from SN 1999ee.  Five spectra from this series obtained between -11d and +13d are displayed with twenty-four spectra from our sample in Figure \ref{uni_23}.  Our data include a series of four spectra from SN 2005am (-4d to +14d).  The progression of feature development in a time series of spectra obtained from individual SNe Ia matches well with the sequence of spectra obtained from diverse objects.  It is clear that spectral development of normal SNe Ia in the NIR is very consistent.  

Although the overall NIR spectral evolution for SNe Ia is uniform in basic shape and features, there is obviously a great deal of diversity in the class of ``normal'' SNe Ia.  For example: our most confident line identification (apart from the \ca\ triplet) is \mg\ at 1.0927 \mum.  Yet spectra obtained within a day or two of each other sometimes exhibit dramatically different depths and widths for the absorption features produced by this line (see Figures \ref{mg_fig} and \ref{mg_fwl}).  Other examples of diversity include the anomalous velocities of SNe 2002cr and 2001bf, the weak features in the -4d spectrum from SN 2004bv, the variations in size and shape of the notch at 1.6 \mum\ over a range of epochs, and the vagaries of the unidentified bump at 1.25 \mum.  The presence of such variability provides motivation for further research and data collection efforts.  Despite these variations in individual features, the overall spectral shape and the general behavior of most features permits placing a spectrum in temporal order within a few days relative to $V_{max}$. 

\subsection{Measured Doppler Velocities}
\label{vtalk}

The Doppler shift from the rest wavelength of a line measures the velocity of the absorbing material along the line of sight to the observer.  As discussed in \S \ref{phlfr}, when the photosphere is close to the line-forming region, the observed Doppler-shift of the absorption minimum of the spectral feature corresponds to the velocity of the photosphere. At later times expansion moves the line-forming region away from the photosphere.  If there is a distinct discontinuity in the composition of the line-forming region, then a boundary of the absorbing layer for a particular ion is clearly defined and the velocity of the absorption minimum for an ion is determined by the inner edge of this layer.  The absorption feature will decrease in strength as the column density drops, but the velocity of the absorption minimum will no longer change until the ion is undetectable.

Table \ref{vtable} and Figure \ref{vplot} show the measured Doppler velocities for \oi, \mg, \si, \ion{S}{1}, \ca, and \fe.   Distinct minimum velocities, as measured from the absorption minima of spectral features, exist for all of these intermediate mass ions.  These velocity minima correspond to the location of the inner edge of a layer of line-forming matter in velocity space and physical space (\S \ref{phlfr}).   Figure \ref{vplot} shows that the line-forming regions from two \oi\ lines and three \mg\ lines occupy similar velocity spaces at $\approx$ 10,000--15,000 \kms. The velocity spaces for \si\ and \ion{S}{1} are slightly lower but overlap the \oi\ and \mg\ region at $\approx$ 8,000--12,000 \kms.  The \ca\ velocities are measured for the IR-triplet and may not be produced by fresh burning products (\S \ref{phlfr}).  This may account for the fact that minimum velocities for \ca\ are not as consistent as they are for some other species.  \fe\ appears to have no minimum velocity as the detected velocity descends from about 12,000 \kms\ at 10 days before maximum brightness to about 2,000 \kms\ at 20 days after maximum.

The distribution of burning products and radially-stratified chemical structure is consistent with the explosion occurring in a C/O WD with a radial density gradient and lower densities on the outside (\S \ref{chemstruct}).  The data appear to eliminate the possibility of large scale mixing of lower mass elements to the center or Fe-group elements to the extreme outer layers.

\subsection{Non-detection of Carbon}
\label{noCI}

Figure \ref{noci} displays the eighteen spectra from Figure \ref{mg_fig} with the locations of strong \ci\ lines marked at Doppler velocities of 11,000 \kms.  The strongest lines at 5,000 K, as listed in Table \ref{5K}, are marked in red on the figure and the weaker lines appear in black. Small features appear in some spectra near the expected location for lines from 0.9406, 1.1330, and 1.1754 \mum, but if these lines produce absorption features then the strongest lines at 1.0693 and 0.9093 \mum\ should be detected and they are not.  The possible feature for the 0.9406 \mum\ line is more likely attributed to \si\ at 0.9413 \mum\ (A complete discussion of this group of spectra is found in \S \ref{mg} and \ref{mg_featanal}.)

 Carbon is not detected in these data, a result that agrees with previous results with NIR data establishing very low limits on carbon abundance in SNe Ia \citep{m06}.  Similar results have been obtained using \ion{C}{2} lines in optical spectra \citep{tanaka}, but the much higher excitation potentials for \ion{C}{2} lines in the NIR make it unlikely that \ion{C}{2} will be detectable in these NIR data.  It is clear however that NIR data show that carbon burning products, O and Mg, are plentiful in the outer layers and that C is absent.  The line strengths for C and O will be similar for the same ionization stages, and if \oi\ in the data indicates unburned material, then \ci\ should also be present in detectable amounts \citep{m06}.  This result suggests that the entire progenitor is burned in the explosion.

\section{SNe Ia Physics}
\label{phys}
\subsection{Generic Model for SNe Ia}

Many years of optical observations support the widely accepted model that SNe Ia are the result of a thermonuclear explosion in a carbon-oxygen white dwarf star (C/O WD)  \citep{hoyle60}.  The WD mass at the time of explosion is expected to be near the Chandrasekhar limit ($M_{Ch} \approx 1.4 M_{\odot}$).  The progenitor WD begins with a mass well below $M_{Ch}$ and may grow by accreting matter from a red giant or main sequence binary companion by Roche lobe overflow.  Alternative models for SNe Ia create progenitors with a He star or another WD as the donor companion.  These evolutionary tracks will also produce a C/O WD near $M_{Ch}$.   Burning of accreted H or He on the surface of the WD converts it to C and O in approximately equal abundance \citep{dominguez01}.  Since the WD is supported against gravitational collapse by the degenerate electron gas, the final structure does not depend on the history of the progenitor.  Given the similar quantity and composition of materials, it is reasonable to assume that all progenitors will explode with similar properties

The primary fuel in the combustion of SNe Ia is carbon and oxygen which produces an ash that consists predominately of elements between silicon and nickel. The composition of the final burning products does not affect the energy production because very little additional binding energy is released in the final transition from Si-group to Fe-group elements.   If a C/O WD progenitor with mass $\approx 1.4 M_{\odot}$ burns its entire mass to at least Si-group elements, $\approx 2 \times 10^{51}$ ergs of nuclear energy will be produced.  That is sufficient energy to overcome the gravitational binding energy of 5--7 $\times 10^{50}$ ergs and provide the kinetic energy to accelerate the ejecta to observed expansion velocities \citep{hmk93, pah95, pah02}.  

The results described in \S \ref{noCI} suggest that the progenitor is almost entirely consumed in the explosion.  The nearly complete burning of progenitors with very similar characteristics can explain why the total production of nuclear energy is approximately constant in SNe Ia.  

\subsection{Chemical Structure of SNe Ia}
\label{chemstruct}

Optical spectra of SNe Ia reveal a range of elements produced by nuclear burning of carbon and oxygen at densities of $10^6-10^9$ \gcm.  To agree with observations of intermediate mass elements, burning conditions must exist in SNe Ia at densities less than $\approx 10^7$ \gcm.  Since the initial density of a C/O WD near $M_{Ch}$ is $\approx 10^9$ \gcm, it is necessary for early burning to lift the WD in its gravitational potential causing it to expand and reduce the mean density \citep{hwt98}.  The quantities and distribution of different burning products are determined by the relation between the hydrodynamical and nuclear time scales.  The hydrodynamical time scale is $\approx 1$ second, but the nuclear rates vary with density.  The range of densities implied by observations includes three distinct regions that produce different products from thermonuclear combustion \citep{hmk93, pah95, pah02}.

At densities greater than $10^7$ \gcm, temperatures exceed $5 \times 10^9$K.   Thermonuclear burning under these conditions proceeds uniformly to nuclear statistical equilibrium (NSE) and produces Fe-peak elements (Fe/Co/Ni).  In the case that $\rho_{burn} > 10^9$ \gcm, electron capture results and stable isotopes are produced.  For $10^9$ \gcm\ $> \rho_{burn} \ge 10^7$ \gcm\ the final burning product is predominantly $^{56}$Ni.  The existence of low and intermediate mass elements in SNe Ia requires the WD to experience a phase of expansion to reduce densities in the outer layers and prevent the entire WD from burning to Ni.  

Intermediate mass elements such as Si, S, and Ca are produced in a wide range of burning conditions ($10^7$ \gcm\ $> \rho_{burn} \ge 4 \times 10^6$ \gcm\ and $5\times 10^9$ K $> T > 3 \times 10^9$K).  Similar abundance ratios for these elements are produced by both explosive O burning and incomplete Si burning.  At the high end of this temperature and density range, silicon continues to burn after both carbon and oxygen are completely consumed, but the Si does not burn all the way to NSE. 

In regions with the lowest burning temperatures and densities ($\rho_{burn} \le 4 \times 10^6$ \gcm\ and $T \le 3 \times 10^9$K), explosive carbon burning occurs generating ash that consists primarily O, Mg and Ne.  The presence of these elements and the absence of C indicates that nuclear burning has occurred, since unburned regions of the SN will contain only C and O from the progenitor. Multiply-ionized atoms from heavier elements may be found in the outer layers from primordial sources.

\subsection{Location of the Photosphere and Line-forming Region}
\label{phlfr}

We define the photosphere as the region within the expanding envelope where continuum photons are produced at a given time and wavelength.  The photospheric radius is not constant, but is a function of wavelength.  The specific location is a function of opacity in the envelope and the energy deposition rate from radioactive decay.  As expansion reduces the optical depth in the outer layers, sequentially deeper layers of explosion products are revealed.  A time series of spectra will trace the chemical composition from the outer regions toward the core as the photosphere recedes in mass space.  NIR observations are particularly effective for recording the chemical structure of SNe Ia because the optical depth for most lines is less in the NIR than it is at shorter wavelengths.  As a result, a greater radial distance can be probed with each spectrum \citep{Wheeler98}. 

Absorption lines found in spectra from SNe Ia are formed in regions of expanding material that lie between the photosphere and the observer.  These line-forming regions in SNe Ia are extended in radial space and hence in velocity space.  To define the location of the line-forming region in radial space we use the Doppler shifts of the absorption components of individual spectral features to measure the expansion velocities.  The radial extent of each line-forming region covers the full range of Doppler velocities between the absorption minimum and the detection limit in the blue wings of each absorption feature.  

During the first few weeks after the explosion, high temperatures within the envelope insure that the primary opacity source in the NIR is Thomson scattering with some contribution from free-free emission.  In a rapidly expanding, scattering dominated envelope, all lines contain an emission component due to a non-vanishing line source function whether or not they display an obvious P-Cygni profile.  Emission photons from each line will reach the observer from different parts of the expanding envelope with different observed velocities.  These emission photons contaminate the absorption feature and make Doppler measurements unreliable between the rest wavelength of the line and the wavelength of the absorption minimum in each spectral feature.

During this phase, while the photosphere and line-forming region are close, the minimum observed Doppler velocity is determined by the location of the photosphere.  Because expansion velocity is proportional to radius, as the photosphere moves deeper with time, progressively slower material is exposed in the line-forming region. As a result, the measured velocity at the absorption minimum for each line will diminish with time until there is insufficient abundance to produce a detectable signal.  The measured range of velocities from the absorption minimum to the blue wing of the feature represents the physical location of the line-forming region in the ejecta. 

In the case where the line-forming region containing a particular ion does not extend below a certain radius, continued expansion will separate the photosphere from the line-producing layer.  The observed Doppler velocity in all spectra obtained after separation occurs will be the velocity at the inner edge of this layer, and the measured Doppler velocity at the absorption minimum will no longer change. The absorption feature will subsequently decrease in strength as the column density drops due to the expansion of the line-forming region. Since the photosphere is detached from the absorbing region, lines are formed far outside the last scattering radius for continuum photons and the photosphere appears to be nearly point-like.

We note that during very early epochs ($\ge 10$ days before $V_{max}$), the envelope is actually cooler than it will soon become, when it warms due to increased energy deposition from radioactive decay.  The cooler material has a lower optical depth, causing the photosphere to temporarily recede far enough toward the center that some of the intermediate mass materials beneath the extreme outer layers are revealed.  Doppler velocities are relatively low since the photosphere is closer to the center of mass and higher pressures in the interior near the photosphere will produce higher ionizations.  Thus multiple ionizations stages from intermediate mass elements may be detected in a single spectrum due to the large range of radial distance probed by NIR spectra.  

Later than approximately ten days before $V_{max}$, the expanding material continues to be heated by radioactive decay and the increasing temperature  increases the opacity and moves the photosphere away from the center in both mass and velocity space.  From this time forward, the photosphere begins the monotonic recession in mass and velocity space that characterizes spectral evolution during the next 15-20 days, as the data in this paper demonstrate.

A few days after $V_{max}$, line-blanketing from thousands of Fe-group lines begins to dominate the spectrum at certain wavelength regions.  The increase in total opacity moves the radius at which features are formed well above the receding photosphere formed by electron-scattering.  A larger radius increases the effective area from which the observed flux is emitted.  For wavelength regions where this line-blanketing occurs, increased flux levels mimic emission features \citep{Wheeler98}.  There are thousands of these lines and they are not individually resolved but they effectively raise the continuum level through the wavelength regions that they occupy by creating a temporary effective photosphere at larger radii.  Figure \ref{am_bumps} compares model data with four spectra from SN 2005am.  The model resolution creates a series of peaks rather than a smooth profile but Doppler shifting will blend the individual lines and the models clearly predict the behavior observed in the data where a pseudo photosphere is created at extended radii that increases the continuum flux in wavelength regions: 0.9--1.2, 1.5--1.85, and 2.05--2.45 \mum.

At 20-30 days after $V_{max}$, the absorbing material is diluted by expansion to the point where it is no longer possible to define a photosphere.  This is a transitional phase between the photospheric epoch and the true nebular phase.  In some wavelength regions, line-blanketing from the thousands of Fe-group lines continues to distort the continuum.

The \ca\ IR triplet forms the strongest single feature in each of our spectra.  Extremely low excitation values for these lines suggest that they are likely to create absorptions of this magnitude at very low \ca\ abundances.  Since solar values in circumstellar material would provide sufficient calcium to generate these features, it is not expected that the P-Cygni profile features found with absorption minima near 0.82 \mum\ are produced by freshly synthesized \ca.

\section{Summary}
\label{sum}
We present forty-one high quality NIR spectra from SNe Ia obtained between two weeks before and ten weeks after $V_{max}$.  We show that the Fourier transform is an excellent method for smoothing SN Ia data to allow more consistent measurements and better comparisons between spectra without compromising the important features. The procedures we describe for fitting continua and measuring absorption features provide consistency for measurement and analysis of a very large number of spectral features from spectra with a wide range of quality and resolution.  We estimate relative line strengths for the strongest lines from ions expected to be found in spectra from SNe Ia in this wavelength range and provide tables of the estimates as a resource for exploring NIR spectra.

By comparing true time series of spectra with ``snapshot'' spectra from many individual SNe Ia, we show that the evolution of spectral features is consistent for NIR spectra from normal SNe Ia.   Because geometrical dilution of the SN reveals deeper layers within the expanding envelope, our sample of forty-one spectra obtained between -14d and +75d provides an excellent record of the physical properties of SNe Ia from the outer layers to deep within the envelope.  

The spectra disclose that burning products are distributed in well-defined, radial layers.  This is an important result that constrains the progenitor evolution and the burning physics.  Table \ref{vtable} and Figure \ref{vplot} show the measured Doppler velocities for \oi, \mg, \si, \ion{S}{1}, \ca, and \fe.  Burning products from the lowest burning densities appear at the highest velocities (\oi\ and \mg\ at $\approx$ 10,000--15,000 \kms) with well-defined minima.  High velocities correspond to the outer layers of SNe Ia and clearly-defined minimum velocities indicate that the line-forming region containing O and Mg has a distinct inner edge.  \si, \ion{S}{1}, and \mn\ are found at somewhat lower velocities ($\approx$ 8,000--12,000 \kms) and they also have clear velocity boundaries to the line-forming region.  The agreement in spectral features from a large number of SNe Ia eliminates the possibility that observations of finite line-forming regions are due to clumps of material.  

\ca\ is only detected in the IR-triplet despite expectations that other \ca\ lines would be present.  Due to the very low excitation values for these lines, this result implies that detection is not due to freshly synthesized material.  \fe\ is detected at a wide range of velocities ($\approx$ 2,000--12,000 \kms) with no apparent minimum. 

The distribution of burning products and radially stratified chemical structure is consistent with the explosion occurring in a C/O WD with a radial density gradient and lower densities on the outside (\S \ref{chemstruct}).  Our results agree with models that predict explosive carbon burning will occur in the extreme outer layers of SNe Ia.  Oxygen burning and partial silicon burning appear to form intermediate layers and some Si must be completely burned to produce Fe detected at extended radii.  There are small overlaps between adjacent layers, but our data eliminate the possibility of large scale mixing of lower mass elements to the center or Fe-group elements to the extreme outer layers.

C, O, and Mg are only detectable by spectral observations up to a few days after $V_{max}$, so it is useful that there are twenty-two spectra in our sample obtained at +5d or earlier.  We do not find carbon in the spectra at any epoch but we do find carbon burning products (O and Mg) occupying the same physical space in the outer layers of the SNe.  There are strong lines from \ci\ in the NIR that have produced detections in subluminous SNe Ia, but \ci\ is not found in our sample.  (\ion{C}{2} lines are expected to be weaker than \ci\ and are also not detected.)  \oi\ and \mg\ are both products of explosive carbon burning and they are detected in nearly the same velocity space in the outer layers.  The line strengths for C and O will be similar for the same ionization stages, so if the \oi\ in our spectra is due to unburned material, then \ci\ should also be present in detectable amounts and it is not \citep{m06}. 

 These results strongly suggest that the entire progenitor is consumed in normal SNe Ia which implies that nearly the same amount of fuel is used to power each explosion and helps to explain the uniformity in SNe Ia observations.  These results constrain burning physics and models on a physical rather than phenomenological basis. 

 The size of this sample also facilitates the search for secondary relationships with other data libraries that may further constrain SN Ia physics and improve the effectiveness of SNe Ia as cosmological distance indicators.  Well-sampled light curves for the SNe in our sample will allow comparison of well-studied photometric parameters with NIR line strengths.   NIR observables can now be compared to empirical patterns such as the relationship between peak brightness and light curve shape, B-V color differences, secondary peaks in the IR, and host galaxy morphology.   This sample may enable accurate ``K corrections'' for photometric observations in the J, H, and K bands.  NIR spectra may also provide insights into areas of SNe Ia physics such as progenitor ZAMS mass, metallicity, rotation, asymmetry in the explosion, and gravitational influence on the detonation \citep{hwt98, wang98, pah02, yoon04a, yoon04b}.  We hope that this sample will truly open the door to use of NIR data for productive SNe Ia research.

\verb"\ack"
\noindent{\bf Acknowledgments:}
Marion is a visiting Astronomer at the Infrared Telescope Facility, which is operated by the University of Hawaii under Cooperative Agreement no. NCC 5-538 with the National Aeronautics and Space Administration, Science Mission Directorate, Planetary Astronomy Program.  We want to express our appreciation to the individuals at the IRTF for guidance and help with the observations.  In particular, Alan Tokunaga, John Rayner, Mike Cushing,  Bill Golisch,  Dave Griep, and Paul Sears have been most helpful.  We would also like to thank the TAC of the IRTF for support and instructive comments.  GHM would like to acknowledge and thank Mike Cushing for helpful comments.  

We thank R. Rudy of The Aersopace Corporation and R. Puetter of UCSD for providing observations of SNe 1998aq and 1998bu.

We are grateful to W. D. Li and M. Papenkova for providing light curve data for some of the SNe in our sample and we thank J. Vinko for his work compiling light curve data. 

PAH is supported in part by NSF grants 0703902 and 0708855.

We thank the referee for constructive comments that have improved the structure and presentation of the paper.

\clearpage

\appendix
\section{Discussion of Individual Features and Line Identifications}
\label{features}
This section discusses line identifications for specific features in the  spectra.  Line identifications and measured Doppler velocities are provided where possible.  Individual lines are referred to by their rest wavelengths.  The spectra are grouped by epoch using the same groups as in \S \ref{results}.  

Velocities measurements discussed in this appendix can be found in Table \ref{vtable} and Figure \ref{vplot}.

\subsection{A Very Early Spectrum (14 days before $V_{max}$)}
\label{ves_featanal}

Discussion in this section makes reference to Figure \ref{fk} which shows the spectrum from SN 2002fk which was obtained fourteen days before $V_{max}$.   Features are labeled in the figure for which we have confident identifications.  

\paragraph{\ion{He}{1}} is not detected in this spectrum from either of the strongest \ion{He}{1} lines in the NIR found at 1.0830 or 2.0581 \mum.  Possible associations for {\ion{He}{1} include the absorption feature found near 1.05 \mum\ which corresponds to a Doppler velocity of 8,600 \kms\ for the 1.0830 \mum\ line.  This is an unrealistically low velocity because if \ion{He}{1} is present, it is expected to be found only in the extreme outer layers of the supernova at high velocities.  The 1.05 \mum\ feature has a plausible alternative identification as \mg\ from the 1.0927 \mum\ line and it is labeled as such in the figure.  The 2.0581 \mum\ \ion{He}{1} line (which is expected to be $6-12$ times weaker than the 1.083 \mum\ line depending on temperature), can possibly be associated with either of two weak features at 2.00 and 2.04 \mum\ with velocities of 9,700 or 2,500 \kms\ which are inconsistent with 8,600 \kms\ if the strongest \ion{He}{1} line at 1.0830 \mum\ produces the 1.05 \mum\ feature. 

\paragraph{\ci} is predicted to generate strong lines at 1.0693, 0.9093, 0.9406, and 1.1754  \mum\ in order of expected strength (see Table \ref{5K}).  Additional discussion of \ci\ is found in \S \ref{noCI} and illustrated in Figure \ref{noci}.  Very weak features can be seen in this spectrum by optimistic observers at Doppler velocities of 10,000--12,000 \kms, but no confident detections can be made and the velocities are not consistent within each spectrum.  Some models for SNe Ia predict that \ci\ will be present in the extreme outer layers, but we are unable to identify \ci\ in this, or any other spectrum in our sample.

\paragraph{\oi} is likely from the 0.9264 \mum\ line contributing to the absorption feature found near 0.89 \mum\ at 11,500 \kms\ in a blend with \mg\ (0.9227 \mum).  The 0.8446 \oi\ line may also contribute to the absorption from the \ca\ IR triplet near 0.81 but this wavelength is at the extreme edge of the spectrum and it is difficult to evaluate the presence of small features in this region due to the strength of the \ca\ feature.  The \oi\ 0.7773 \mum\ line is expected to be 3-20 times stronger than the 0.9264 line and the 0.7773 \mum\ line is detected in all early spectra from which we have coverage of the wavelength region down to 0.7 \mum\ but this spectrum does not cover that range.

\paragraph{\mg} produces strong features that are clearly detected in this spectrum (and labeled in Figure \ref{fk}) from 0.9227 \mum, found near 0.89 \mum\ at 10,300 \kms, and 1.0927 \mum, found near 1.05 \mum\ at 11,300 \kms. The measured velocity for the 0.9227 \mum\ line is 1,000 \kms\ less than the measured velocity for the 1.0927 \mum\ line due to intrusion on the blue side of the \mg\ absorption from the emission component of the very strong \ca\ IR triplet.  This line is also likely to be blended with \oi\ at 0.9264 \mum.  The high quality of this spectrum reveals a probable detection near 0.97 \mum\ for the \mg\ line at 1.0092 \mum\ with a Doppler velocity of 10,900 \kms.  This line is predicted to be 5-20 times weaker than the 0.9227 \mum\ line which is consistent with the observed feature strengths.  The absorption feature near 1.60 \mum\ receives contributions from \mg\ at 1.6787 \mum\ with 11,600 \kms\ in addition to \si\ (1.6930 \mum) and \co\ (many lines 1.57--1.77 \mum).  Weak evidence can be found for \mg\ from the 0.9632 and 2.1369 \mum\ lines at similar velocities.

\paragraph{\si} from the strong line at 0.9413 \mum\ is the likely source of the feature found near 0.91 \mum\ at a Doppler velocity of 11,300 \kms.  This velocity is similar to the velocities measured for \mg\ which is consistent with models for this epoch.  The 1.6930 \si\ line may be part of the 1.6 \mum\ absorption feature but the velocity to the minimum of the feature is nearly 18,000 \kms\ for this line.  This ion may also contribute to the feature found near 1.09 \mum\ from the 1.1311 \mum\ line in a blend with \ion{Si}{3} and \ion{Fe}{3}.  However, \si\ at 1.1311 \mum\ is a much weaker line than the 0.9413 \mum\ line, and the velocity does not correlate with other \si\ lines.  If there is sufficient \si\ to produce a feature of this magnitude, then detections should also be possible from lines at 1.1737 and 1.7183 \mum.  There is no evidence in this spectrum for these other \si\ lines.

\paragraph{\ion{Si}{3}} can be present in the same spectrum with \si\ if the photosphere is deep within the envelope so that temperature and pressure differences between absorbing regions near the photosphere and in the outer layers contain different ionization stages of the same element (see \S \ref{phlfr}).  That scenario suggests that \ion{Si}{3} will be found in slower and hotter line-forming regions near to the photosphere.  Due to a relatively low excitation minimum and high oscillator strength, the \ion{Si}{3} 0.9324 \mum\ line is predicted to be $\sim$ 40 times stronger at 10,000 K than the next strongest \ion{Si}{3} line in the NIR.  This line may contribute to the feature observed near 0.91 \mum\ at a Doppler velocity of 8,400 \kms\ in a blend with \si\ and \mg.  A table of estimated line strengths using a temperature of 10,000 K shows change in the relative line strengths for \ion{Si}{3} compared to 5,000 K (see Table \ref{10K}).  For example: the fact that the \ion{Si}{3} 1.0526 \mum\ line is not detected in this spectrum is consistent with higher temperatures in the \ion{Si}{3} line-forming region.  Other possible identifications for \ion{Si}{3} are the feature at 1.22 \mum\  from the 1.2541 \mum\ line (a blend of 1.2523 and 1.2601 \mum) at 9,600 \kms\ and 1.09 \mum\  from 1.1341 \mum\ at 10,500 \kms.  

\paragraph{The \ca\ IR triplet} is strong in this spectrum from SN 2002fk with a Doppler velocity of 12,300 \kms.  The velocity measurement is based on a blend of the 0.8498 and 0.8542 \mum\ \ca\ lines, giving a mean wavelength of 0.8538 \mum, and measured with respect to the bluer and  deeper component of the absorption feature.  Resolution is sufficient to separate this pair from the other member of the triplet found at 0.8662 \mum.  The shape of the absorption feature is consistent with estimates that the 0.8538 blend is about twice as strong as the 0.8662 line.  

The very strong P-Cygni profile from the \ca\ IR triplet may not be produced by freshly synthesized Ca (\S \ref{phlfr}).  There is no evidence in the spectrum for \ca\ from the next strongest lines at 1.1876 or 0.9906 \mum\ and this may indicate that \ca\ produced in the explosion is not detectable.

This SXD spectrum lacks coverage in the wavelength region to detect a possible high-velocity component from \ca.

\paragraph{\ion{Fe}{2}} lines are more difficult to compare based on our simple line strength estimates because the actual line strengths are affected by multiple transitions. As a consequence, the simple calculations in our tables for the relative estimated strength of each \ion{Fe}{2} line may be less accurate than the estimates for smaller atoms.  The \ion{Fe}{2} 0.9998 \mum\ line is a possible source for the feature found near 0.97 \mum\ in this spectrum, but the velocity would be only 8,100 \kms.  This velocity is similar to some of the conjectured \ion{Si}{3} velocities. The 1.0500 \mum\ line from \fe\ is not detected, but the 1.0863 line may be a part of the feature near 1.05 \mum\ in a blend with \mg.  The 1.1126 \mum\ \fe\ line is a possible contributor near 1.09 \mum\  in a blend with \ion{Si}{3}.

\paragraph{\ion{Fe}{3}} at 0.9124 is possible in a blend with \oi\ and \mg\ but the Doppler velocity at the absorption minimum is very low at 7,100 \kms.  As noted in the discussion for \mg\ in this section, this absorption feature is distorted by emission from the \ca\ triplet and the true velocity may be higher.  This ion has several lines from 1.60--1.67 \mum\ that may add to the 1.6 \mum\ feature with the strongest line at 1.6697 \mum\ having a velocity of 9,900 \kms\ at the minimum.  The 1.1323 \mum\ line may also form, or contribute to, the absorption near 1.09 \mum\ at a velocity of 10,000 \kms.

\paragraph{\ion{Co}{2}} is detected at $\sim$ 12,000 \kms\ from individual lines at 1.7772, 2.2205, 2.4596 , and 2.3613 \mum. The 1.7462 \mum\ line is not found in this spectrum. Numerous \co\ lines between 1.57--1.77 \mum\ are likely contributors to the 1.60 \mum\ feature.

\subsection{The Photospheric Phase (Outer Layers)\\
(-14d to +5d with respect to $V_{max}$)}
\label{mg_featanal}
During this epoch the photosphere is expected to be in the outer layers of the SN and the line-forming regions will be close to the photosphere.  Discussion in this section makes reference to Figure \ref{mg_fig} showing eighteen spectra obtained between -14 and +5 days with respect to to $V_{max}$ and also to Figure \ref{mg_fwl} that displays four representative spectra from this group: SNe 2004bw (LRS obtained at -9d), 2005am (SXD obtained at -4d), 2000dm (SXD obtained at +0d), and 2001bf (SXD obtained at +4d) with features labeled for which we have confident identifications.  These figures only display the spectra in the wavelength region 0.65 -- 1.35 \mum\ to permit detailed inspection of the region where most of the features occur in this epoch.  The complete spectra from 0.7--2.5 \mum\ can be viewed in Figure \ref{gh}. 

\paragraph{\ci} is not detected from any of these spectra.  The strongest \ci\ lines are found at 1.0693, 0.9087, 0.9406 and 0.9639 \mum, listed in order of expected strength. (see Figure \ref{noci} and \S \ref{noCI} for additional discussion of \ci\ in our sample).

\paragraph{\oi} at 0.7773 forms the feature near 0.75 \mum\ detected in all LRS spectra in this group (the SXD spectra do not cover the wavelength region of this line).  Table \ref{vtable} shows the Doppler velocity of this line is measured at 11,300--13,500 \kms\ except for velocities of 9,300 and 9,400 measured in the two spectra from SN 2002cr. (Note that the spectra from SN 2002cr exhibit lower velocities than other spectra for all explosion products.)  The \oi\ feature is strong in all spectra from this group that include the appropriate wavelength region. 

The 0.9263 \mum\ line from \oi\ is expected to be 3--17 times weaker than the 0.7773 \mum\ line but it is likely to contribute to the 0.89 \mum\ feature in a blend with \mg\ 0.9227 \mum\ at velocities of 11,000--15,000.  Possible evidence for the 1.1287 \mum\ \oi\ line (which should be half as strong as the 0.9263 \mum\ line) appears near 1.09 \mum\ in some of these spectra.  

The strongest features at 1.09 \mum\ occur in spectra from SNe 2002hw (-1d) and 2000dm (+1d) and correlate with the strongest absorptions at 0.89 \mum\ which is suggestive that \oi\ contributes to both features. 

\paragraph{\mg} is found in this group producing strong features from lines at 0.9227 and 1.0927 \mum\ and weaker but detectable features in higher S/N spectra from lines at 1.0092 and 1.6787 \mum.  The Doppler velocities for the 1.0927 line are 11,100--14,500 \kms\ while the measured velocities for the 0.9227 line are $\sim$ 1000 \kms\ slower as a result of the enormous P-Cygni emission from \ca\ that impinges on the blue side of this absorption feature (Table \ref{vtable}).  Stronger \mg\ features are often correlated with higher measured Doppler velocities.  As they were for \oi, the lowest velocities are found in the spectrum of SN 2002cr.  Figures \ref{all} and \ref{mg_fig} show that there is an obvious correlation between strong spectral features from \mg\ at 1.0927 \mum\ and the strongest features from the 0.9227 \mum\ line.  The final spectrum (in temporal order) to show evidence of \mg\ (SN 2000dk, +5d) clearly shows features from both of these strong lines.

The \mg\ line at 0.7890 \mum\ may contribute to the 0.75 \mum\ feature in a blend with \oi\ but the velocities to the absorption minima of this feature would be 14,000--18,000 \kms\ and that is inconsistent with other \mg\ measurements.  The 0.7890 \mum\ line may flatten the P-Cygni emission component of the \oi\ line as in the spectrum from SN 2002cr.

The 1.0092 \mum\ \mg\ line is expected to be 2 orders of magnitude weaker than 1.0927 \mum\ but it is detectable in most spectra in this group, forming a weak absorption near 0.97 \mum\ at velocities comparable to the other \mg\ lines.  This feature is detected in all but the noisiest spectra from this group and it often exhibits a double or triple bottom containing minima near 0.95, 0.97 and 0.99 \mum\ with 0.97 \mum\ being the strongest.  The 0.95 and 0.99 \mum\ features are smaller than the one sigma noise but are consistent from spectrum to spectrum in magnitude and location relative to the central feature.  The central feature is probably \mg\ at 1.0092 \mum, while candidates for the others are \ca\ at 0.9906 \mum\ and \fe\ at 0.9998 \mum.  The fact that the feature near 0.97 \mum\ becomes stronger after maximum light suggests that the primary source of absorption at this wavelength may be shifting from \mg\ to \fe. 

Velocity measurements of the \mg\ line at 1.6787 \mum\ and the 1.60 \mum\ absorption feature correspond closely to the velocities of the other \mg\ lines.  This broad feature probably receives contributions from other ions, notably \si\ and \co.  The absorption feature is particularly strong in spectra that exhibit other strong \mg\ features, such as: SN 2005am (at -2d and +2d), SNe 2002hw (-1d) and 2000dm (+1d).  That may indicate increased abundance of \mg\ near max light.  \mg\ lines at 0.9632 and 2.1369 \mum\  are also possible in weak features or blends. 

Inspection of the complete sample (Figure \ref{all}) shows that all twenty-two spectra in our sample obtained at or before +2d show strong \mg\ features near 0.89 and 1.05 \mum.  Even spectroscopically unusual SNe Ia such as 2004bv exhibit distinct \mg\ signatures.  Measured velocities for \mg\ are given in Table \ref{vtable} and plotted in Figure \ref{vplot}.

\paragraph{\si} from the 0.9413 \mum\ line is the probable source of the feature found near 0.91 \mum. Note that this feature is distinct from the absorption found near 0.89 \mum\ and it does not appear in all spectra.  The measured velocities are between 10,200 and 14,100 \kms. As noted in the discussion of the \mg\ line at 0.9227 \mum, impingement from the emission component of the \ca\ triplet may affect the location of this feature, pushing it toward the red.  The 0.91 \mum\ feature, and thus the 0.9413 \mum\ \si\ line, is not detected in spectra from SNe 2004bw (-9d), 2003W (-7d and -6d), 2002hw (-1d), 2000dm (+0d), or 2000dk (+5d).  In general, the spectra without this \si\ detection show very strong \mg\ features that may obscure the \si\ signal due to P-Cygni emission from \mg\ at 0.9227 \mum.  

The 1.6930 \mum\ line from \si\ is likely to contribute to the 1.60 \mum\ feature but velocities to the absorption minima are 14,000--18,000 \kms\ which is higher than other measured \si\ velocities including measurements during this epoch of the strong \si\ features at optical wavelengths.  The line at 0.7846 \mum\ may contribute to the 0.75 \mum\ feature in a blend with \oi\ but velocities to the absorption minima are 12,200--16,400 \kms\ which is about 2000 \kms\ higher than for the 0.9413 \mum\ line.  Possible evidence for the 1.1748 \mum\ line appears near 1.13 \mum\ in four of the spectra at plausible velocities of 10,100--11,900.  However three of the four spectra with 1.13 \mum\ features are the ones that do not have a 0.91 \mum\ feature, making the \si\ association unlikely.

\paragraph{\ion{Si}{3}} is not detected in this group except for the earliest spectrum at -14d and discussed in the previous section.

\paragraph{\ion{S}{1}} probably begins to contribute to the 0.89 \mum\ feature after maximum light from 0.9223 \mum\ line at velocities of 10,000--12,000 \kms.  But the \ion{S}{1} 1.0457 \mum\ line, which is only 3 times weaker than the 0.9223 \mum\ line, is not detected.

\paragraph{\ion{S}{2}} is not detected in these data but all of the strongest lines for this ion are on the blue side of 0.83 \mum\ so their Doppler shifted positions will not be included in the SXD spectra that comprise the majority of spectra (and highest resolution) in this group.

\paragraph{\ca\ IR-triplet} dominates the spectra in this group, forming the 0.82 \mum\ feature with velocities near 12,000 \kms\ for most SNe in our sample. The only exception is the spectrum from SN 2004bv (-4d) which may be a SN 1991T like object in which the \ca\ is absent from early spectra but present, as in spectra from normal SNe Ia, at later times.  The spectrum in this sample from SN 2004bv at +60d shows the \ca\ and seems to be normal.   

The square bottom of the \ca\ absorption region in these features is due to separately resolved contributions from two parts of the triplet. On the blue side of the absorption feature, we use a mean wavelength of 0.8538 \mum\ for the blend, weighted by oscillator strength, for the pair of lines at 0.8498 and 0.8542 \mum.  The 0.8662 \mum\ line can be easily identified in these spectra as a separate part of the feature at about one half the predicted strength for the 0.8538 \mum\ line.  To measure Doppler velocities, we use the location of the minimum at the blue side of the feature and compare it to the 0.8538 \mum\ rest wavelength of the blend.  

The shape of the emission region forming the P-Cygni profile apparently reflects the shape of the absorption region.  Flat or round shapes in the absorption component correspond to similar but inverted profiles in the emission component.  

\paragraph{High Velocity \ca} The spectrum of SN 2004bw (-9d) shows two distinct minima for \ca\ with one at 13,400 \kms\ and a high-velocity component at 24,000 \kms.  Both spectra from SN 2003W (-7d and -6d) have \ca\ velocities above 22,100 \kms\ without a lower velocity component.  The spectra from SN 2002cr (-7d and -5d) appear to have a fading high velocity \ca\ component at 18,800 \kms\ and a lower velocity \ca\ absorption at 9,700 \kms.  (Both spectra from SN 2002cr exhibit lower velocities for all measured features than other spectra at similar epochs.)  SN 2000dn (-6d) also has a weaker \ca\ feature near 22,000 \kms\ and a strong \ca\ absorption near 10,400 \kms.  All LRS spectra obtained at -5d or earlier appear to have high velocity \ca\ components with velocities measured between 18,800--24,000 \kms.  Since the high-velocity features have rounded bottoms, we determine the Doppler velocities for the high-velocity \ca\ feature using 0.8579 \mum\ as the mean wavelength for a blend of all three strong lines in the \ca\ IR-triplet. 

\paragraph{Other \ca} may be detected in some spectra near 1.14 \mum\ from the 1.1876 \mum\ line at velocities of 12,000--16,300 \kms.  The spectra in which this feature is the most obvious also have the flattest tops to the IR-triplet emission and that could be due to the next strongest \ca\ line at 0.8921 \mum\ at similar velocities.  The feature attributed to \mg\ near 0.97 \mum\ may have a contribution from \ca\ at 0.9906 \mum\ but the velocity would be $\sim$ 8,000 \kms\ which is low. The weak absorption feature near 0.97 \mum\ often exhibits a double or triple bottom with minima near 0.95, 0.97 and 0.99 \mum\ with 0.97 being the strongest.  The 0.95 and 0.99 \mum\ features are smaller than the one sigma noise but are present in most spectra.  An absorption near 0.95 \mum\ created by \ca\ at 0.9906 \mum\ would have a Doppler velocity of 12,300 \kms. 

\paragraph{\mn} produces no unambiguous detections but the \mn\ lines at 0.7414 and 0.7342 \mum\ are a possible source for the feature found near 0.70 \mum\ in LRS spectra.  This identification would require velocities of 14,000-17,000 \kms\ which are higher than those found for most of the other species.  The strong \mn\ line at 0.9406 \mum\ may contribute to the feature near 0.91 \mum\ either alone or in a blend with \si\ at velocities that are $\sim$ 300 \kms\ less than values given for \si\ at 0.9413 in Table \ref{vtable}.  The 0.8410 line is also possible in the blend dominated by the \ca\ IR-triplet.

\paragraph{The steep increase in flux found near 1.55 \mum} is used by some models to define the transition from partial to complete silicon burning.  This feature is suggested in the spectra from this group as early as +0d but it becomes distinctive by +4d.

\paragraph{\ion{Fe}{2}} at 0.9998 \mum\ probably contributes to the 0.97 \mum\ absorption that we have associated with \mg\ (1.0092 \mum).  The fact that this feature becomes stronger after max light suggests that the primary source for this absorption is shifting from \mg\ to \fe.  Doppler velocities for the \fe\ association with this feature begin above 10,000 at -4d and follows as the photosphere recedes into regions with lower expansion velocities to about 7,000 \kms\ +2d.  

 The velocities required to match the 0.7712 \mum\ \fe\ line with the 0.75 \mum\ absorption feature in these spectra are inconsistent with \fe\ velocities measured for the 0.9998 \mum\ line and with other species found in this group.  There is no evidence of \fe\ from lines at  0.7462 or 1.0500 \mum.  The 1.0863 \mum\ line may contribute to the 1.05 \mum\ feature in a blend with \mg\ but this line is expected to be $\sim$ 20 times weaker than some of the other \fe\ lines that are not detected.

Fe-group lines are complicated by many potential transitions at the same wavelength and nearly the same strength within each multiplet so the simple formula we have been using to estimate line strength may not be accurate. In this group, the 0.9998 \mum\ line is the only \fe\ line detected, although the calculated line strengths suggest that lines at 0.7462, 0.7712, and 1.0500 \mum\ should also be detected.

\paragraph{\ion{Co}{2}} probably contributes to the 1.60 \mum\ feature with numerous \ion{Co}{2} lines between 1.57--1.77 \mum. In a few spectra (SNe 2002hw (-1d), 2000dm (+1d), and 2005am (+2d)) the strongest \ion{Co}{2} line (1.7772 \mum) may be separately resolved at about 11,000 \kms.

\paragraph{Fe-group blends} begin to affect the spectra in this group as line-blanketing from thousands of Fe-Group lines results in pseudo-emission features found near 1.52 \mum\ and between 1.6--2.1 \mum.  Model predictions for this effect are compared to the multiple spectra obtained from SN 2005am in Figure \ref{am_bumps} and discussed in Section \ref{phlfr}.  Broad humps, that we attribute to unresolved Fe-group lines raising the continuum, are also observed in the spectra from 1.6-2.1 \mum\ and from 2.1-2.3 \mum.  The bump near 1.25 \mum\ is very strong early in this group (possibly \ion{Si}{3}), weaker through the middle and strong again in the later spectra.  These features appear in all spectra from this group.

\subsection{The Extended Photospheric Phase (Intermediate Layers)\\
(+4d to +24d with respect to $V_{max}$)}
\label{fe_featanal}

During the phase represented by this group, the shape of the continuum becomes influenced by line-blanketing from Fe-group lines which extends the effective photospheric radius (see \S \ref{phlfr}).  Discussion in this section refers to Figure \ref{fe_fig} showing ten spectra from SNe Ia obtained between +4d and +24d with respect to to $V_{max}$ and also to Figure \ref{fe_fwl} that displays four representative spectra from SNe 2000dk (+5d), 2001bg (+10d), 2004ab (+18d), and 2004da (+24d) with the features labeled for which we have confident identifications.   

\paragraph{\oi} produces the strong absorption near 0.75 \mum\ in the LRS spectrum of SN 2002ha (+8d) and this line may also be responsible for the small features at this wavelength up to +18d.  The 0.9263 \mum\ line from \oi\ is expected to be 3--17 times weaker than the 0.7773 \mum\ line but it is likely to be a part of the 0.89 \mum\ feature in a blend with \ion{S}{1} at least early in the epoch represented by this group. The measured Doppler velocities for the strongest detections of \oi\ are all 9,800 \kms\ or higher.

Alternative explanations for the 0.75 \mum\ feature are \fe\ or \mn, but the Doppler velocities for these lines do not fit velocities measured for other possible lines from these ions.  

\paragraph{\mg} is found in this only in the spectrum of SN 2000dk at +5d which is the oldest spectrum int the sample to show strong \mg\ absorption.  \mg\ is the primary contributor forming the distinctive notch near 1.60 \mum. This spectrum also has strong \mg\ absorption from the 0.9227 and 1.0927 \mum\ lines.  The 1.0092 \mum\ \mg\ line is expected to be 2 orders of magnitude weaker than 1.0927 \mum\ but at +5d or earlier it may contribute to the absorption near 0.97 \mum\ either alone or in a blend with \fe. 

\paragraph{\ion{Si}{1}} is not detected in our data although this ion produces many strong lines in the NIR including lines at 1.2032, 1.0827, and 1.0585 \mum.  These lines may help form the features found near 1.03 and 1.06 \mum\ in the oldest spectrum from this group (SN 2004da at +24d).

\paragraph{\si} is possible with the 0.9413 \mum\ line contributing to the 0.92 \mum\ absorption found in some spectra in this group.  The \si\ 1.6930 \mum\ line may also be part of the 1.60 \mum\ feature.  The 1.1737 \mum\ line from \si\ is not detected so it is unlikely that the 1.1311 \mum\ line is responsible for the feature found near 1.12 \mum\ in some of the spectra.  The 1.12 \mum\ feature does not occur in the same spectra as the 0.92 \mum\ feature which eliminates the possible association of the 1.12 \mum\ feature with \si.

\paragraph{\ion{S}{1}} at 0.9223 \mum\ contributes to the 0.90 \mum\ feature in all spectra in this group and probably replaces \mg\ as the primary source of the absorption after +5d post-maximum. Velocities for this ion begin at 10,300--13,700 \kms\ at +0d and diminish to 8,900 at +18d.  After +10 days, some spectra do not have any absorption feature at this location.  The 1.3809 \mum\ \ion{S}{1} line is too far into the region of high atmospheric opacity to be detected.  Other possible associations can be made for \ion{S}{1} at 0.9423 and 1.1349 \mum, but they both have serious questions.  The feature found near 1.10 \mum\ is possibly associated with the 1.1349 \mum\ line and the velocity shift is plausible of 8,500 at +5d and 7,000 at +18d.  However this feature is near an order transition in SXD spectra which casts suspicion on any features at this location from SXD spectra (in particular the spectrum from SNe 2000dk (+5d)).  Since LRS spectra do not have order transitions, the weak feature near 1.1 \mum\ is more likely to be real in the LRS spectra.  The \ion{S}{1} 0.9423 \mum\ line may also produce features in a few spectra, but the 0.92 \mum\ absorption feature does not appear in the same spectra that produce other probable \ion{S}{1} identifications.  
 
\paragraph{\ca\ IR triplet} velocities remain well above 10,000 \kms\ for the spectra in this group.  Separate features can be detected in the spectra from this group for both the 0.8358 \mum\ blend and the 0.8662 \mum\ line as described in the previous section.  The shape of the absorption feature is consistent with estimates that the 0.8538 \mum\ blend is about twice as strong as the 0.8662 \mum\ line.  

\paragraph{Other \ca} is not detected from the lines at 1.1876 or 0.9906 \mum\ or from other \ca\ line. The weak features found in some spectra near 1.10 \mum\ are possibly associated with the 1.1876 \mum\ line but the velocity  would be more than 21,000 \kms\ which is inconsistent with other photospheric velocities during this phase.  That makes it unlikely that the \ca\ line at 0.9931 \mum, which is estimated to be $\sim 20$ times weaker, provides a contribution to the absorption found near 0.97 \mum.

\paragraph{\ion{Mn}{2}} at 0.9447 \mum\ is a possible detection in the spectra from this group that contain features near 0.92\mum .  This feature may be attributed to either \mn\ or \si\ alone or in a blend of both.  A rest wavelength of 0.9447 \mum\ would produce velocities of 8,500--12,500 \kms\ for \mn, or about 1000 \kms\ greater than velocities listed for \si\ at 0.9413 \mum\ in Table \ref{vtable}. However \mn\ produces more than 90 strong lines between 0.9309--0.9509 \mum\ so it is not possible to measure a definitive velocity.  These lines do not appear in Tables \ref{5K} or \ref{10K} because they are slightly weaker than the 0.9447, 0.9388, and 0.9486 \mum\ lines that do appear in the tables.   The strongest \mn\ lines at 0.7414 and 0.7342 \mum\ may form the features found with a similar velocity separation near 0.72 \mum, but there are only two spectra with which to compare these lines.  An alternative identification for those features is \fe.

\paragraph{The steep increase in flux found near 1.55 \mum} may define the transition from partial to complete silicon burning.  This feature is present in all spectra from this group but it is less well defined in the spectra from SNe 2005am (+14d) and 2004da (+24d).  This does not appear to be an evolutionary effect because the features is present in many later spectra.

\paragraph{\ion{Fe}{2}} at 0.9998 \mum\ is the likely source of the absorption found near 0.97 \mum\ in all spectra from this group through +18d.  Before +5d (see the previous section) this feature may receive a contribution from \mg\ at 1.0092 \mum, but the absorption is strongest from +5d to +10d when it is likely to be entirely due to \fe.  Velocities for \ion{Fe}{2} measured from the 0.9998 \mum\ line are near 9,000 \kms\ at +0d and diminish to less than 3,000 by +18d. \fe\ is also likely to be the source of absorptions near 0.75 and 0.72 \mum\ after +10d from the two strong lines at 0.7712 and 0.7462 \mum\ at about 8,500 \kms\ in the spectra from SNe 2004bk (+12) and 2002ef (+16). The \fe\ line from 0.9998 \mum\ appears to diminish in strength after about +10d but it becomes strong again about one month post-maximum.

\paragraph{\ion{Co}{2}} forms distinct features in the later spectra in this group. Spectra obtained at +10d or later show distinct absorption at the location of many \co\ lines, most prominently: 1.7772, 1.7462, 2.2205, 2.4596, and 2.3613 \mum\ at velocities near 9,000 \kms\ (see Figure \ref{trans_coii}).  Possible identification of \co\ lines from 1.7772 and 1.7462 \mum\ can be made in the spectrum from SN 2000dk as early as +5d.

\subsection{Transitional Phase (Inner Layers)\\
(+37d to +75d with respect to $V_{max}$)}
\label{transs_featanal}

This epoch is a transitional phase between the photospheric era and a true nebular phase.   Discussion in this section makes reference to Figure \ref{transfig} showing seven spectra from SNe Ia obtained between +37d and +75d with respect to $V_{max}$ and Figure \ref{trans_coii} which includes 13 spectra from our sample and two additional spectra from other sources.   In Figure \ref{trans_coii} the positions of strong \co\ and \fe\ lines are marked so that they correspond to 11,000 \kms\ at the top spectrum and 6,000 \kms\ at the bottom spectrum.  This reflects to the reduction in measured velocities for these spectra.  The data plotted in yellow (SNe 1998aq and 1998bu) were obtained by \citeauthor{Rudy} and have not been previously published.  \co\ lines are easily detected in absorption as early as two weeks post-maximum and \fe\ lines (in addition to the 0.9998 \mum\ which appears as early as +0d) begin to appear about one month post-maximum (see \S \ref{fe} and \ref{transs}). 

This group displays a lot of new structure relative to earlier NIR spectra and the features are remarkably similar from one spectrum to the next.  The spectrum of SN 2004bv at +60d fits well with the others in this group although the spectrum from this SN at -5d was unusual (see Figure \ref{mg_fig}).  This is the same time frame in which unreddened SNe Ia have a very low dispersion for B-V colors (Lira, 1995). The common assumption about this era is that expansion has reduced the opacity beyond the point where a distinct photosphere can be defined and individual absorption features can no longer be produced.  Recent work by \citet{Branch08} suggests spectra from this epoch can
 be modeled by absorption features from permitted lines.  

\paragraph{\ca\ IR triplet} velocities remain near 10,000 \kms\ which is consistent with this feature being produced by \ca\ that was not synthesized in the explosion (see \S \ref{phlfr}).  Separate features can be detected in many spectra from this group for both the 0.8358 \mum\ blend and the 0.8662 \mum\ line.  

\paragraph{The steep increase in flux found near 1.55 \mum} may define the transition from partial to complete silicon burning.  This feature is present is all spectra from this group but is more sharply defined in some spectra compared to others.  As shown in Figure \ref{transfig}, the spectrum from SN 2004bv at +60 displays this feature but it is smeared out in the spectrum of SN 2001fe, also obtained at +60d.  

\paragraph{\ion{Fe}{2}} lines at 0.7712, 0.9998, and 1.0863 \mum\ appear in absorption beginning about 30 days after maximum light and in all spectra in this group (Figure \ref{trans_coii}).  The \ion{Fe}{2} 0.9998 \mum\ line is present is spectra from about +0d to +18d, it is not detected from +18d to +24d, and then reappears after +30d (see Figure \ref{gh}).  No other detections from individual \ion{Fe}{2} lines are reliable from NIR spectra prior to this later phase.

\paragraph{\co} lines are clearly detected in wavelength regions where a pseudo photosphere is created at extended radii (1.5--1.85 and 2.05--2.45 \mum; see \S \ref{phlfr}) and Figure \ref{trans_coii}.  Strong absorption features are detected from many \co\ lines.  The LRS spectra from SNe 2003cd (+45) and 2004bv (+60) as shown in Figures \ref{transfig} only marginally resolve \co\ absorptions for 1.7772, 1.7462, 2.2205, 2.4596, and 2.3613 lines and the group of \co\ lines from 1.57--1.67 are not resolved at all.  Figure \ref{trans_coii} shows how the locations of these lines shift to longer wavelengths with increasing time of observation relative to maximum light.  The positions of strong \co\ and \fe\ lines are drawn on the figure so that the indicated velocities are 11,000 \kms\ at the top spectrum and 6,000 \kms\ at the bottom spectrum.

\subsection{Summary of Line Identification by Ion for the Entire Sample}
\label{idbyion}
Velocity references are compiled in Table \ref{vtable} and Figure \ref{vplot}.

\paragraph{\ion{He}{1}} is not detected from either of the strong lines at 1.0830 or 2.0581 \mum.  Unrealistically low Doppler velocities are required to align the 1.0830 \mum\ line with the absorption feature found near 1.05 \mum.  The 2.0581 line (which is expected to be 6-12 times weaker than the 1.083 line depending on temperature), is not observed.

\paragraph{\ci} is not detected.  Speculative identifications can be made for lines at 0.9406, 1.1331, 1.1756 and 1.6890 \mum\ in the very early spectra, but if these lines produce absorption features, the much stronger lines at 1.0693 and 0.9087 should also be detected and they are not.  At each location of possible \ci\ detection there are plausible alternative identifications (see Figure \ref{noci}).

\paragraph{\oi} forms a strong feature from the 0.7773 \mum\ line producing an absorption near 0.75 \mum.  This feature is found in all LRS spectra up to +10d and possibly as late as +18d (SXD spectra lack the wavelength coverage to include this feature).  \oi\ is also likely as a contributor to the 0.89 feature from the line at 0.9264 \mum.  Doppler velocities are 9,300--15,700 \kms\ with most measurements near 11,000 \kms. 

\paragraph{\ion{C}{2} \& \ion{O}{2}} are not detected.  All \ion{C}{2} lines in the NIR are factors of $10^2 - 10^4$ weaker than the \ion{C}{2} at 0.6580 \mum\ line which has been tentatively identified in a few optical spectra.  

\paragraph{\mg} is found in all spectra between -14d and +5d, producing strong features from lines at 0.9227 and 1.0927 \mum.  Velocities are 9,800--14,700 \kms\ with a few anomalies near 9,000 \kms.  Measured velocities for the 0.9227 line are $\approx 1000$ \kms\ slower than for the 1.0927 \mum\ line as a result of the enormous P-Cygni emission from \ca\ impinging on the absorption feature.  Many spectra also contain features from weaker \mg\ lines at 1.0092 and 1.6787 \mum.

\paragraph{\ion{Si}{1}} is not detected although some spectra exhibit weak features that could possibly be attributed to this ion from lines at 1.2032, 1.0827, or 1.0585 \mum.  

\paragraph{\si} from the 0.9413 \mum\ line produces the feature near 0.91 \mum\ at velocities from 10,200--14,100 \kms\ between -14d and +5d.  After +5d the feature appears intermittently as late as +18d, but this may be a blend with \mn\ or entirely due to \mn. The \si\ 1.6930 line contributes to the 1.60 feature.  The 1.1737 line is not detected so it is unlikely that the 1.1311 line is responsible for the feature found near 1.12 in some of the spectra.  The 1.12 feature does not occur in the same spectra as the 0.91 feature so the \si\ identification is unlikely.

\paragraph{\ion{Si}{3}} is possible in the earliest spectra from the 0.9324 \mum\ line at the very low velocity of 8,500 \kms.  Other possible identifications for this ion are the feature at 1.22 (from 1.2541, a blend of 1.2523 and 1.2601, at 9,600 \kms) and 1.09 (from 1.1341 at 10,500 \kms).  \ion{Si}{3} can be present in the same spectrum with \si\ if the photosphere is deep within the envelope so that temperature and pressure differences between absorbing regions near the photosphere and in the outer layers contain different ionizations of the same element (\S \ref{phlfr}).  

\paragraph{\ion{S}{1}} at 0.9223 \mum\ replaces \mg\ as the primary source of the 0.90 feature after maximum light. Velocities begin at 10,300--13,700 \kms\ at +0d and diminish to 8,900 at +18d.  After +10 days, there are a few spectra with no feature at this location.  Other possible associations can be made for \ion{S}{1} at 1.3809, 0.9423, and 1.1349, but they all have serious questions.   

\paragraph{\ca\ IR triplet} is the strongest feature in most spectra.  Velocities are greater than 10,000 \kms\ for most spectra, although there are exceptions such as SN 2001bf (+4d) for which the velocity is 7,600 \kms\ and this feature is very weak in the spectrum from SN 2004bv at -5d.  SN 2001bf has very low velocities for all identified lines and SN 2004bv may be 91T-like.  Distinct individual features can be detected in most of the SXD spectra for both the 0.8358 blend and the 0.8662 line.  The shape of the absorption feature is consistent with estimates that the 0.8538 blend is $\sim$ twice as strong as the 0.8662 line.  Very low excitation values for these lines suggest that absorption features from the \ca\ IR triplet may not come from freshly synthesized material.

\paragraph{High Velocity \ca} All LRS spectra obtained at -5d or earlier appear to have high velocity \ca\ components with velocities measured between 18,800--24,000 \kms.

\paragraph{Other \ca} is not detected from the next strongest lines at 1.1876 or 0.9906 \mum\ or from other \ca\ lines in any spectrum in our sample. \ca\ appears to be absent from the freshly synthesized explosion products.

\paragraph{\ion{Mn}{2}} at 0.9447 \mum\ is possibly detected after maximum light.  The strongest \mn\ lines at 0.7414 and 0.7342 \mum\ form features at similar velocities near 0.72 \mum, but there are only two spectra with which to compare this line.

\paragraph{\ion{Fe}{2}} is detected in post-maximum spectra from the 0.9998 \mum\ line at velocities near 9,500 \kms\ at +0d, diminishing to less than 3,000 by +18d. This line may contribute to the feature in pre-maximum spectra attributed to the \mg\ line at 1.0092 \mum.  \fe\ is likely to be the source of absorptions near 0.75 and 0.72 after +10d.    Fe lines are more difficult to compare based on our simple line strength estimates because the actual line strengths are affected by multiple transitions. As a consequence, the simple calculations in our tables for the relative estimated strength of each \ion{Fe}{2} line may be less accurate than the estimates for smaller atoms.  In spectra from our sample obtained before +37d, the 0.9998 \mum\ line is the only \fe\ line detected, although the calculated line strengths suggest that lines at 0.7462, 0.7712, 1.0500, and 1.0863 \mum\ lines should also be detected.

\paragraph{\ion{Co}{2}} may be detected soon after maximum light in some of the high resolution spectra.  After two weeks post maximum, \co\ lines are clearly detected at velocities beginning near 10,000 \kms\ and diminishing to 6,000 \kms.  \co\ is found in wavelength regions where a pseudo photosphere is created at extended radii (1.5--1.85 and 2.05--2.45 \mum).  \co\ is most obvious in the higher resolution (SXD) spectra and individual \co\ lines are marginally resolved in many other later time spectra.

\clearpage
\begin{deluxetable}{ccccccc}
\tabletypesize{\scriptsize}
\tablecolumns{7} \tablewidth{0pc}
\tablecaption{Comparing the Effects of Fourier Transform Smoothing Parameters \label{fttable}}
\tablehead{\colhead{Smoothing} & 
\colhead{Minimum (\mum)} & 
\colhead{Velocity} & 
\colhead{Line Depth} & 
\colhead{Minimum (\mum)} & 
\colhead{Velocity}& 
\colhead{Line Depth} \\ 
\colhead{Parameter} & 
\colhead{$\lambda_{rest}=1.0927$ \mum} &
\colhead{\kms} & 
\colhead{normalized} & 
\colhead{$\lambda_{rest}=0.9227$ \mum} &
\colhead{\kms} & 
\colhead{normalized} }

\startdata
39 &  1.0463 & 12,700 & 1.00 & 0.8859 & 11,900 & 1.00 \\
32 &  1.0467 & 12,500 & 0.99 & 0.8864 & 11,800 & 0.99 \\
27 &  1.0474 & 12,400 & 0.97 & 0.8866 & 11,700 & 0.96 \\
22 &  1.0479 & 12,300 & 0.95 & 0.8873 & 11,500 & 0.82 \\
16 &  1.0479 & 12,300 & 0.87 & 0.8894 & 10,800 & 0.86 \\
11 &  1.0471 & 12,500 & 0.78 & 0.8921 &  9,900 & 0.82 \\

\enddata

\tablecomments{These data are plotted in Figure \ref{ft3} and discussed in \S \ref{ftx3}.}

\end{deluxetable}

\clearpage

\begin{deluxetable}{lclrccccc}
\rotate
\tabletypesize{\footnotesize}
\tablecolumns{9} \tablewidth{0pc}
\tablecaption{List of Spectra by Estimated Epoch Relative to $V_{max}$ \label{snelist_epoch}}
\tablehead{\colhead{SN} &
\colhead{Est. Epoch} &
\colhead{Obs.} &
\colhead{Avg.} &
\colhead{S/N} &
\colhead{S/N} &
\colhead{S/N}&
\colhead{S/N}&
\colhead{Redshift}\\
\colhead{Name} &
\colhead{wrt $V_{max}$} &
\colhead{Date (UT)} &
\colhead{Res.} &
\colhead{$0.95-1.10\mu$m} &
\colhead{$1.10-1.30\mu$m} &
\colhead{$1.45-1.70\mu$m}  &
\colhead{$1.95-2.35\mu$m} &
\colhead{Host (z)}}

\startdata
2002fk & -14d & Sep 19.4 &  1200 &  74 &  67 &  47 &  25 &  0.007125 \\
2004bw & -9d & May 30.4 &  200 &  74 &  54 &  17 &  12 &  0.021408 \\
2003W & -8d & Feb 2.4 &  200 &  104 &  29 &  47 &  22 &  0.020071 \\
2003W & -7d & Feb 3.4 &  200 &  144 &  132 &  41 &  25 &  0.020071 \\
2002cr & -7d & May 8.4 &  200 &  153 &  133 &  77 &  40 &  0.009590 \\
2000dn & -6d & Oct 1.5 &  120 &  37 &  34 &  7 &  3 &  0.032019 \\
2004bv & -5d & May 30.5 &  1200 &  91 &  69 &  54 &  37 &  0.010544 \\
2002cr & -5d & May 10.3 &  200 &  53 &  37 &  11 &  4 &  0.009590 \\
2005am & -4d & Mar 5.3 &  1200 &  20 &  21 &  18 &  12 &  0.007889 \\
2001en & -4d & Oct 8.4 &  1200 &  15 &  19 &  10 &  7 &  0.015871 \\
2002el & -4d & Aug 20.5 &  750 &  6 &  7 &  4 &  4 &  0.023300 \\
2001br & -2d & May 22.6 &  750 &  13 &  18 &  9 &  7 &  0.020628 \\
2004bl & -2d & May 8.3 &  1200 &  7 &  7 &  4 &  3 &  0.017319 \\
2001dl & -1d & Aug 12.5 &  750 &  9 &  12 &  7 &  5 &  0.020694 \\
2002hw & -1d & Nov 14.3 &  1200 &  12 &  12 &  6 &  4 &  0.017535 \\
2005am & +0d & Mar 9.4 &  1200 &  30 &  32 &  22 &  16 &  0.007889 \\
2000dm & +0d & Oct 1.2 &  750 &  15 &  21 &  10 &  9 &  0.015304 \\
2003du & +1d & May 9.3 &  1200 &  47 &  31 &  24 &  15 &  0.016764 \\
2001bf & +4d & May 21.5 &  1200 &  50 &  51 &  36 &  29 &  0.015501 \\
2000do & +4d & Oct 2.2 &  750 &  15 &  10 &  7 &  6 &  0.010864 \\
2004da & +4d & Jul 8.5 &  750 &  8 &  8 &  6 &  5 &  0.016308 \\
2000dk & +5d & Oct 1.5 &  750 &  19 &  14 &  11 &  7 &  0.017439 \\
2005am & +7d & Mar 16.4 &  750 &  3 &  2 &  2 &  3 &  0.007889 \\
2002ha & +8d & Nov 14.2 &  200 &  60 &  23 &  13 &  10 &  0.014046 \\
2001bg & +10d & May 22.2 &  1200 &  18 &  10 &  9 &  7 &  0.007122 \\
2002ef & +11d & Aug 20.6 &  200 &  22 &  7 &  5 &  2 &  0.023977 \\
2005am & +14d & Mar 23.3 &  1200 &  26 &  15 &  19 &  13 &  0.007889 \\
2004da & +14d & Jul 18.4 &  750 &  4 &  2 &  5 &  5 &  0.016308 \\
2004ab & +18d & Mar 7.5 &  750 &  41 &  27 &  45 &  36 &  0.005807 \\
2001en & +18d & Oct 30.3 & 200 &  45 &  21 &  18 &  10 &  0.015871 \\
2004da & +18d & Jul 22.5 &  750 &  35 &  15 &  30 &  23 &  0.016308 \\
2004bk & +19d & May 8.5 &  200 &  12 &  8 &  2 &  0 &  0.023036 \\
2004da & +24d & Jul 28.4 &  750 &  18 &  10 &  15 &  10 &  0.016308 \\
2004E & +35d & Feb 21.5 &  200 &  42 &  13 &  6 &  1 &  0.029807 \\
2003cg & +37d & May 9.2 &  200 &  66 &  56 &  80 &  47 &  0.004130 \\
2001gc & +40d & Jan 14.5 & 200 & 28 & 12 & 8 & 6 &  0.019290 \\
2002fk & +42d & Nov 14.4 &  750 &  37 &  13 &  15 &  10 &  0.007125 \\
2004ca & +46d & Jul 28.6 &  200 &  56 &  24 &  13 &  5 &  0.017806 \\
2004bv & +53d & Jul 28.3 &  200 &  63 &  45 &  38 &  16 &  0.010544 \\
2001fe & +60d & Jan 14.6 & 120 & 45 & 15 & 12 & 6 &  0.013539 \\
2003du & +75d & Jul 22.3 &  200 &  41 &  7 &  8 &  2 &  0.016764 \\
\enddata

\end{deluxetable}

\clearpage

\begin{deluxetable}{lclrccccc}
\rotate
\tabletypesize{\footnotesize}
\tablecolumns{9} \tablewidth{0pc}
\tablecaption{List of Spectra in Order of Discovery \label{snelist_disc}}
\tablehead{\colhead{SN} &
\colhead{Est. Epoch} &
\colhead{Obs.} &
\colhead{Avg.} &
\colhead{S/N} &
\colhead{S/N} &
\colhead{S/N}&
\colhead{S/N}&
\colhead{Redshift}\\
\colhead{Name} &
\colhead{wrt $V_{max}$} &
\colhead{Date (UT)} &
\colhead{Res.} &
\colhead{$0.95-1.10\mu$m} &
\colhead{$1.10-1.30\mu$m} &
\colhead{$1.45-1.70\mu$m}  &
\colhead{$1.95-2.35\mu$m} &
\colhead{Host (z)}}

\startdata
2000dk & +5d & Oct 1.5 &  750 &  19 &  14 &  11 &  7 &  0.017439 \\
2000dm & +0d & Oct 1.2 &  750 &  15 &  21 &  10 &  9 &  0.015304 \\
2000dn & -6d & Oct 1.5 &  120 &  37 &  34 &  7 &  3 &  0.032019 \\
2000do & +4d & Oct 2.2 &  750 &  15 &  10 &  7 &  6 &  0.010864 \\
2001bf & +4d & May 21.5 &  1200 &  50 &  51 &  36 &  29 &  0.015501 \\
2001bg & +10d & May 22.2 &  1200 &  18 &  10 &  9 &  7 &  0.007122 \\
2001br & -2d & May 22.6 &  750 &  13 &  18 &  9 &  7 &  0.020628 \\
2001dl & -1d & Aug 12.5 &  750 &  9 &  12 &  7 &  5 &  0.020694 \\
2001en & -4d & Oct 8.4 &  1200 &  15 &  19 &  10 &  7 &  0.015871 \\
2001en & +18d & Oct 30.3 & 200 &  45 &  21 &  18 &  10 &  0.015871 \\
2001fe & +60d & Jan 14.6 & 120 & 45 & 15 & 12 & 6 &  0.013539 \\
2001gc & +40d & Jan 14.5 & 200 & 28 & 12 & 8 & 6 &  0.019290 \\
2002cr & -7d & May 8.4 &  200 &  153 &  133 &  77 &  40 &  0.009590 \\
2002cr & -5d & May 10.3 &  200 &  53 &  37 &  11 &  4 &  0.009590 \\
2002ef & +11d & Aug 20.6 &  200 &  22 &  7 &  5 &  2 &  0.023977 \\
2002el & -4d & Aug 20.5 &  750 &  6 &  7 &  4 &  4 &  0.023300 \\
2002fk & -14d & Sep 19.4 &  1200 &  74 &  67 &  47 &  25 &  0.007125 \\
2002fk & +42d & Nov 14.4 &  750 &  37 &  13 &  15 &  10 &  0.007125 \\
2002ha & +8d & Nov 14.2 &  200 &  60 &  23 &  13 &  10 &  0.014046 \\
2002hw & -1d & Nov 14.3 &  1200 &  12 &  12 &  6 &  4 &  0.017535 \\
2003W & -8d & Feb 2.4 &  200 &  104 &  29 &  47 &  22 &  0.020071 \\
2003W & -7d & Feb 3.4 &  200 &  144 &  132 &  41 &  25 &  0.020071 \\
2003cg & +37d & May 9.2 &  200 &  66 &  56 &  80 &  47 &  0.004130 \\
2003du & +1d & May 9.3 &  1200 &  47 &  31 &  24 &  15 &  0.016764 \\
2003du & +75d & Jul 22.3 &  200 &  41 &  7 &  8 &  2 &  0.016764 \\
2004E & +35d & Feb 21.5 &  200 &  42 &  13 &  6 &  1 &  0.029807 \\
2004ab & +18d & Mar 7.5 &  750 &  41 &  27 &  45 &  36 &  0.005807 \\
2004bk & +19d & May 8.5 &  200 &  15 &  8 &  2 &  0 &  0.023036 \\
2004bl & -2d & May 8.3 &  1200 &  7 &  7 &  4 &  3 &  0.017319 \\
2004bw & -9d & May 30.4 &  200 &  74 &  54 &  17 &  12 &  0.021408 \\
2004bv & -5d & May 30.5 &  1200 &  91 &  69 &  54 &  37 &  0.010544 \\
2004bv & +53d & Jul 28.3 &  200 &  63 &  45 &  38 &  16 &  0.010544 \\
2004ca & +46d & Jul 28.6 &  200 &  56 &  24 &  13 &  5 &  0.017806 \\
2004da & +4d & Jul 8.5 &  750 &  8 &  8 &  6 &  5 &  0.016308 \\
2004da & +14d & Jul 18.4 &  750 &  4 &  2 &  5 &  5 &  0.016308 \\
2004da & +18d & Jul 22.5 &  750 &  35 &  15 &  30 &  23 &  0.016308 \\
2004da & +24d & Jul 28.4 &  750 &  18 &  10 &  15 &  10 &  0.016308 \\
2005am & -4d & Mar 5.3 &  1200 &  20 &  21 &  18 &  12 &  0.007889 \\
2005am & +0d & Mar 9.4 &  1200 &  30 &  32 &  22 &  16 &  0.007889 \\
2005am & +7d & Mar 16.4 &  750 &  3 &  2 &  2 &  3 &  0.007889 \\
2005am & +14d & Mar 23.3 &  1200 &  26 &  15 &  19 &  13 &  0.007889 \\
\enddata

\end{deluxetable}

\clearpage

\begin{deluxetable}{lrrrrrrrrrr}
\tabletypesize{\footnotesize}
\tablecolumns{11} \tablewidth{0pc}
\tablecaption{Measured Doppler Velocities for Identified Lines --
                         Blueshift (\kms)  \label{vtable}}
\tablehead{\colhead{Name} & \colhead{Epoch} & \colhead{\oi} & \colhead{\oi} & 
\colhead{\mg} & \colhead{\mg} & \colhead{\mg}& \colhead{\si}& \colhead{\ion{S}{1}} & 
\colhead{\ca}& \colhead{\ion{Fe}{2}}\\ 
  & \colhead{$V_{max}$} & \colhead{0.7773} & \colhead{0.9264} & \colhead{0.9227} & \colhead{1.0092} &
\colhead{1.0927} & \colhead{0.9413}  & \colhead{0.9223} & \colhead{0.8538}& \colhead{0.9998}}

\startdata
 2002fk & -14 &        &  11100 &   9900 &  10800 &  11300 &  10800 &        &  12600 &        \\
 2004bw &  -9 &  11300 &  13000 &  11800 &  13400 &  12500 &        &        &  13000 &  10600 \\
  2003W &  -8 &  13500 &  11000 &   9800 &  12600 &  14500 &        &        &        &   9700 \\
  2003W &  -7 &  13300 &  11000 &   9800 &  14800 &  11000 &        &        &        &  12000 \\
 2002cr &  -7 &   9300 &  11000 &   9800 &  12600 &  11100 &  11600 &        &   9800 &   9700 \\
 2000dn &  -6 &  11800 &  11300 &  10100 &  12800 &  12500 &  10400 &        &  10400 &  10000 \\
 2004bv &  -5 &        &  11600 &  10400 &  14400 &  12400 &  11500 &  10200 &   8900 &        \\
 2002cr &  -5 &   9400 &  10700 &   9500 &  12200 &  11200 &        &        &   9600 &   9400 \\
 2005am &  -4 &        &  13200 &  12000 &  13400 &  12900 &  12100 &        &  12100 &  10600 \\
 2001en &  -4 &        &  14800 &  13600 &  13300 &  13500 &  12800 &        &  11400 &  10500 \\
 2001br &  -2 &        &  15700 &  14500 &  11900 &  13700 &  14100 &        &  14200 &   9100 \\
 2002hw &  -1 &        &  11900 &  10700 &  12900 &  11900 &  10200 &        &  12200 &        \\
 2005am &  +0 &        &  13100 &  11900 &  11000 &  13100 &  10400 &  11700 &  12500 &   7700 \\
 2000dm &  +0 &        &  11600 &  10400 &   9500 &  11500 &        &  10300 &  10900 &   6700 \\
 2003du &  +1 &        &  15000 &  13800 &  13500 &  14900 &  10800 &  13700 &  14200 &        \\
 2001bf &  +4 &        &   9800 &   8600 &   9700 &   9600 &   7400 &   8500 &   7600 &   6800 \\
 2000do &  +4 &        &  11700 &  10700 &        &  11600 &   7300 &   9400 &  10400 &   3700 \\
 2000dk &  +5 &        &  10800 &   9600 &   9000 &  10800 &        &   9500 &  11600 &   2500 \\
 2002ha &  +8 &  10200 &  11300 &        &        &        &        &  10000 &  11300 &   3100 \\
 2001bg & +10 &        &  12300 &        &        &        &  10500 &  11000 &  12300 &   3100 \\
 2002ef & +11 &   9800 &        &        &        &        &        &        &  11400 &   2300 \\
 2005am & +14 &        &  10500 &        &        &        &        &   9200 &  11400 &   2900 \\
 2004ab & +18 &        &  10300 &        &        &        &   8200 &   8900 &  12200 &   2600 \\
 2002en & +18 &  12300 &        &        &        &        &   7500 &        &  13000 &        \\

\enddata
\tablecomments{These data are plotted in Figure \ref{vplot}.}
\end{deluxetable}

\clearpage

\begin{deluxetable}{crcrrrr}
\tablecolumns{7} 
\tablewidth{0pc}
\tabletypesize{\scriptsize}
\tablecaption{Estimated Line Strengths at 5,000K in the Region $0.7 - 2.4$ \mum \label{5K}}
\tablehead{\colhead{Ion} & 
\colhead{$\lambda_{rest}$(air)} & 
\colhead{No. lines} & 
\colhead{$\log(gf)$} & 
\colhead{Excitation} & 
\colhead{Line Strength} & 
\colhead{Line Strength}\\
 & 
\colhead{\mum} &
\colhead{in blend \tablenotemark{a}} & 
 &
\colhead{Min. (eV)} & 
\colhead{5,000K \tablenotemark{b}} & 
\colhead{10,000K \tablenotemark{b}}}

\startdata
\hline
  \ion{He}{1} &  1.0830 &   3 &  -0.047 &   19.82096 &     15.1 &      6.0 \\
  \ion{He}{1} &  2.0581 &   1 &  -0.424 &   20.61718 &      1.000 &      1.000 \\
  \ion{He}{1} &  0.7065 &   3 &  -0.460 &   20.96551 &      0.410 &      0.614 \\
\hline
   \ci\ &  1.0693 &   7 &   0.348 &    7.48830 &      1.000 &      1.000 \\
   \ci\ &  0.9093 &   2 &   0.142 &    7.48830 &      0.622 &      0.622 \\
   \ci\ &  0.9406 &   1 &   0.225 &    7.68529 &      0.477 &      0.599 \\
   \ci\ &  0.9658 &   1 &  -0.264 &    7.48830 &      0.244 &      0.244 \\
   \ci\ &  1.4543 &   1 &  -0.110 &    7.68529 &      0.221 &      0.277 \\
   \ci\ &  1.1754 &   2 &   0.661 &    8.64774 &      0.139 &      0.535 \\
   \ci\ &  0.8335 &   1 &  -0.420 &    7.68529 &      0.108 &      0.136 \\
   \ci\ &  1.1330 &   1 &   0.280 &    8.53767 &      0.075 &      0.253 \\
\hline
  \ion{C}{2} &  0.6580 &   2 &   0.118 &   14.44980 &     88,393 &   253 \\
  \ion{C}{2} &  0.7235 &   3 &   0.330 &   16.33422 &   1,815 &     46.3 \\
  \ion{C}{2} &  1.8905 &   2 &   0.258 &   19.49586 &      1.000 &      0.959 \\
  \ion{C}{2} &  1.7846 &   3 &   0.550 &   20.15184 &      0.427 &      0.878 \\
  \ion{C}{2} &  0.9903 &   3 &   1.010 &   20.95206 &      0.192 &      1.000 \\
  \ion{C}{2} &  0.9230 &   3 &   0.620 &   20.84618 &      0.100 &      0.461 \\
\hline
   \oi\ &  0.7773 &   3 &   0.324 &    9.14671 &      1.000 &      1.000 \\
   \oi\ &  0.8446 &   3 &   0.170 &    9.52201 &      0.294 &      0.454 \\
   \oi\ &  0.9264 &   9 &   0.690 &   10.74166 &      0.057 &      0.365 \\
   \oi\ &  1.1290 &   9 &   0.500 &   10.98960 &      0.021 &      0.177 \\
   \oi\ &  0.7990 &   6 &   0.280 &   10.98960 &      0.013 &      0.106 \\
   \oi\ &  1.3165 &   3 &  -0.095 &   10.98960 &      0.005 &      0.045 \\
\hline
  \ion{O}{2} &  0.6695 &   2 &  -0.594 &   23.44307 &      2,155 &      97.6  \\
  \ion{O}{2} &  2.1085 &   2 &  -1.690 &   25.66290 &      1.000 &      0.595 \\
  \ion{O}{2} &  1.3811 &   3 &  -1.730 &   25.66290 &      0.912 &      0.543 \\
  \ion{O}{2} &  1.4008 &   2 &   0.530 &   29.62095 &      0.017 &      1.000 \\
  \ion{O}{2} &  0.7114 &   3 &  -0.170 &   29.07061 &      0.012 &      0.378 \\
  \ion{O}{2} &  1.3029 &   2 &   0.460 &   29.82223 &      0.009 &      0.674 \\
  \ion{O}{2} &  1.2500 &   1 &   0.210 &   29.82223 &      0.005 &      0.379 \\
  \ion{O}{2} &  1.1667 &   2 &   0.870 &   30.50576 &      0.005 &      0.784 \\
\hline
  \ion{Mg}{1}&  0.8807 &   1 &  -0.137 &    4.34610 &      1.000 &      0.807 \\
  \ion{Mg}{1}&  1.1828 &   1 &  -0.290 &    4.34610 &      0.703 &      0.568 \\
  \ion{Mg}{1}&  1.5033 &   3 &   0.340 &    5.10817 &      0.511 &      1.000 \\
  \ion{Mg}{1}&  1.7109 &   1 &   0.140 &    5.39409 &      0.166 &      0.453 \\
  \ion{Mg}{1}&  1.4878 &   9 &   0.660 &    5.94632 &      0.153 &      0.790 \\
  \ion{Mg}{1}&  1.2083 &   4 &   0.450 &    5.75363 &      0.147 &      0.609 \\
\hline
 \mg\ &  0.9227 &   2 &   0.270 &    8.65529 &      1.000 &      1.000 \\
 \mg\ &  1.0927 &   3 &   0.020 &    8.86425 &      0.346 &      0.441 \\
 \mg\ &  0.7890 &   3 &   0.650 &   10.00000 &      0.106 &      0.504 \\
 \mg\ &  0.8228 &   2 &   0.030 &   10.00000 &      0.025 &      0.121 \\
 \mg\ &  1.0092 &   3 &   1.020 &   11.63047 &      0.006 &      0.178 \\
 \mg\ &  0.9632 &   3 &   0.660 &   11.56982 &      0.003 &      0.083 \\
 \mg\ &  2.1369 &   1 &   0.390 &   11.50533 &      0.002 &      0.048 \\
 \mg\ &  1.6787 &   3 &   0.730 &   12.08542 &      0.001 &      0.054 \\
\hline
  \ion{Si}{1} &  1.2032 &   1 &   0.440 &    4.95413 &      1.000 &      1.000 \\
  \ion{Si}{1} &  1.0827 &   1 &   0.220 &    4.95413 &      0.603 &      0.603 \\
  \ion{Si}{1} &  1.0585 &   1 &  -0.020 &    4.95413 &      0.347 &      0.347 \\
  \ion{Si}{1} &  1.5888 &   1 &  -0.030 &    5.08269 &      0.251 &      0.292 \\
\hline
 \si\ &  0.6355 &   2 &   0.297 &    8.12157 &   10,126 &       49.6 \\
 \si\ &  1.6930 &   2 &   0.350 &   12.14781 &      1.000 &      0.524 \\
 \si\ &  0.9413 &   3 &   0.980 &   12.84018 &      0.855 &      1.000 \\
 \si\ &  0.7849 &   3 &   0.490 &   12.52627 &      0.573 &      0.466 \\
 \si\ &  1.1737 &   3 &   0.620 &   12.88100 &      0.340 &      0.416 \\
 \si\ &  1.3681 &   2 &   0.130 &   12.88100 &      0.110 &      0.135 \\
 \si\ &  1.7183 &   3 &   0.890 &   14.10541 &      0.037 &      0.187 \\
 \si\ &  1.1311 &   3 &  -0.460 &   12.84018 &      0.031 &      0.036 \\
\hline
\ion{Si}{3} &  0.9324 &   1 &  -0.120 &   20.55363 &   7,223 &     41.6 \\
\ion{Si}{3} &  1.0526 &   1 &  -2.950 &   20.55363 &     10.7 &      0.062 \\
\ion{Si}{3} &  1.2523 &   3 &   0.500 &   24.99684 &      1.000 &      1.000 \\
\ion{Si}{3} &  0.7465 &   6 &   0.360 &   24.99684 &      0.724 &      0.724 \\
\ion{Si}{3} &  1.2601 &   3 &   0.160 &   24.99662 &      0.457 &      0.457 \\
\ion{Si}{3} &  0.9800 &   1 &   0.230 &   25.33550 &      0.245 &      0.363 \\
\ion{Si}{3} &  1.1341 &   3 &   0.430 &   25.56430 &      0.228 &      0.441 \\
\hline
   \ion{S}{1} &  0.9223 &   3 &   0.420 &    6.52494 &      1.000 &      1.000 \\
   \ion{S}{1} &  1.0457 &   3 &   0.260 &    6.86061 &      0.317 &      0.469 \\
   \ion{S}{1} &  1.8940 &   6 &   0.450 &    8.04620 &      0.031 &      0.183 \\
   \ion{S}{1} &  2.2694 &   3 &   0.240 &    7.87043 &      0.029 &      0.139 \\
   \ion{S}{1} &  1.3809 &   3 &   0.110 &    7.87043 &      0.022 &      0.103 \\
\hline
  \ion{S}{2} &  0.8315 &   1 &  -0.470 &   14.06839 &      1.000 &      1.000 \\
  \ion{S}{2} &  0.7967 &   1 &  -0.820 &   14.00342 &      0.519 &      0.482 \\
  \ion{S}{2} &  0.7590 &   1 &  -0.880 &   14.23489 &      0.264 &      0.321 \\
  \ion{S}{2} &  0.8855 &   2 &  -1.110 &   14.16002 &      0.185 &      0.206 \\
  \ion{S}{2} &  1.4501 &   2 &  -0.216 &   16.59204 &      0.005 &      0.096 \\
  \ion{S}{2} &  1.3529 &   1 &  -0.425 &   16.53575 &      0.004 &      0.063 \\
\hline
  \ion{Ca}{1} &  0.6149 &   3 &   0.100 &    1.89906 &     5.1 &      2.0 \\
  \ion{Ca}{1} &  0.7148 &   1 &   0.208 &    2.70919 &      1.000 &      1.000 \\
  \ion{Ca}{1} &  1.9753 &   4 &  -0.831 &    1.89906 &      0.599 &      0.234 \\
  \ion{Ca}{1} &  0.7326 &   1 &   0.073 &    2.93271 &      0.436 &      0.565 \\
  \ion{Ca}{1} &  1.9453 &   1 &  -1.094 &    1.88593 &      0.337 &      0.130 \\
\hline
 \ca\ &  0.8538 &   2 &  -0.362 &    1.70005 &  13,944 &     40.3 \\
 \ca\ &  0.8662 &   1 &  -0.623 &    1.69252 &   7,780 &     22.3 \\
 \ca\ &  1.1839 &   1 &   0.300 &    6.46831 &      1.000 &      0.731 \\
 \ca\ &  0.8921 &   3 &   0.729 &    7.05003 &      0.696 &      1.000 \\
 \ca\ &  1.1950 &   1 &   0.000 &    6.46831 &      0.501 &      0.367 \\
 \ca\ &  0.8249 &   2 &   0.621 &    7.51535 &      0.184 &      0.454 \\
 \ca\ &  0.8202 &   1 &   0.315 &    7.50564 &      0.093 &      0.227 \\
 \ca\ &  0.9931 &   1 &   0.072 &    7.51535 &      0.052 &      0.128 \\
\hline
  \ion{Mn}{1} &  1.2900 &   1 &  -1.059 &    2.11436 &      1.000 &      0.541 \\
  \ion{Mn}{1} &  1.3310 &   2 &  -1.360 &    2.14284 &      0.468 &      0.262 \\
  \ion{Mn}{1} &  1.3630 &   3 &  -1.516 &    2.16386 &      0.311 &      0.178 \\
  \ion{Mn}{1} &  1.3859 &   2 &  -1.638 &    2.17836 &      0.227 &      0.133 \\
  \ion{Mn}{1} &  1.5184 &   2 &   0.606 &    4.88919 &      0.074 &      1.000 \\
  \ion{Mn}{1} &  0.8740 &   3 &  -0.055 &    4.43521 &      0.046 &      0.370 \\
  \ion{Mn}{1} &  1.5263 &   1 &   0.379 &    4.88919 &      0.044 &      0.593 \\
  \ion{Mn}{1} &  0.7309 &   3 &  -0.084 &    4.43521 &      0.043 &      0.346 \\
\hline
 \ion{Mn}{2} &  0.7414 &   3 &  -2.202 &    3.70608 &      1.000 &      1.000 \\
 \ion{Mn}{2} &  0.7342 &   2 &  -2.713 &    3.70979 &      0.306 &      0.307 \\
 \ion{Mn}{2} &  0.9447 &   1 &  -2.389 &    4.06545 &      0.282 &      0.428 \\
 \ion{Mn}{2} &  0.9388 &   2 &  -2.554 &    4.07405 &      0.189 &      0.290 \\
 \ion{Mn}{2} &  0.8695 &   1 &   0.577 &    9.24489 &      0.002 &      0.971 \\
 \ion{Mn}{2} &  0.8769 &   1 &   0.491 &    9.24489 &      0.001 &      0.797 \\
 \ion{Mn}{2} &  0.8820 &   1 &   0.347 &    9.24489 &      0.001 &      0.572 \\
 \ion{Mn}{2} &  0.9486 &   2 &   0.491 &    9.46988 &      0.001 &      0.614 \\
 \hline
  \ion{Fe}{1} &  0.8677 &   2 &  -1.212 &    2.17609 &      1.000 &      0.417 \\
  \ion{Fe}{1} &  0.8824 &   1 &  -1.364 &    2.19801 &      0.670 &      0.286 \\
  \ion{Fe}{1} &  1.1973 &   1 &  -1.476 &    2.17609 &      0.545 &      0.227 \\
  \ion{Fe}{1} &  0.8388 &   1 &  -1.493 &    2.17609 &      0.524 &      0.218 \\
  \ion{Fe}{1} &  0.7511 &   5 &   0.107 &    4.17798 &      0.200 &      0.851 \\
  \ion{Fe}{1} &  0.8220 &   2 &   0.249 &    4.32039 &      0.199 &      1.000 \\
  \ion{Fe}{1} &  0.7196 &   3 &  -0.120 &    4.10365 &      0.141 &      0.550 \\
  \ion{Fe}{1} &  0.7941 &   2 &   0.154 &    4.38676 &      0.137 &      0.744 \\
  \ion{Fe}{1} &  0.7495 &   2 &  -0.102 &    4.22065 &      0.112 &      0.500 \\
  \ion{Fe}{1} &  0.7999 &   1 &   0.048 &    4.37164 &      0.111 &      0.593 \\
\hline
 \ion{Fe}{2} &  0.7712 &   1 &  -2.543 &    3.90368 &      1.000 &      1.000 \\
 \ion{Fe}{2} &  0.7462 &   1 &  -2.734 &    3.89187 &      0.662 &      0.653 \\
 \ion{Fe}{2} &  0.9998 &   1 &  -1.826 &    5.48450 &      0.133 &      0.832 \\
 \ion{Fe}{2} &  1.0500 &   2 &  -1.997 &    5.54914 &      0.077 &      0.521 \\
 \ion{Fe}{2} &  1.0863 &   1 &  -2.121 &    5.58957 &      0.053 &      0.374 \\
 \ion{Fe}{2} &  1.1126 &   1 &  -2.236 &    5.61560 &      0.038 &      0.278 \\
 \ion{Fe}{2} &  0.7515 &   1 &  -2.362 &    5.82361 &      0.018 &      0.163 \\
 \ion{Fe}{2} &  0.9550 &   1 &  -2.027 &    6.21915 &      0.015 &      0.223 \\
\hline
\ion{Fe}{3} &  0.7261 &   1 &  -2.910 &   13.13443 &      1.000 &      0.354 \\
\ion{Fe}{3} &  0.7265 &   1 &  -1.708 &   14.62496 &      0.501 &      1.000 \\
\ion{Fe}{3} &  0.9124 &   1 &  -2.577 &   14.62496 &      0.068 &      0.135 \\
\ion{Fe}{3} &  1.6722 &   1 &   0.235 &   22.87010 &      0.000 &      0.006 \\
\ion{Fe}{3} &  0.7456 &   1 &  -1.905 &   20.88295 &      0.000 &      0.000 \\
\ion{Fe}{3} &  1.6672 &   1 &   0.057 &   22.86637 &      0.000 &      0.004 \\
\ion{Fe}{3} &  1.2039 &   2 &   0.663 &   23.61137 &      0.000 &      0.007 \\
\ion{Fe}{3} &  1.2786 &   1 &   0.723 &   23.67243 &      0.000 &      0.007 \\
\hline
  \ion{Co}{1} &  0.7085 &   1 &  -1.018 &    1.88271 &      1.000 &      0.589 \\
  \ion{Co}{1} &  0.7053 &   1 &  -1.440 &    1.95586 &      0.319 &      0.205 \\
  \ion{Co}{1} &  0.8098 &   4 &   0.290 &    4.02115 &      0.142 &      1.000 \\
  \ion{Co}{1} &  0.8027 &   6 &   0.116 &    4.14628 &      0.071 &      0.579 \\
  \ion{Co}{1} &  0.8374 &   2 &  -0.040 &    4.07216 &      0.059 &      0.441 \\
\hline
 \co\ &  1.7772 &   1 &  -2.087 &    5.04595 &      1.000 &      0.736 \\
 \co\ &  1.7462 &   1 &  -2.284 &    5.12265 &      0.532 &      0.428 \\
 \co\ &  2.2205 &   1 &  -2.374 &    5.04595 &      0.516 &      0.380 \\
 \co\ &  2.4596 &   1 &  -2.416 &    5.12265 &      0.392 &      0.316 \\
 \co\ &  2.3613 &   1 &  -2.491 &    5.17524 &      0.292 &      0.250 \\
 \co\ &  1.6064 &   1 &  -2.568 &    5.12265 &      0.276 &      0.223 \\
 \co\ &  1.5759 &   1 &  -2.725 &    5.04595 &      0.230 &      0.169 \\
 \co\ &  1.6361 &   1 &  -2.625 &    5.17524 &      0.215 &      0.184 \\
 \co\ &  1.7239 &   1 &  -2.649 &    5.17524 &      0.203 &      0.174 \\
 \co\ &  2.2497 &   1 &  -2.697 &    5.20879 &      0.168 &      0.150 \\
 \co\ &  2.1347 &   1 &  -2.906 &    5.04595 &      0.152 &      0.112 \\
 \co\ &  0.8342 &   2 &   0.753 &   10.41714 &      0.003 &      1.000 \\
 \co\ &  0.8737 &   1 &   0.663 &   10.41714 &      0.002 &      0.813 \\
\enddata

\tablenotetext{a}{If more than one line is included in the blend, the indicated wavelength is a mean value, weighted by oscillator strength.  Some lines have been ommitted from the table if their rest wavelengths are very close to that of a stronger line.}

\tablenotetext{b}{The estimated line strength is computed using $N_n \propto g_n f_n*10^{-\chi_n/kT}$, where $f_n$ is the oscillator strength, $g_n$ is the statistical weight, $\chi_n$ is the excitation potential, and $T$ is the temperature.  The values are normalized by dividing by the value for one of the strongest lines for each ion. This is usually the strongest line at that temperature, but in some cases (for example \ca) the strongest line is so much stronger that the normalization is nor relevant.  In that case we chose the second or third strongest line. (see \S \ref{lineid})}

\end{deluxetable}

\clearpage
\begin{deluxetable}{crcrrrr}
\tablecolumns{7} 
\tablewidth{0pc}
\tabletypesize{\scriptsize}
\tablecaption{Estimated Line Strengths at 10,000K in the Region $0.7 - 2.4$ \mum \label{10K}}
\tablehead{\colhead{Ion} & 
\colhead{$\lambda_{rest}$(air)} & 
\colhead{No. lines} & 
\colhead{$\log(gf)$} & 
\colhead{Excitation} & 
\colhead{Line Strength} & 
\colhead{Line Strength}\\
 & 
\colhead{\mum} &
\colhead{in blend \tablenotemark{a}} & 
 &
\colhead{Min. (eV)} & 
\colhead{5,000K \tablenotemark{b}} & 
\colhead{10,000K \tablenotemark{b}}}

\startdata
\hline
  \ion{He}{1} &  1.0830 &   3 &  -0.047 &   19.82096 &     15.1 &      6.0 \\
  \ion{He}{1} &  2.0581 &   1 &  -0.424 &   20.61718 &      1.000 &      1.000 \\
  \ion{He}{1} &  0.7065 &   3 &  -0.460 &   20.96551 &      0.410 &      0.614 \\
\hline
   \ci\ &  1.0693 &   7 &   0.348 &    7.48830 &      1.000 &      1.000 \\
   \ci\ &  0.9093 &   2 &   0.142 &    7.48830 &      0.622 &      0.622 \\
   \ci\ &  0.9406 &   1 &   0.225 &    7.68529 &      0.477 &      0.599 \\
   \ci\ &  1.1754 &   2 &   0.661 &    8.64774 &      0.139 &      0.535 \\
   \ci\ &  1.7325 &   8 &   1.030 &    9.70241 &      0.028 &      0.368 \\
   \ci\ &  1.6890 &   1 &   0.568 &    9.00319 &      0.049 &      0.286 \\
   \ci\ &  1.4543 &   1 &  -0.110 &    7.68529 &      0.221 &      0.277 \\
   \ci\ &  1.1330 &   1 &   0.280 &    8.53767 &      0.075 &      0.253 \\
\hline
  \ion{C}{2} &  0.6580 &   2 &   0.118 &   14.44980 &     88,393 &   253 \\
  \ion{C}{2} &  0.7235 &   3 &   0.330 &   16.33422 &   1,815 &     46.3 \\
  \ion{C}{2} &  0.9903 &   3 &   1.010 &   20.95206 &      0.192 &      1.000 \\
  \ion{C}{2} &  1.8905 &   2 &   0.258 &   19.49586 &      1.000 &      0.959 \\
  \ion{C}{2} &  1.7846 &   3 &   0.550 &   20.15184 &      0.427 &      0.878 \\
  \ion{C}{2} &  0.9230 &   3 &   0.620 &   20.84618 &      0.100 &      0.461 \\
\hline
   \oi\ &  0.7773 &   3 &   0.324 &    9.14671 &      1.000 &      1.000 \\
   \oi\ &  0.8446 &   3 &   0.170 &    9.52201 &      0.294 &      0.454 \\
   \oi\ &  0.9264 &   9 &   0.690 &   10.74166 &      0.057 &      0.365 \\
   \oi\ &  1.1290 &   9 &   0.500 &   10.98960 &      0.021 &      0.177 \\
   \oi\ &  1.8021 &  18 &   0.880 &   12.07943 &      0.004 &      0.120 \\
   \oi\ &  0.7990 &   6 &   0.280 &   10.98960 &      0.013 &      0.106 \\
\hline
 \mg\ &  0.9227 &   2 &   0.270 &    8.65529 &      1.000 &      1.000 \\
 \mg\ &  0.7890 &   3 &   0.650 &   10.00000 &      0.106 &      0.504 \\
 \mg\ &  1.0927 &   3 &   0.020 &    8.86425 &      0.346 &      0.441 \\
 \mg\ &  1.0092 &   3 &   1.020 &   11.63047 &      0.006 &      0.178 \\
 \mg\ &  0.8228 &   2 &   0.030 &   10.00000 &      0.025 &      0.121 \\
 \mg\ &  0.9632 &   3 &   0.660 &   11.56982 &      0.003 &      0.083 \\
 \mg\ &  1.8606 &   9 &   1.220 &   12.85873 &      0.001 &      0.068 \\
 \mg\ &  1.6787 &   3 &   0.730 &   12.08542 &      0.001 &      0.054 \\
 \mg\ &  2.1369 &   1 &   0.390 &   11.50533 &      0.002 &      0.048 \\
\hline
 \si\ &  0.6355 &   2 &   0.297 &    8.12157 &   10,126 &       49.6 \\
 \si\ &  0.9413 &   3 &   0.980 &   12.84018 &      1.000 &      1.000 \\
 \si\ &  1.6930 &   2 &   0.350 &   12.14781 &      1.169 &      0.524 \\
 \si\ &  0.7849 &   3 &   0.490 &   12.52627 &      0.671 &      0.466 \\
 \si\ &  1.1737 &   3 &   0.620 &   12.88100 &      0.397 &      0.416 \\
 \si\ &  1.7183 &   3 &   0.890 &   14.10541 &      0.043 &      0.187 \\
 \si\ &  1.3681 &   2 &   0.130 &   12.88100 &      0.128 &      0.135 \\
 \si\ &  2.1967 &   3 &   0.680 &   14.13227 &      0.025 &      0.112 \\
\hline
\ion{Si}{3} &  0.9324 &   1 &  -0.120 &   20.55363 &   7,223 &     41.6 \\
\ion{Si}{3} &  1.2523 &   3 &   0.500 &   24.99684 &      0.094 &      1.000 \\
\ion{Si}{3} &  0.7465 &   6 &   0.360 &   24.99684 &      0.068 &      0.724 \\
\ion{Si}{3} &  1.2601 &   3 &   0.160 &   24.99662 &      0.043 &      0.457 \\
\ion{Si}{3} &  1.1341 &   3 &   0.430 &   25.56430 &      0.021 &      0.441 \\
\ion{Si}{3} &  0.9800 &   1 &   0.230 &   25.33550 &      0.023 &      0.363 \\
\ion{Si}{3} &  0.8266 &   6 &   0.830 &   26.65707 &      0.004 &      0.311 \\
\ion{Si}{3} &  0.7612 &   1 &   0.660 &   26.60039 &      0.003 &      0.225 \\
\ion{Si}{3} &  1.5015 &   1 &   0.030 &   25.77484 &      0.005 &      0.137 \\
\ion{Si}{3} &  0.9375 &   1 &  -0.490 &   25.33550 &      0.004 &      0.069 \\
\ion{Si}{3} &  0.7729 &   2 &  -0.680 &   24.99669 &      0.006 &      0.066 \\
\ion{Si}{3} &  1.0526 &   1 &  -2.950 &   20.55363 &      1.000 &      0.062 \\
\hline
  \ion{S}{2} &  0.8315 &   1 &  -0.470 &   14.06839 &      1.000 &      1.000 \\
  \ion{S}{2} &  0.7967 &   1 &  -0.820 &   14.00342 &      0.519 &      0.482 \\
  \ion{S}{2} &  0.7590 &   1 &  -0.880 &   14.23489 &      0.264 &      0.321 \\
  \ion{S}{2} &  0.7721 &   1 &  -0.990 &   14.29456 &      0.179 &      0.232 \\
  \ion{S}{2} &  0.8855 &   2 &  -1.110 &   14.16002 &      0.185 &      0.206 \\
  \ion{S}{2} &  1.4501 &   2 &  -0.216 &   16.59204 &      0.005 &      0.096 \\
  \ion{S}{2} &  1.3529 &   1 &  -0.425 &   16.53575 &      0.004 &      0.063 \\
\hline
 \ca\ &  0.8538 &   2 &  -0.362 &    1.70005 &  13,944 &     40.3 \\
 \ca\ &  0.8662 &   1 &  -0.623 &    1.69252 &   7,780 &     22.3 \\
 \ca\ &  0.8921 &   3 &   0.729 &    7.05003 &      0.696 &      1.000 \\
 \ca\ &  1.1839 &   1 &   0.300 &    6.46831 &      1.000 &      0.731 \\
 \ca\ &  0.8249 &   2 &   0.621 &    7.51535 &      0.184 &      0.454 \\
 \ca\ &  1.1950 &   1 &   0.000 &    6.46831 &      0.501 &      0.367 \\
 \ca\ &  0.9891 &   3 &   0.943 &    8.43855 &      0.045 &      0.327 \\
 \ca\ &  0.8202 &   1 &   0.315 &    7.50564 &      0.093 &      0.227 \\
 \ca\ &  1.8850 &   2 &   0.878 &    9.01809 &      0.010 &      0.144 \\
 \ca\ &  0.9931 &   1 &   0.072 &    7.51535 &      0.052 &      0.128 \\
\hline
 \ion{Mn}{2} &  0.7414 &   3 &  -2.202 &    3.70608 &      1.000 &      1.000 \\
 \ion{Mn}{2} &  0.8695 &   1 &   0.577 &    9.24489 &      0.002 &      0.971 \\
 \ion{Mn}{2} &  0.8769 &   1 &   0.491 &    9.24489 &      0.001 &      0.797 \\
 \ion{Mn}{2} &  0.9486 &   2 &   0.491 &    9.46988 &      0.001 &      0.614 \\
 \ion{Mn}{2} &  0.8820 &   1 &   0.347 &    9.24489 &      0.001 &      0.572 \\
 \ion{Mn}{2} &  0.9447 &   1 &  -2.389 &    4.06545 &      0.282 &      0.428 \\
 \ion{Mn}{2} &  0.7342 &   2 &  -2.713 &    3.70979 &      0.306 &      0.307 \\
 \ion{Mn}{2} &  0.9388 &   2 &  -2.554 &    4.07405 &      0.189 &      0.290 \\
 \ion{Mn}{2} &  0.7221 &   3 &   0.768 &   10.67048 &      0.000 &      0.288 \\
 \ion{Mn}{2} &  0.9444 &   1 &   0.129 &    9.46988 &      0.000 &      0.267 \\
 \ion{Mn}{2} &  0.8110 &   1 &  -2.002 &    5.37762 &      0.033 &      0.228 \\
 \ion{Mn}{2} &  1.5401 &   3 &   0.237 &    9.86600 &      0.000 &      0.216 \\
\hline
 \fe\ &  0.8288 &   1 &   0.693 &    9.65426 &      0.003 &      1.000 \\
 \fe\ &  0.9297 &   1 &   0.412 &    9.65426 &      0.001 &      0.524 \\
 \fe\ &  0.7712 &   1 &  -2.543 &    3.90368 &      1.000 &      0.460 \\
 \fe\ &  0.9998 &   1 &  -1.826 &    5.48450 &      0.133 &      0.382 \\
 \fe\ &  0.9095 &   1 &   0.269 &    9.65426 &      0.001 &      0.377 \\
 \fe\ &  0.9183 &   2 &   0.245 &    9.70089 &      0.001 &      0.338 \\
 \fe\ &  0.7462 &   1 &  -2.734 &    3.89187 &      0.662 &      0.300 \\
 \fe\ &  1.0500 &   2 &  -1.997 &    5.54914 &      0.077 &      0.239 \\
 \fe\ &  1.0863 &   1 &  -2.121 &    5.58957 &      0.053 &      0.172 \\
 \hline
\ion{Fe}{3} &  0.7265 &   1 &  -1.708 &   14.62496 &      0.501 &      1.000 \\
\ion{Fe}{3} &  0.7261 &   1 &  -2.910 &   13.13443 &      1.000 &      0.354 \\
\ion{Fe}{3} &  0.9124 &   1 &  -2.577 &   14.62496 &      0.068 &      0.135 \\
\ion{Fe}{3} &  1.2786 &   1 &   0.723 &   23.67243 &      0.000 &      0.007 \\
\ion{Fe}{3} &  1.2039 &   2 &   0.663 &   23.61137 &      0.000 &      0.007 \\
\ion{Fe}{3} &  1.6722 &   1 &   0.235 &   22.87010 &      0.000 &      0.006 \\
\ion{Fe}{3} &  1.2955 &   2 &   0.602 &   23.67243 &      0.000 &      0.006 \\
\hline
 \co\ &  0.8342 &   2 &   0.753 &   10.41714 &      0.003 &      1.000 \\
 \co\ &  0.8737 &   1 &   0.663 &   10.41714 &      0.002 &      0.813 \\
 \co\ &  1.7772 &   1 &  -2.087 &    5.04595 &      1.000 &      0.736 \\
 \co\ &  0.8582 &   1 &   0.598 &   10.48815 &      0.002 &      0.644 \\
 \co\ &  0.9070 &   1 &   0.433 &   10.41714 &      0.001 &      0.479 \\
 \co\ &  1.7462 &   1 &  -2.284 &    5.12265 &      0.532 &      0.428 \\
 \co\ &  0.8803 &   2 &   0.521 &   10.70637 &      0.001 &      0.419 \\
 \co\ &  2.2205 &   1 &  -2.374 &    5.04595 &      0.516 &      0.380 \\
 \co\ &  0.8512 &   2 &   0.426 &   10.61330 &      0.001 &      0.375 \\
 \co\ &  2.4596 &   1 &  -2.416 &    5.12265 &      0.392 &      0.316 \\
 \co\ &  2.3613 &   1 &  -2.491 &    5.17524 &      0.292 &      0.250 \\
\enddata

\tablenotetext{a}{If more than one line is included in the blend, the indicated wavelength is a mean value, weighted by oscillator strength.  Some lines have been ommitted from the table if their rest wavelengths are very close to that of a stronger line.}

\tablenotetext{b}{The estimated line strength is computed using $N_n \propto g_n f_n*10^{-\chi_n/kT}$, where $f_n$ is the oscillator strength, $g_n$ is the statistical weight, $\chi_n$ is the excitation potential, and $T$ is the temperature.  The values are normalized by dividing by the value for one of the strongest lines for each ion. This is usually the strongest line at that temperature, but in some cases (for example \ca) the strongest line is so much stronger that the normalization is nor relevant.  In that case we chose the second or third strongest line. (see \S \ref{lineid})}

\end{deluxetable}

\clearpage

\begin{figure}
\plotone{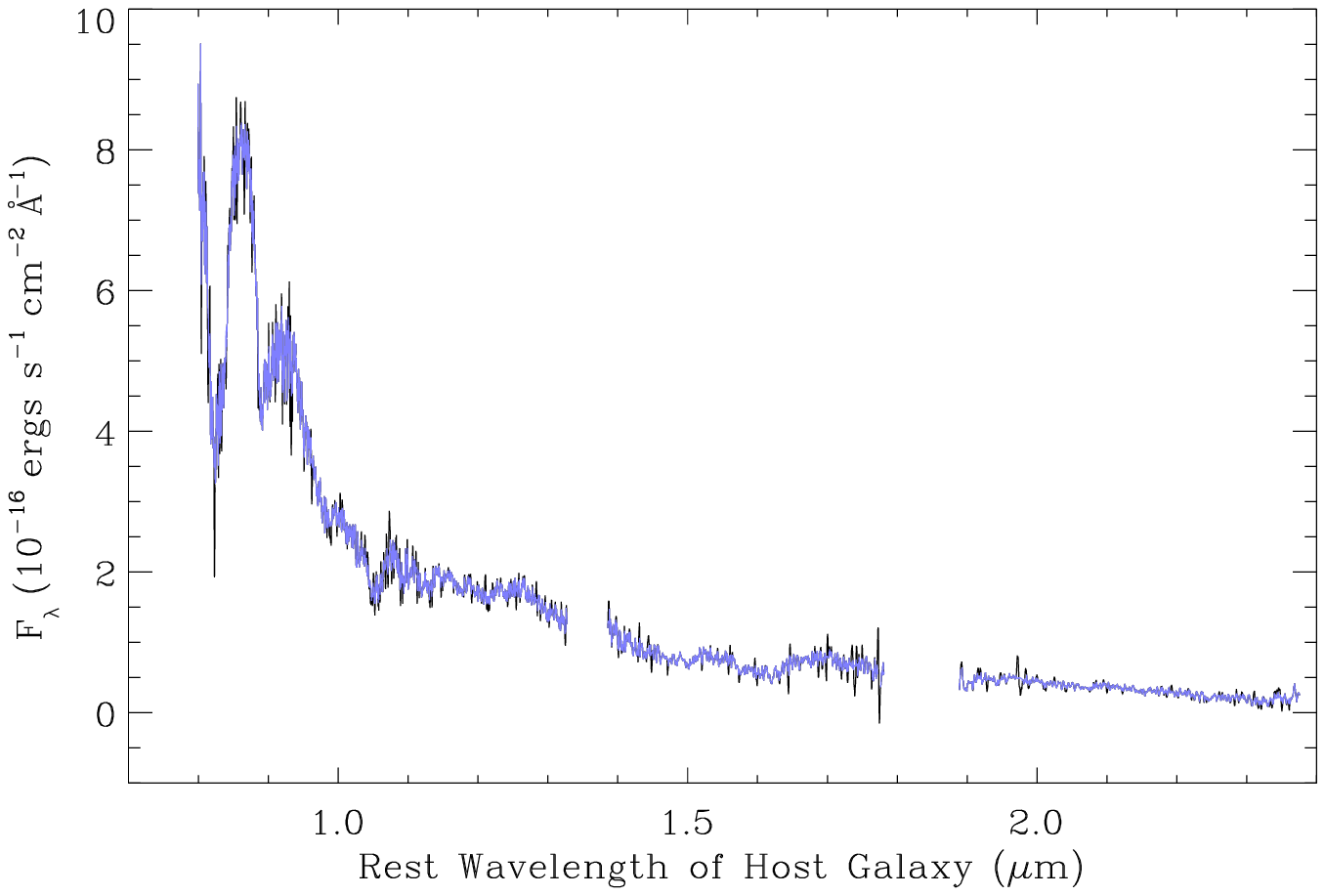}
\caption[]{The spectrum from SN 2000dm at +0 day provides an example of our spike elimination routine (see text \S \ref{obs}).  The original data is plotted in black and the cleaned spectrum is superimposed in blue (see discussion \S \ref{obs}). \label{spikes}}
\end{figure}

\begin{figure}
\plottwo{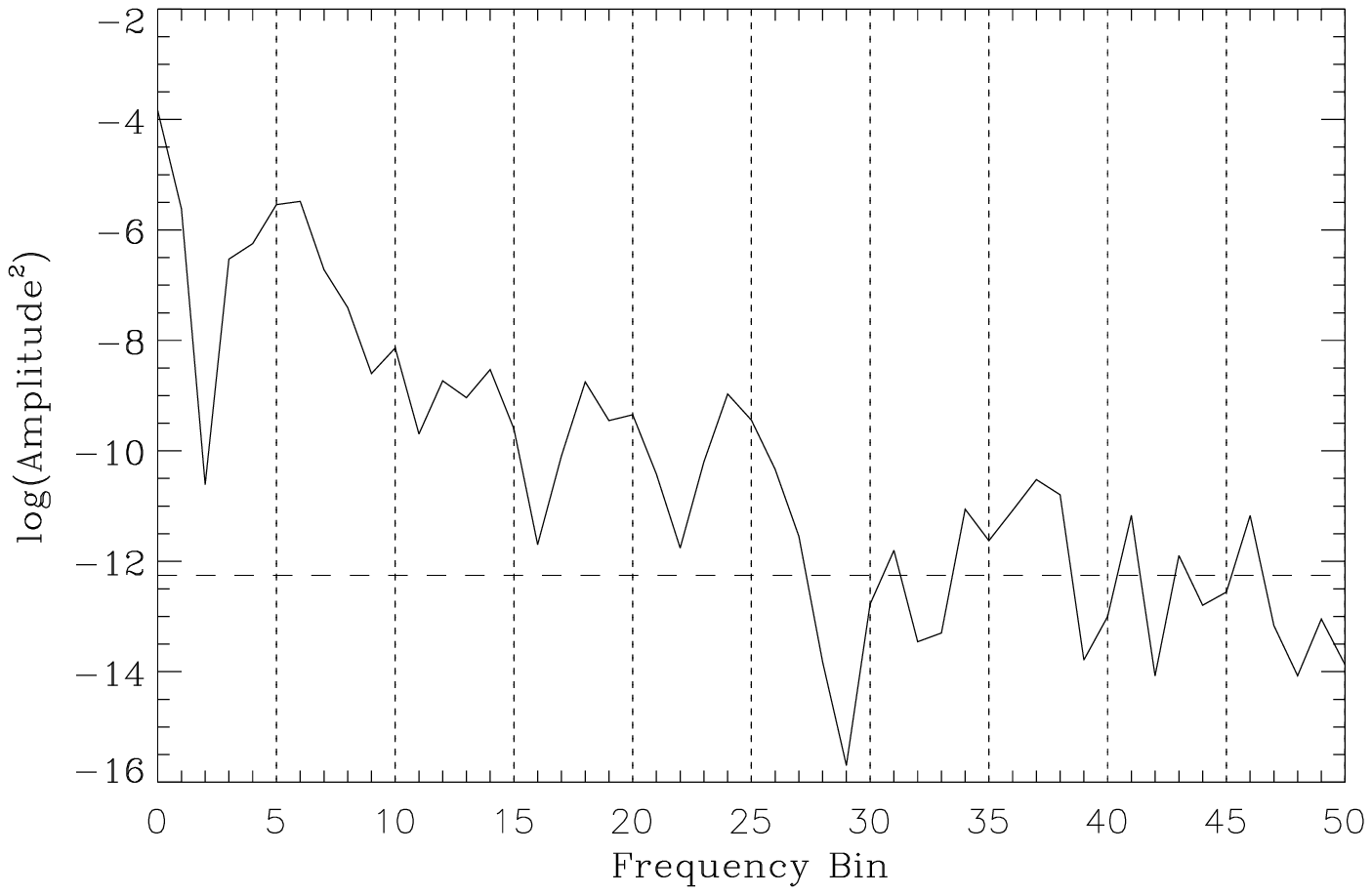}{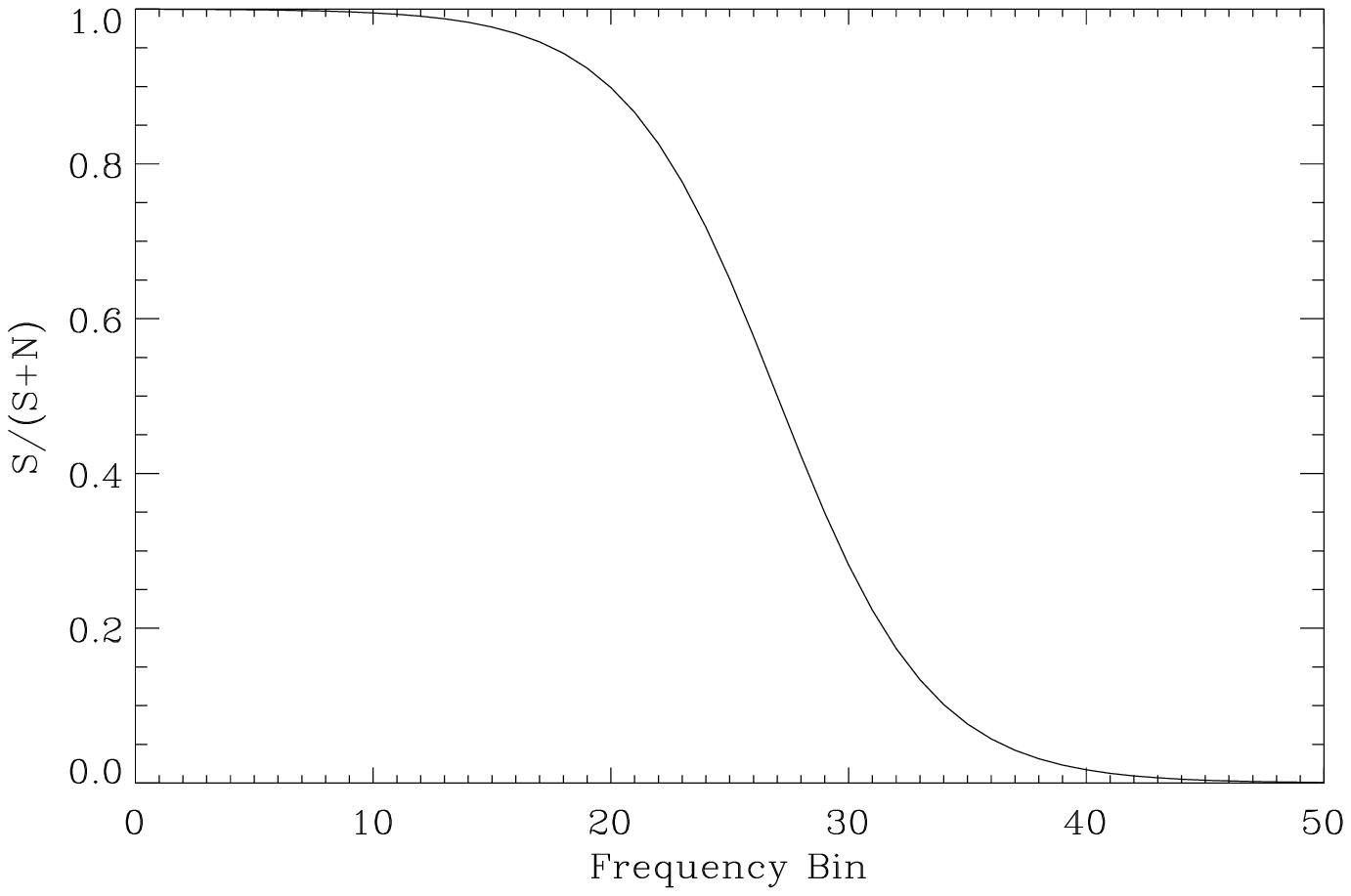}
\caption[]{The power spectrum and filter used in our reduction package (see text \S \ref{ftx3}).  The left panel is a log plot of the first fifty points (out of 1901) in one half of the power spectrum for the wavelength region 0.80--1.35 \mum\ from the SXD data from SN 2005am at -4 days. The noise level (N) is taken to be a constant equal to the mean of the next 100 points (50:150).  The location of N is indicated in the figure by a horizontal dashed line.  The signal (S) descends from near -4 on the left edge and intersects the noise level at point 27.  The right panel shows the filter produced by the formula: S/(S+N)  (see discussion \S \ref{ftx3}). \label{power}}
\end{figure}

\begin{figure}
\plotone{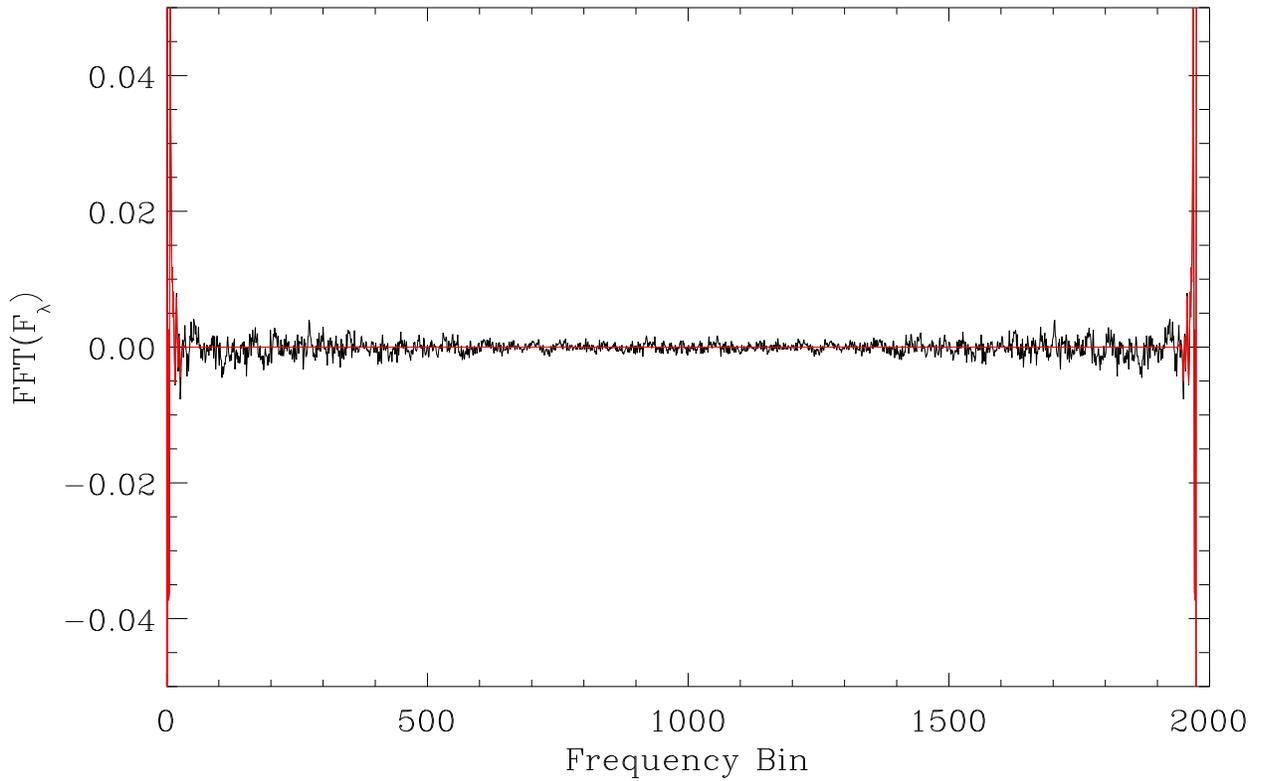}
\caption[]{The Fourier transform for the wavelength region 0.80--1.35 \mum\ from SN 2005am at -4 days (see text \S \ref{ftx3}). The initial frequency spectrum is plotted in black.  The same spectrum after processing with the filter as described in the text is superimposed in red (see discussion \S \ref{ftx3}). \label{ft1}}
\end{figure}

\begin{figure}
\plotone{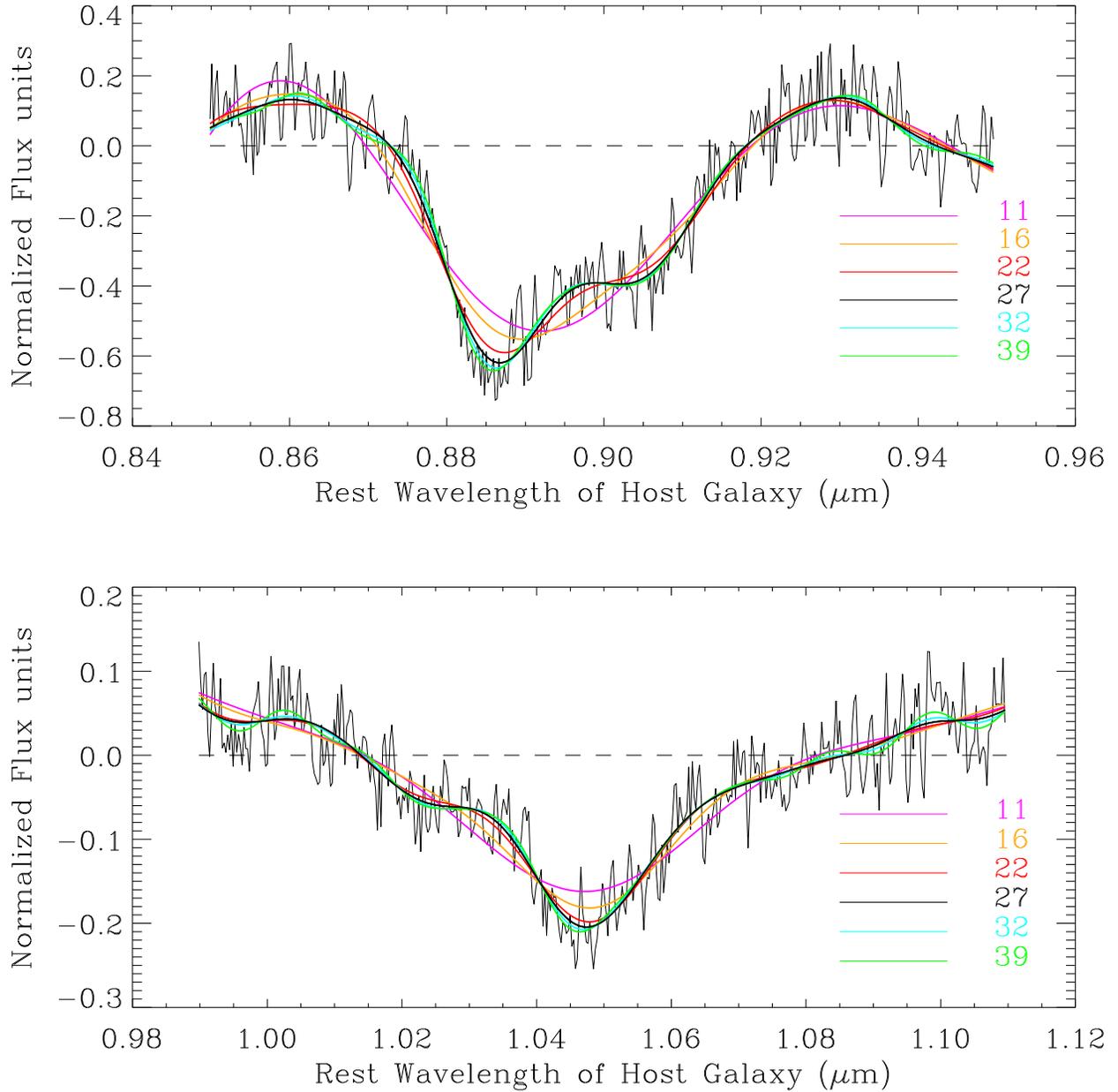}
\caption[] {Two absorption features from the high S/N SXD spectrum of SN 2005am at -4d demonstrate that the FT smoothing technique is robust and the measured velocities do not change significantly due to fine interpretations of the smoothing parameter by the user (see text \S \ref{ftx3} and Table \ref{fttable}).  The raw data are black and the smoothed spectrum in black is the Fourier transform spectrum as with a smoothing parameter of 27 as shown in Figure \ref{power}a. The spectra produced by other smoothing parameters are plotted in different colors (see discussion \S \ref{ftx3}). On this scale, the distance between ticks on the wavelength axis (0.005 \mum) is about 1,500 \kms. \label{ft3}}
\end{figure}

\begin{figure}
\plotone{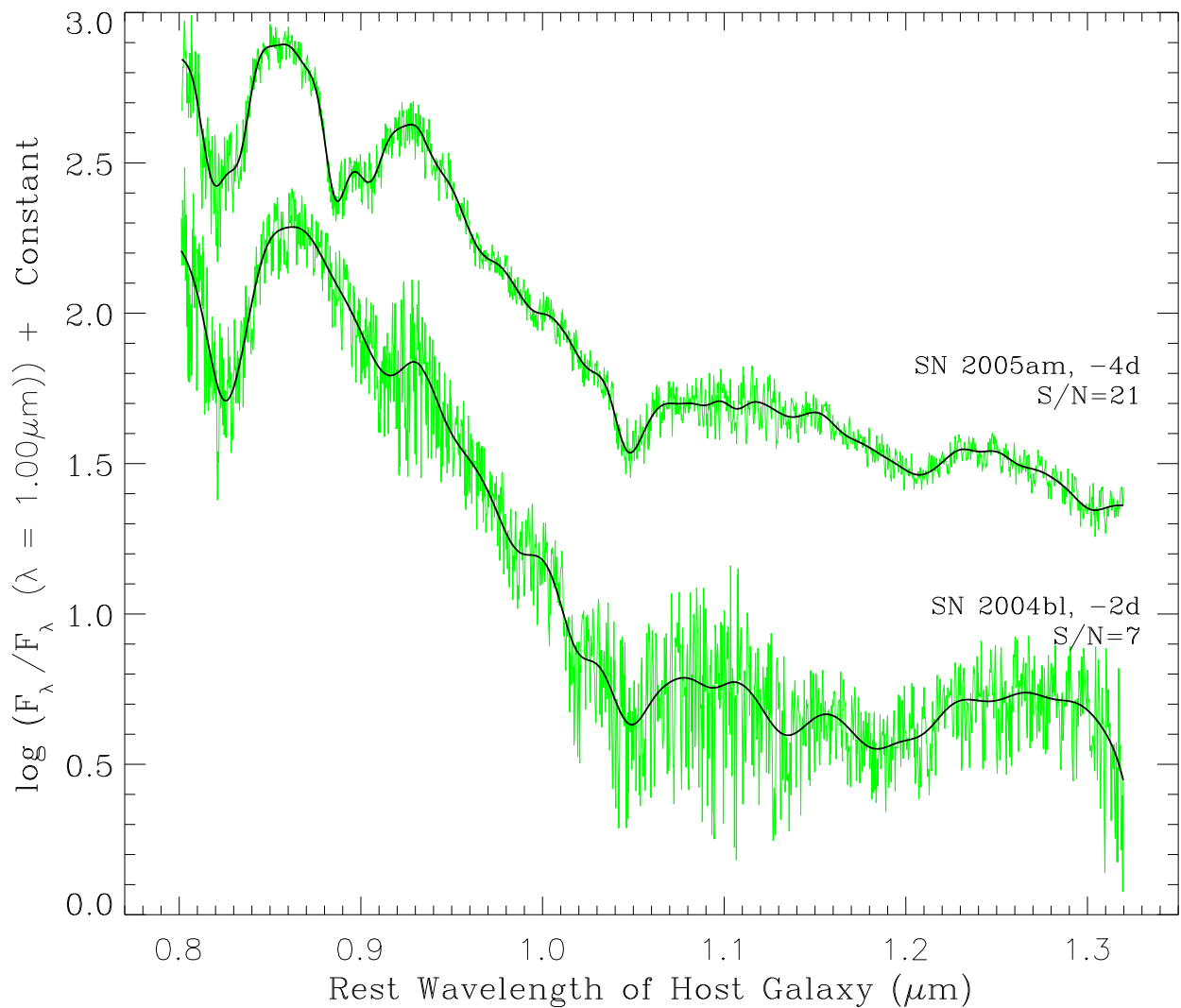}
\caption[]{Two spectra with different S/N show how use of the FT smoothed spectra facilitate understanding of the overall spectral shape and identification of the precise locations of maxima and minima (see text \S \ref{ftx3}).  The raw data are plotted in green and the FT smoothed spectra are superimposed in black (see discussion \S \ref{ftx3}).  \label{ft2}}
\end{figure}

\begin{figure}
\plotone{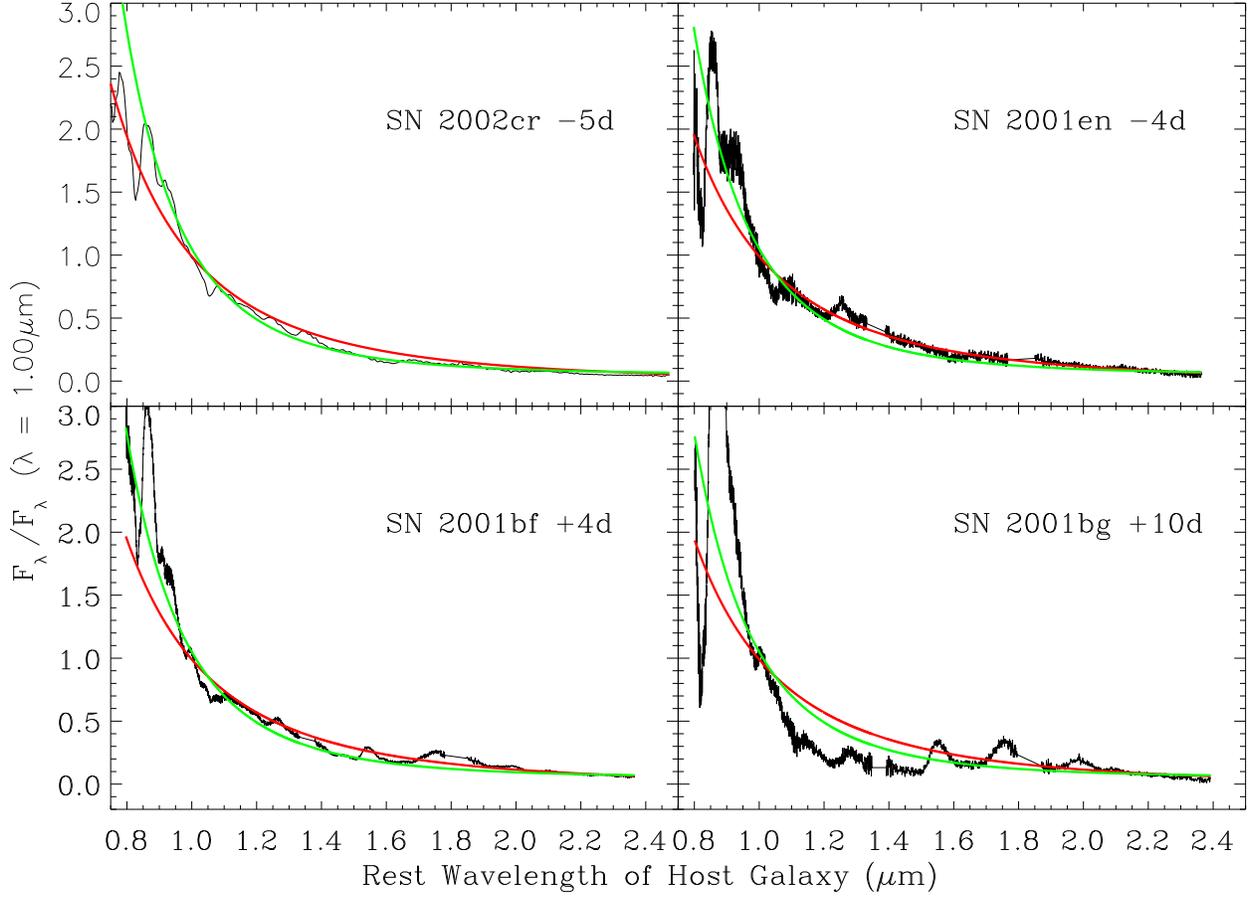}
\caption[]{Raw data from four spectra showing possible fits to the continua (see text \S \ref{flc23}).  The top left spectrum is LRS, the others are SXD.  The red line is a -3.0 power law and the green line is a -4.5 power law.  It is clear that a single power law does not fit any of the spectra (see discussion \S \ref{flc23}). \label{confit1}}
\end{figure}

\begin{figure}
\epsscale{0.8}
\plotone{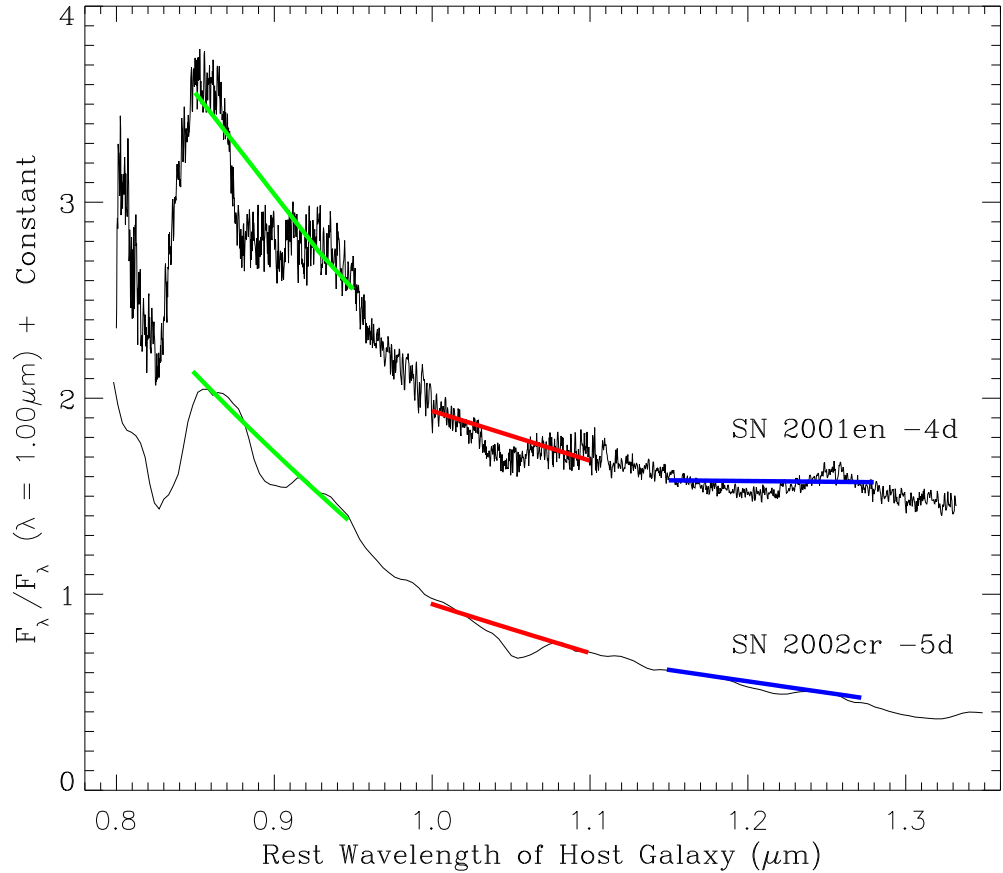}
\caption[]{Estimated fits for local continua in the vicinity of absorption features are plotted on the figure in red, green, and blue (see text \S \ref{flc23}).  The top spectrum is a raw data SXD spectrum from SN 2001en at -4 days.  On the bottom is a raw LRS spectrum from SN 2002cr at -5 days.  These are the same spectra displayed in the top two panels of Figure \ref{confit1} and the fitted regions are expanded in Figure \ref{confit3} (see discussion \S \ref{flc23}).   \label{confit2}}
\end{figure}

\begin{figure}
\plotone{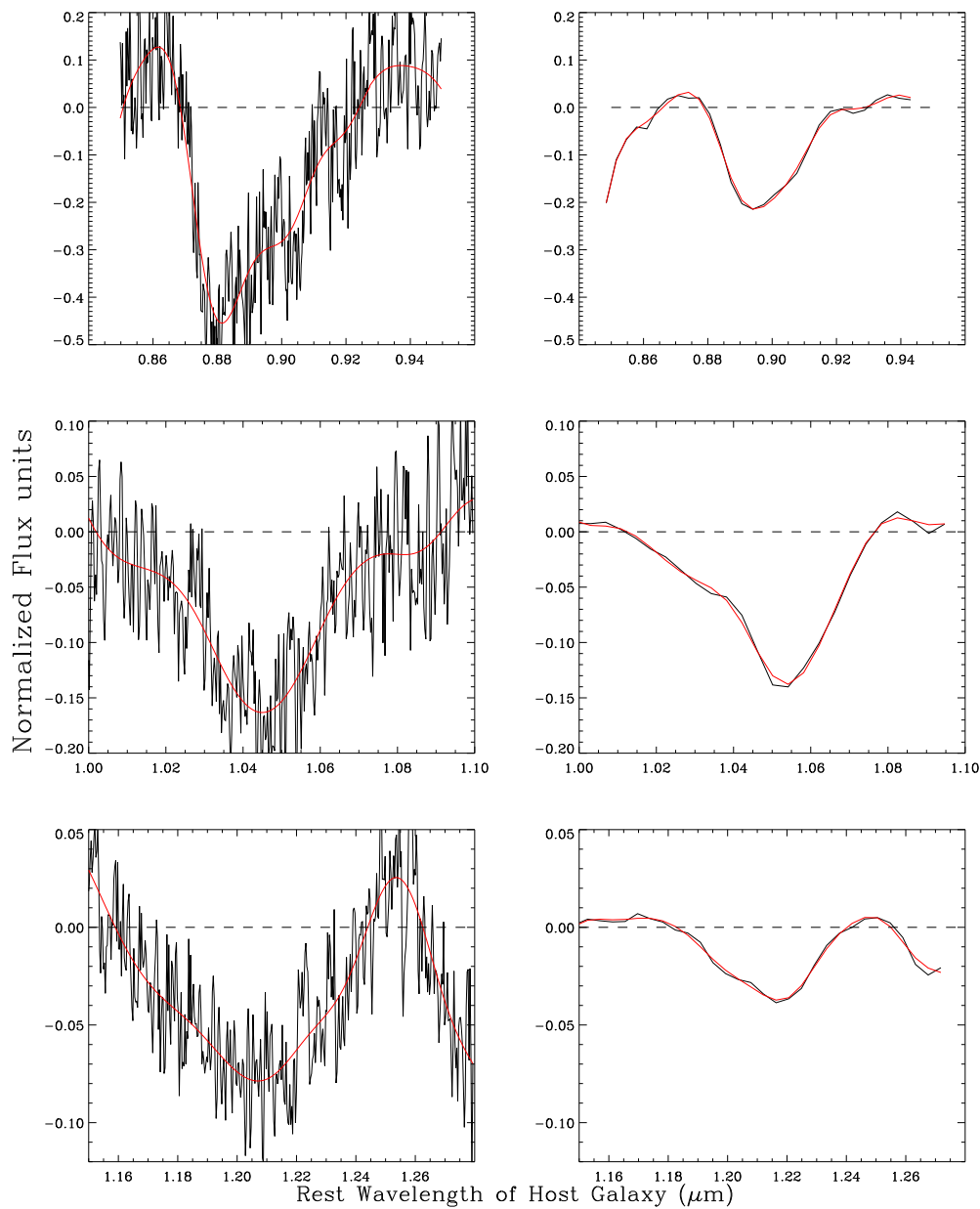}
\caption[]{Expanded spectral regions from Figure \ref{confit2} with the continuum normalized to zero (see text \S \ref{flc23}).  The left column displays SXD data from SN 2001en (-4d) and the right column is LRS data from SN 2002cr (-5d). The top row corresponds to the regions in Figure \ref{confit2} marked with a green continuum, the middle row is from the red regions and the bottom row is from the blue regions.  The FT smoothed spectrum is superimposed on the raw data in red (see discussion \S \ref{flc23}). \label{confit3}}
\end{figure}

\clearpage

\begin{figure}
\epsscale{0.6}
\plotone{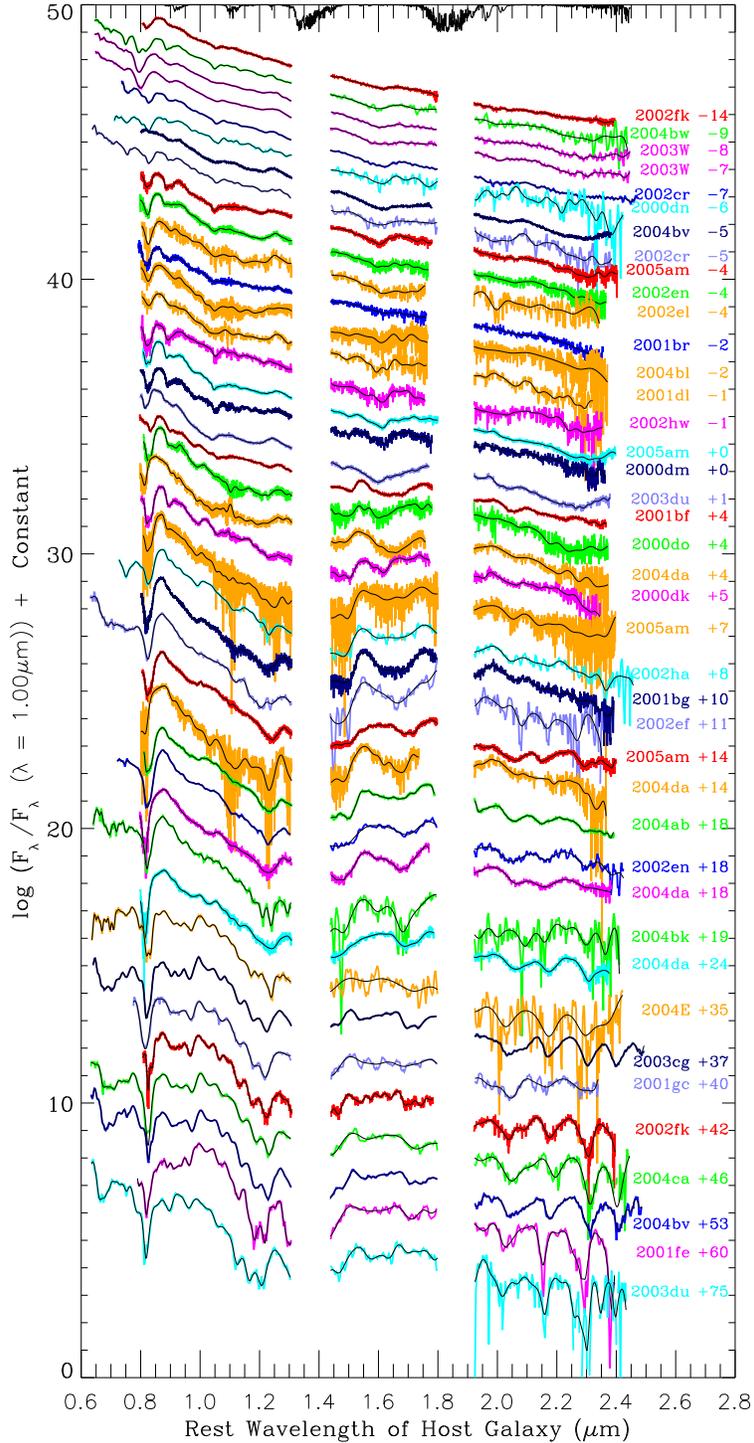}
\caption[]{Forty-one spectra from normal SNe Ia obtained between -14d and +75d with respect to $V_{max}$ (see text \S \ref{spectra}).  The abscissa is wavelength in microns, shifted to the rest frame of the host galaxy and the ordinate is log flux. All spectra have been normalized to 1.0 at 1.0 \mum\ and shifted by a constant for clarity. Spectra plotted in yellow have low S/N and will be omitted for the analysis of spectral features.  Observational and spectral details are provided in Tables \ref{snelist_epoch} and \ref{snelist_disc}. \label{all}}
\end{figure}

\begin{figure}
\plotone{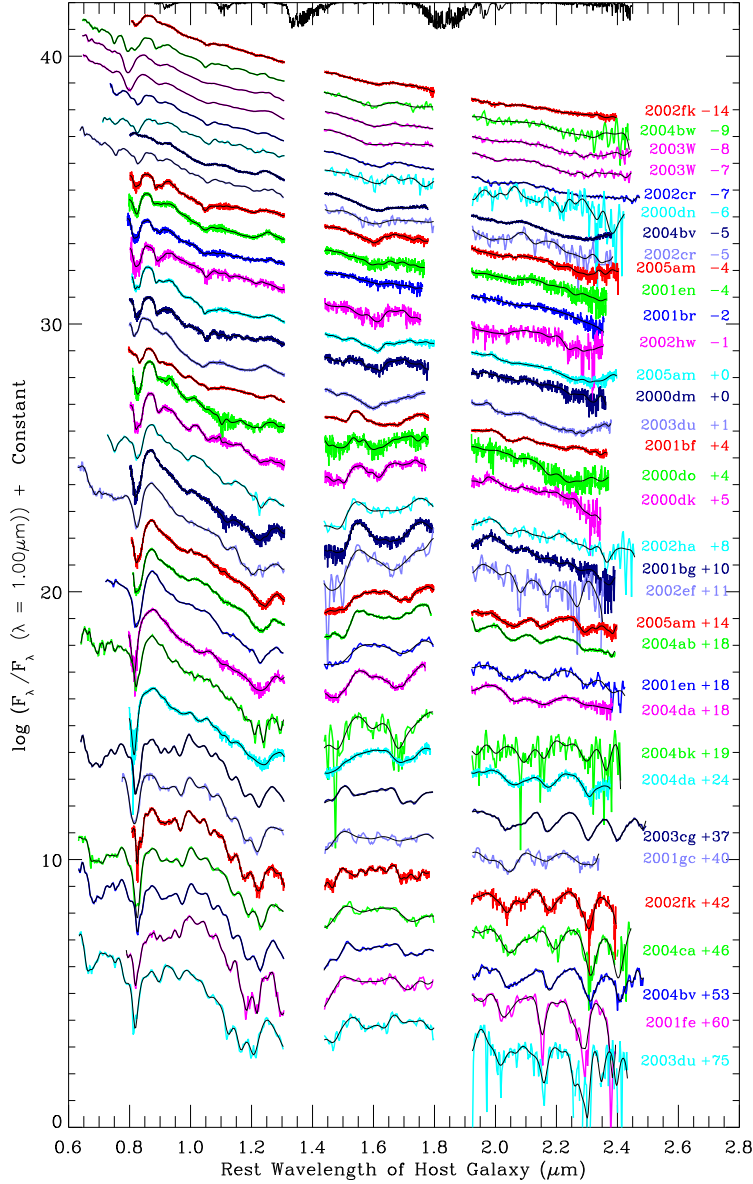}
\epsscale{0.7}
\caption[]{Thirty-four spectra selected from our sample for S/N ratio greater than 10 in the region $1.10-1.30$ \mum\ (see Tables \ref{snelist_epoch} and \ref{snelist_disc}).  These spectra cover the entire temporal range of our sample, from -14 to +75 days with respect to $V_{max}$ (see text \S \ref{spectra}).  The abscissa is wavelength, in microns, shifted to the rest frame of the host galaxy and the ordinate is log flux. All spectra have been normalized to 1.0 at 1.0 \mum\ and shifted by a constant for clarity. \label{gh}}
\end{figure}

\begin{figure}
\plotone{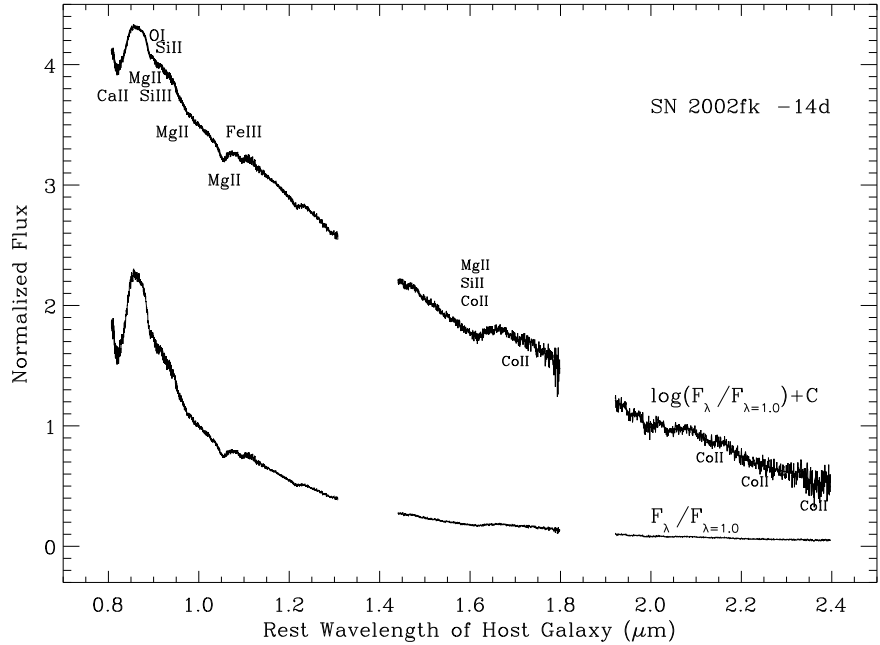}
\caption[]{A very early spectrum of SN Ia 2002fk obtained fourteen days before $V_{max}$; only 5 or 6 days after the explosion. The data are displayed in both linear and log space (plus a constant) to demonstrate how the sizes of features at longer wavelengths are exaggerated by the log scale.  Features are labeled for which we have reasonably confident identifications.  A general discussion of this spectrum can be found in \S \ref{ves} and a more detailed analysis is found in \ref{ves_featanal}. \label{fk}}
\end{figure}

\begin{figure}
\plotone{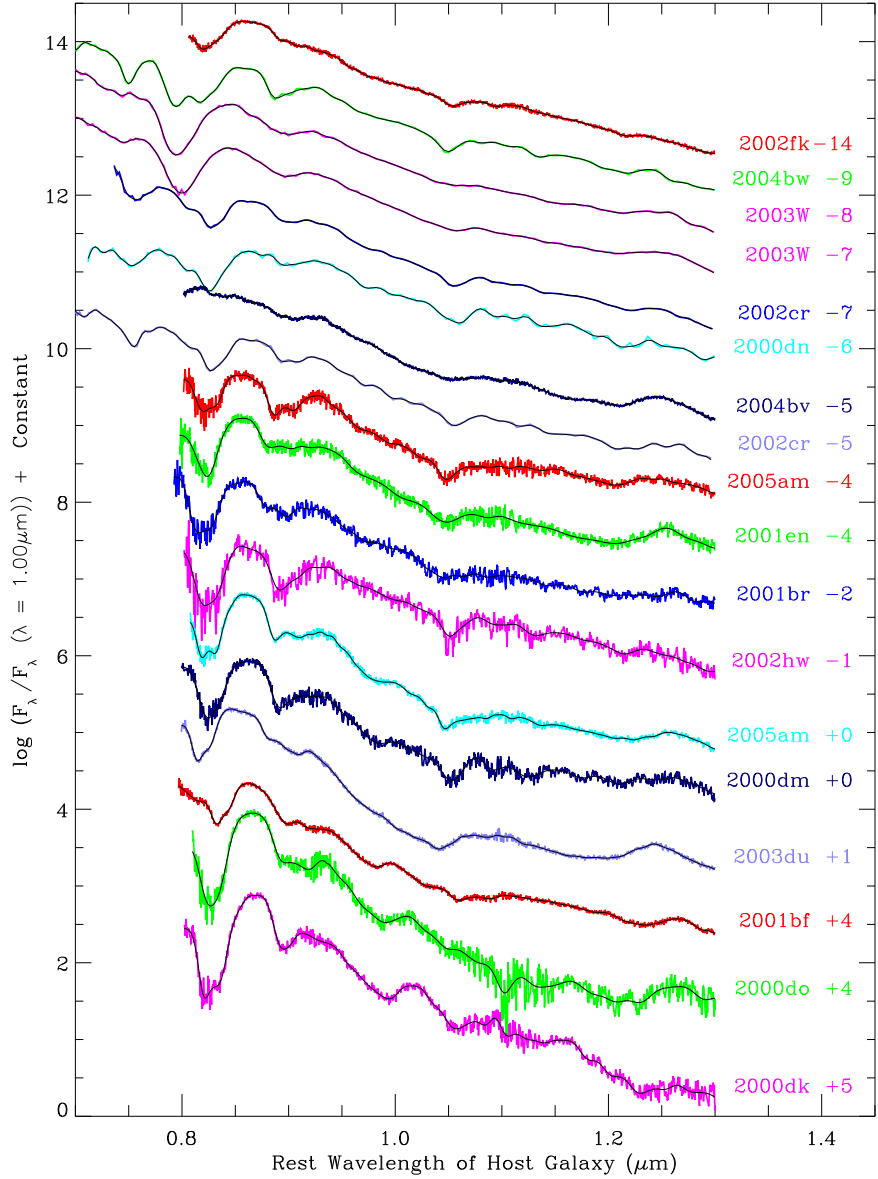}
\caption[]{Eighteen spectra from the \emph{Photospheric Phase} of SNe Ia obtained between -14d and +5d relative to $V_{max}$.  During this epoch the photosphere is expected to be in the outer layers of the SN and the line-forming regions will be close to the photosphere.  The figure displays only the wavelength region 0.65 -- 1.35 \mum\ for detail. The complete spectra are found in Figure \ref{gh}.  A general discussion of features detected during the \emph{Photospheric Phase} can be found in \S \ref{mg} and a more detailed analysis in found in \ref{mg_featanal}. \label{mg_fig}}
\end{figure}

\begin{figure}
\plotone{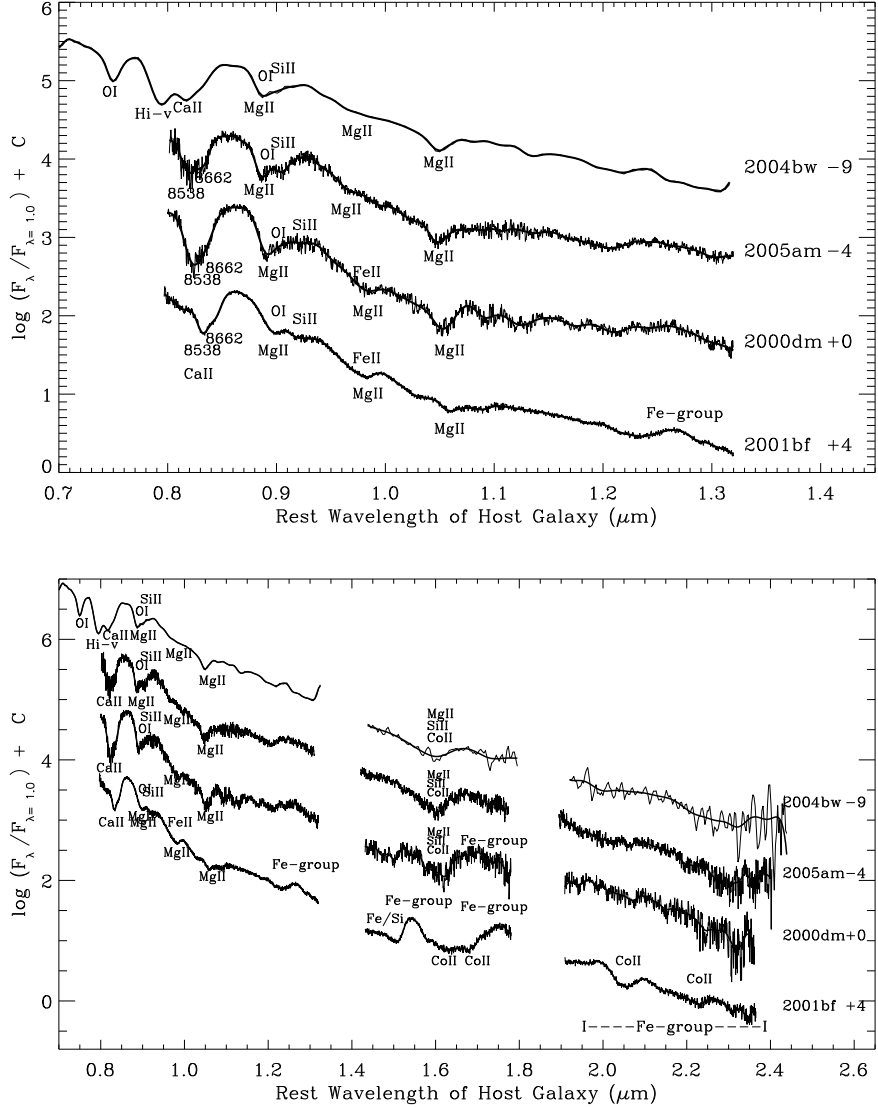}
\caption[]{Four representative spectra from the \emph{Photospheric Phase} of spectral development with features labeled.  The spectra are from SNe 2004bw (LRS obtained at -9d), 2005am (SXD obtained at -4d), 2000dm (SXD obtained at +0d), and 2001bf (SXD obtained at +4d).  The top panel displays only the wavelength region 0.65 -- 1.35 \mum\ for detail while the bottom panel contains the complete spectra.  A general discussion of these features can be found in \S \ref{mg} and a more detailed analysis in found in \ref{mg_featanal}. \label{mg_fwl}}
\end{figure}

\begin{figure}
\plotone{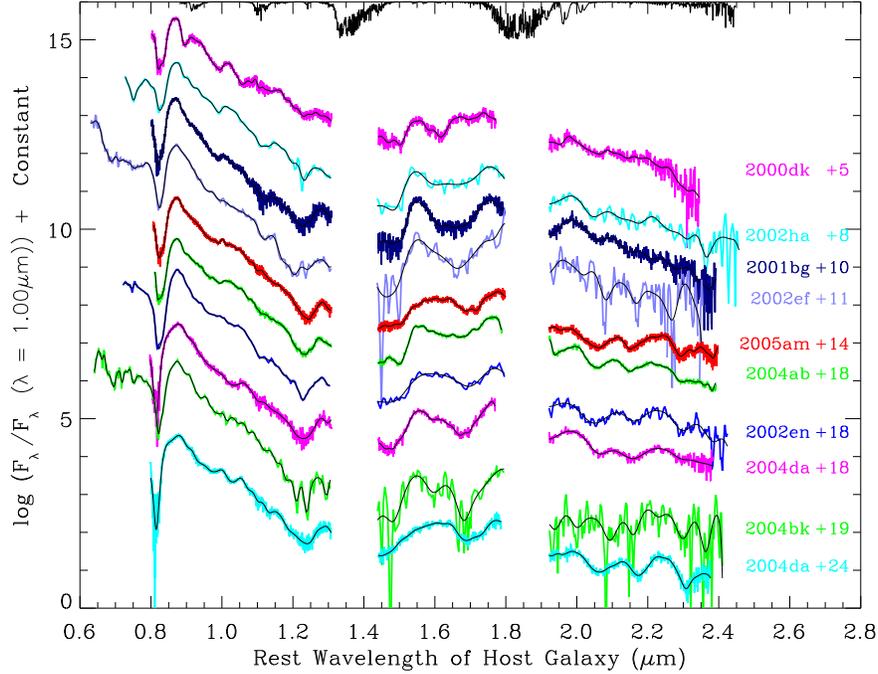}
\caption[]{Ten spectra from the \emph{Extended Photospheric Phase} of SNe Ia obtained between +4d and +24d with respect to to $V_{max}$.  During the phase represented by this group, the shape of the continuum becomes influenced by line-blanketing from Fe-group lines which extends the effective photospheric radius (see \S \ref{phlfr}).  A general discussion of features detected during the \emph{Extended Photospheric Phase} can be found in \S \ref{fe} and a more detailed analysis is found in \S \ref{fe_featanal}.  \label{fe_fig}}
\end{figure}

\begin{figure}
\plotone{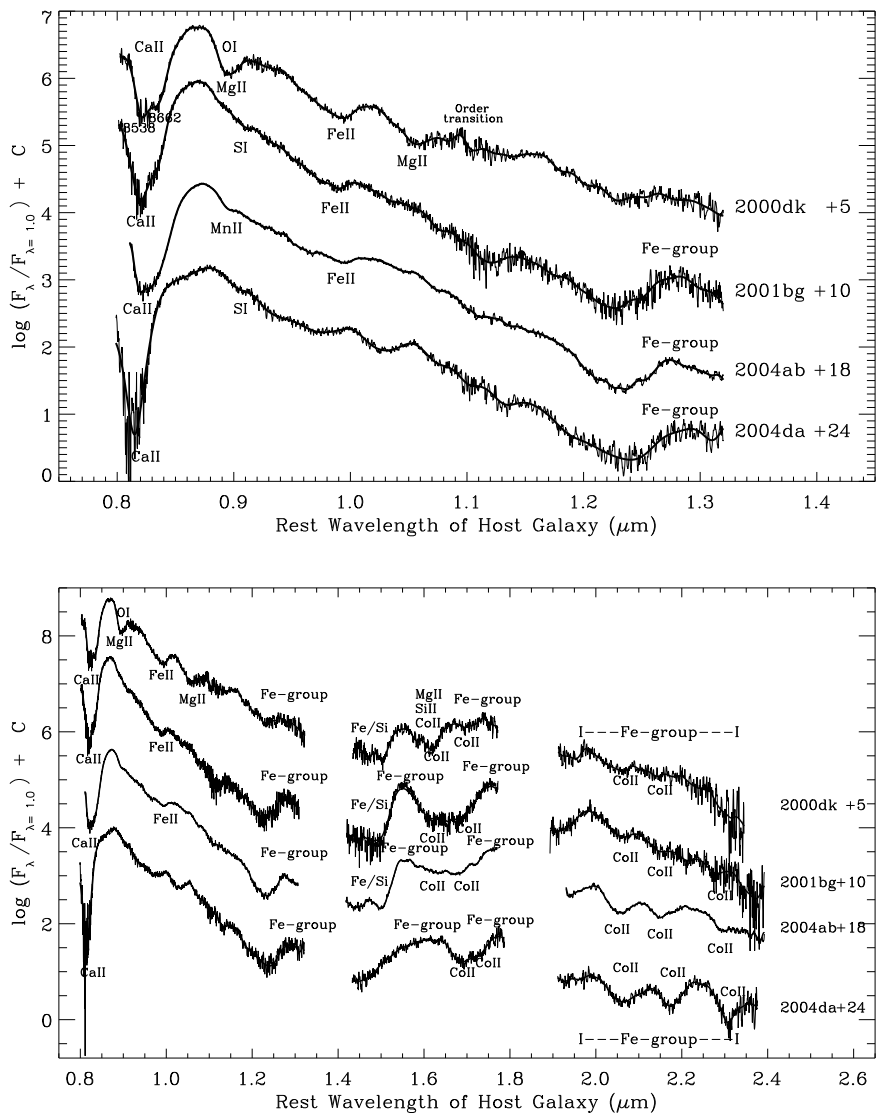}
\caption[]{Four representative spectra from the the \emph{Extended Photospheric Phase} of spectral development with features labeled.  The spectra are from SNe 2000dk (+5d), 2001bg (+10d), 2004ab (+18d), and 2004da (+24d).  The top panel displays only the wavelength region 0.65 -- 1.35 \mum\ for detail while the bottom panel contains the complete spectra.  A general discussion of these features can be found in \S \ref{fe} and a more detailed analysis in found in \ref{fe_featanal}. \label{fe_fwl}}
\end{figure}

\begin{figure}
\plotone{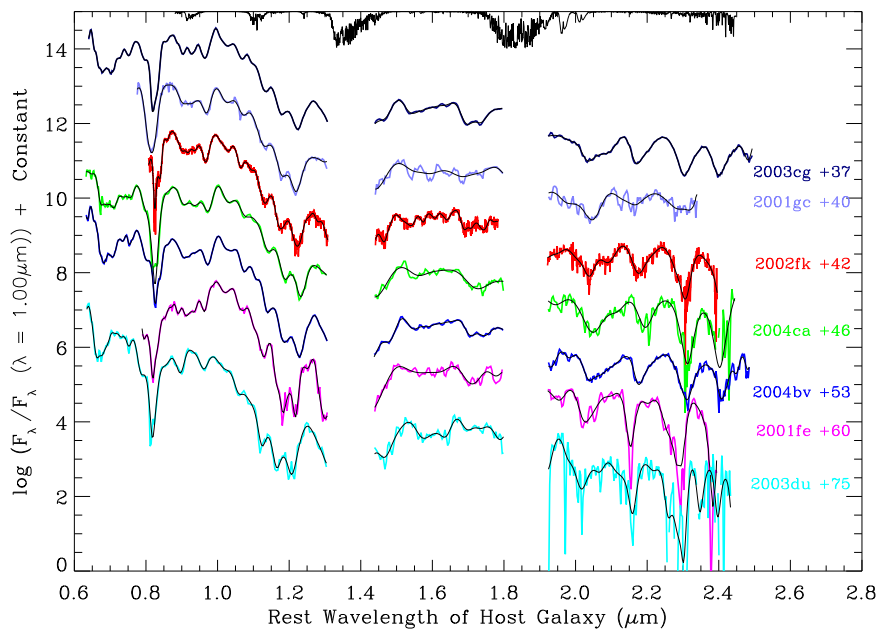}
\caption[]{Seven spectra from the \emph{Transitional Phase} SNe Ia obtained between +37d and +75d with respect to to $V_{max}$.  This epoch is a transitional phase between the photospheric era and a true nebular phase.  This group displays a lot of new structure relative to earlier NIR spectra and the features are remarkably similar from one spectrum to the next.   A general discussion of features detected during the \emph{\co\ Phase} can be found in \S \ref{transs} and a more detailed analysis is found in \S \ref{transs_featanal}.  \label{transfig}}
\end{figure}

\clearpage

\begin{figure}
\plotone{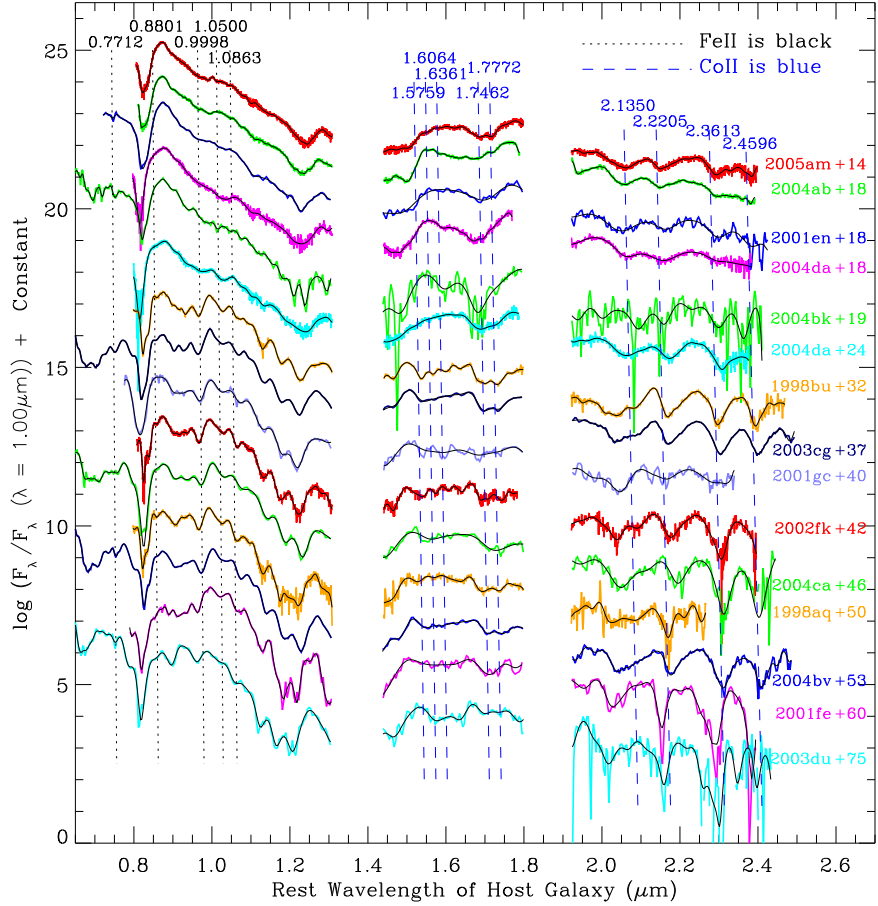}
\caption[]{Fifteen spectra between +14d and +75d are presented with the positions of strong \co\ and \fe\ lines drawn on the figure so that the indicated velocities are 11,000 \kms\ for the top spectrum and 6.000 \kms\ for the bottom spectrum.  This corresponds to the measured velocities for features from spectra during this epoch.  The data plotted in yellow (SNe 1998aq and 1998bu) were obtained by \citeauthor{Rudy} and have not been previously published.  \co\ lines are easily detected in the all spectra in this group and and \fe\ lines begin to appear about one month post-maximum (see \S \ref{fe} and \ref{transs}). \label{trans_coii}}
\end{figure}

\begin{figure}
\plotone{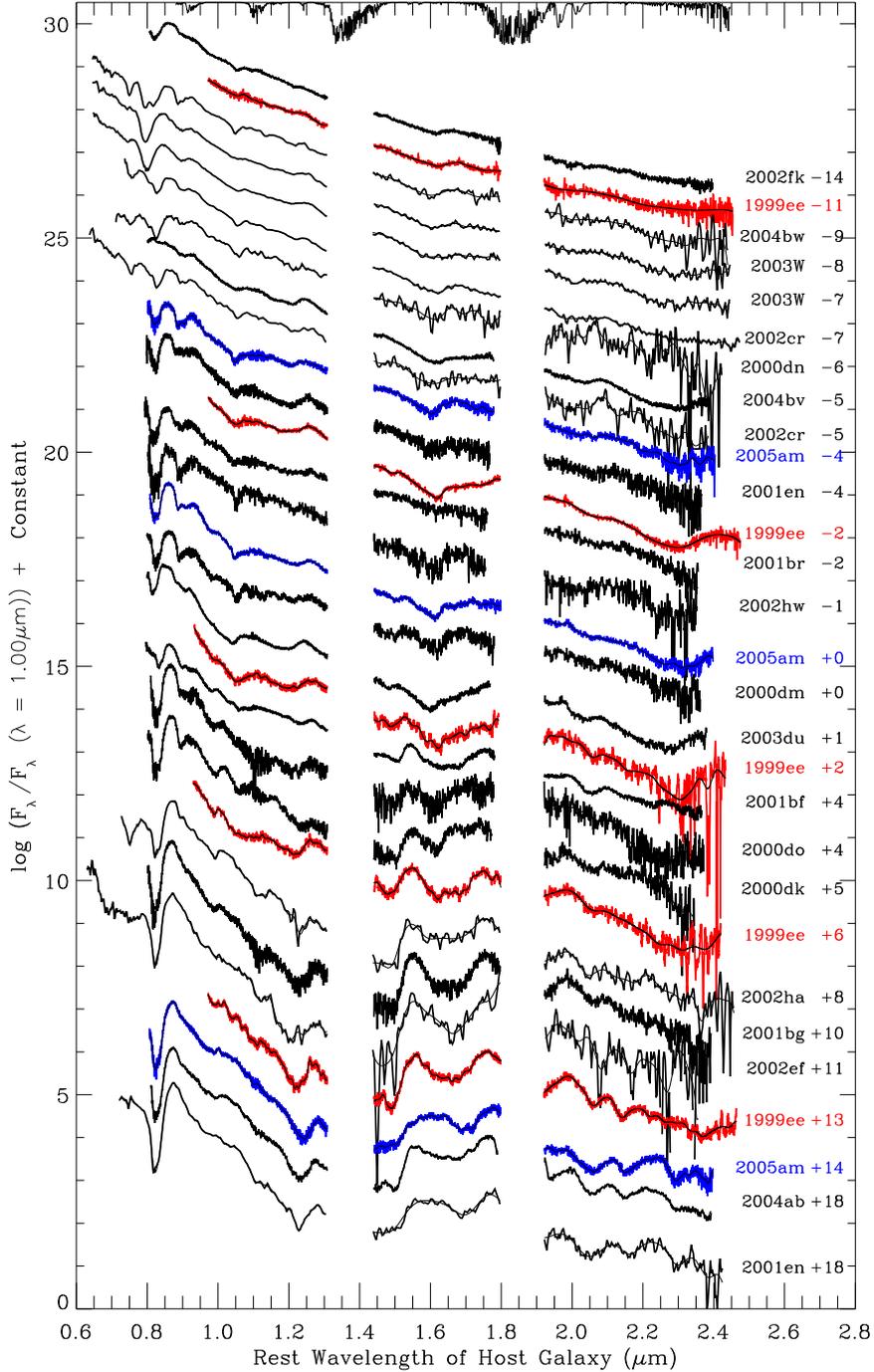}
\epsscale{0.7}
\caption[]{The uniformity of feature evolution in NIR spectra of SNe Ia within a few weeks of maximum light is demonstrated by interleaving ``snapshot'' data from 19 SNe Ia in our sample with two time sequences of spectra obtained from individual SNe Ia (see text \S \ref{use}).  Spectral shape and feature development with time are found to be consistent in individual events and also between diverse SNe Ia.  Five spectra from SN 1999ee obtained between -11d and +13d are shown in red \citep{Hamuy02}.  The four spectra in blue are from our sample and show SN 2005am between -4d and +14d.  \label{uni_23}}
\end{figure}

\begin{figure}
\epsscale{0.7}
\plotone{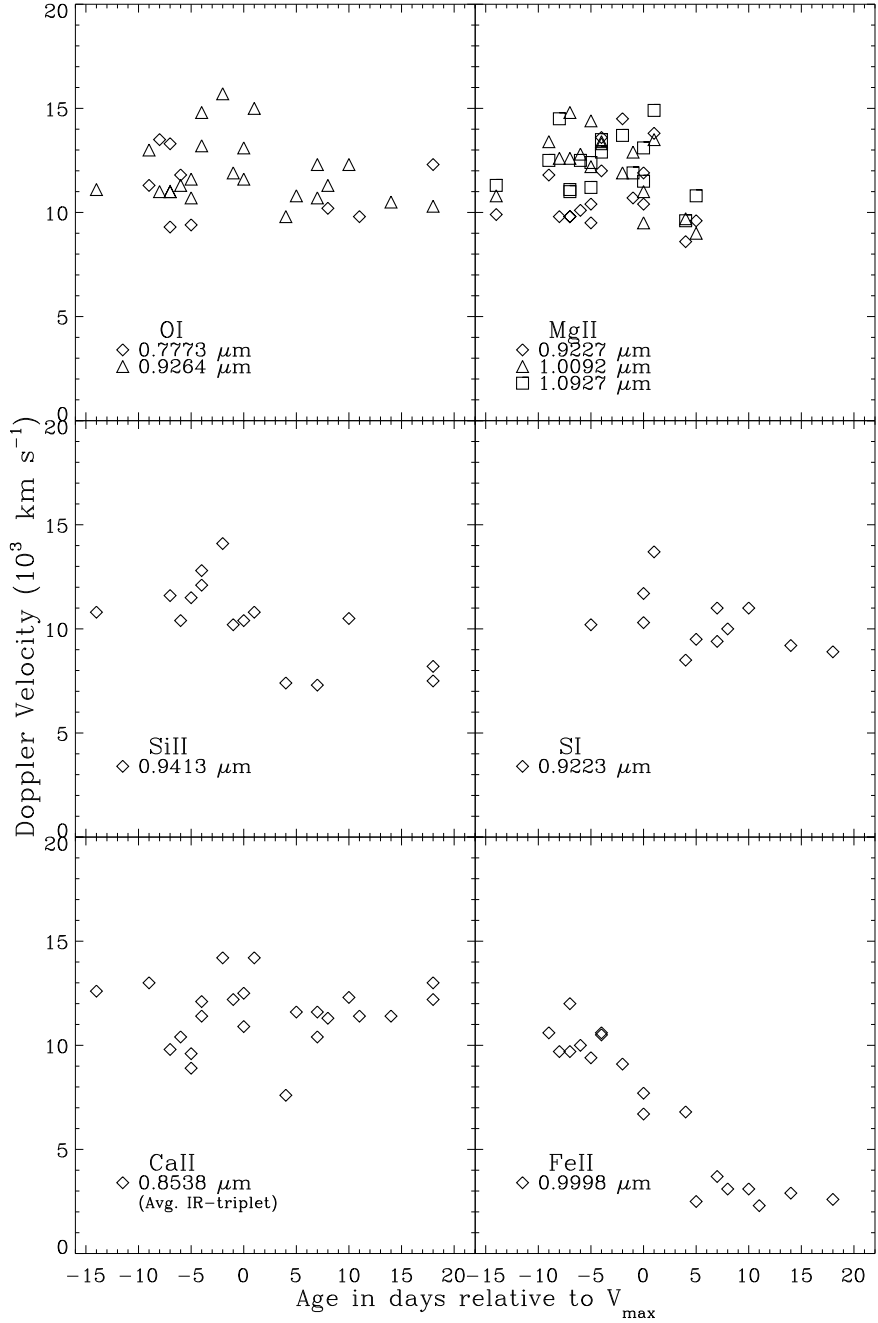}
\caption[]{Plot of measured Doppler velocities for ions with confident identifications in spectra obtained before +20d.  In all panels, the ordinate is the time of observation in days relative to $V_{max}$, and the abscissa is the Doppler velocity in $10^3$ \kms.  The \oi\ data include measurements from two lines (0.7774 and 0.9264 \mum), the \mg\ data include three lines (0.9227, 1.0092, and 1.0927 \mum), and there is one line each from \si\ (0.9413 \mum), \ion{S}{1} (0.9223 \mum), the \ca\ IR-triplet (0.8538 \mum\ blended line), and \ion{Fe}{2} (0.9998 \mum).  The measurements are compiled in Table \ref{vtable} and discussed in \S \ref{vtalk}. \label{vplot}}
\end{figure}

\begin{figure}
\plotone{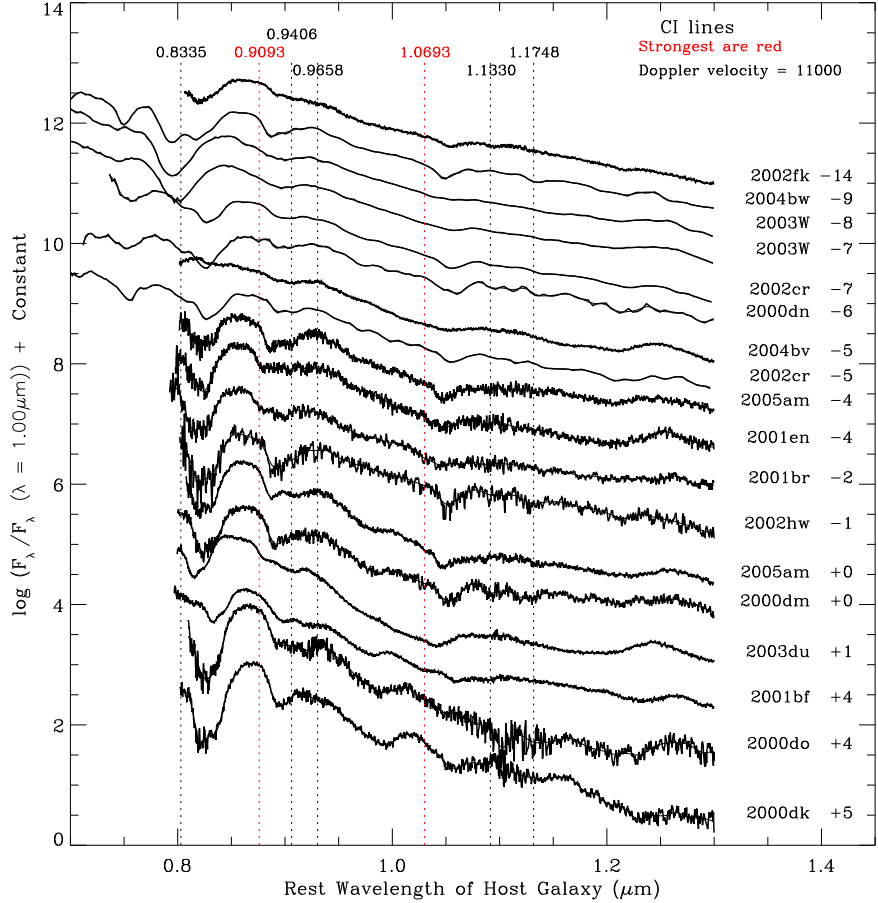}
\caption[]{Eighteen NIR spectra from SNe Ia as in Figure \ref{mg_fig} (-14d to +5d).  The locations of strong \ci\ lines marked on the figure at Doppler velocities of 11,000 \kms\ which is typical for \oi\ during this epoch.  The strongest lines, as indicated in Table \ref{10K}, are marked in red and the weaker lines appear in black. Small features appear in some spectra near the expected location for lines from 0.9406, 1.1330, and 1.1754 \mum, but if these lines produce absorption features then the stronger lines at 1.0693 and 0.9093 \mum\ should be detected and they are not.  The possible feature for the 0.9406 \mum\ line is more likely attributed to \si\ at 0.9413 \mum\ (see \S \ref{noCI}). \label{noci}}
\end{figure}

\begin{figure}
\plotone{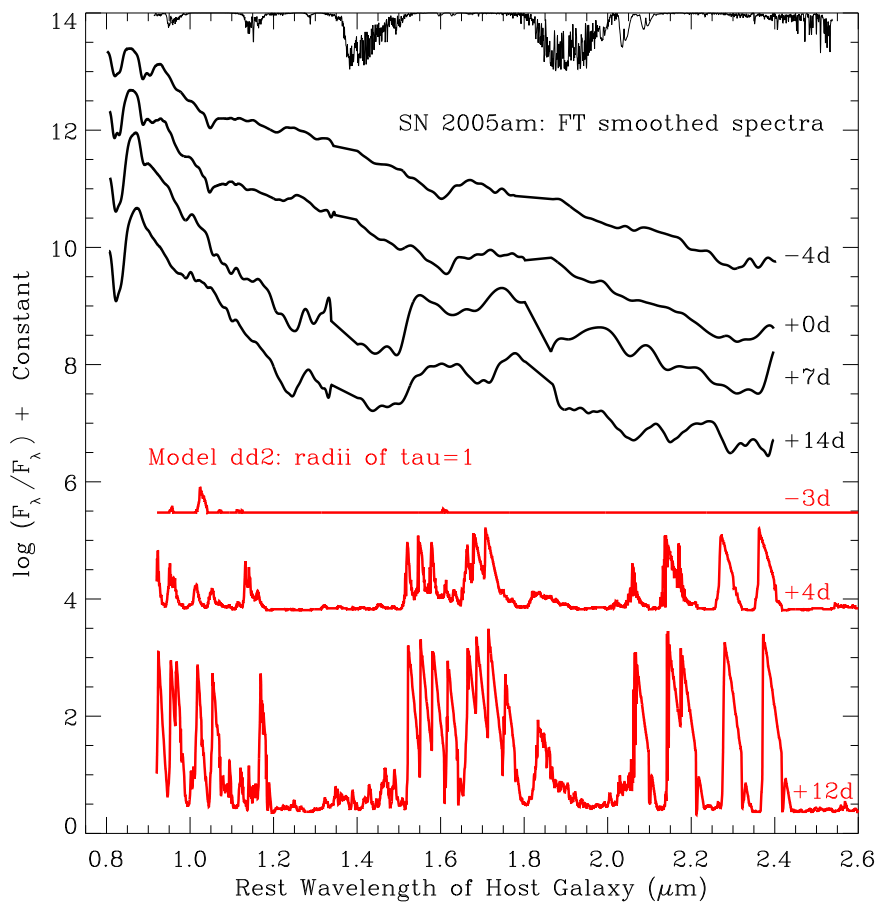}
\caption[]{Model data for radius at $\tau=1.0$ as a function of wavelength compared to spectra obtained from SN 2005am at epochs: -4d, +0d, +7d, and +14d.  The models clearly predict the observed behavior of a pseudo photosphere created at extended radii that increases the continuum flux in wavelength regions: 0.9--1.2, 1.5--1.85, and 2.05--2.45 \mum.  Increased opacity due to line-blanketing from thousands of Fe-group lines increases the effective radius of the photosphere at certain wavelengths.  The model resolution creates individual peaks but groups of lines will be blended by Doppler shifting to smear out the sharp features. (see text \S \ref{phlfr}). \label{am_bumps}}
\end{figure}


\clearpage

\begin{thebibliography}{}

\bibitem[Branch, et al.(2008)]{Branch08} Branch, D., Jeffery, D. J., Parrent, J., Baron, E., Troxel, M. A., Stanishev, V.,  Keithley, M., Harrison, J., and  Bruner, C., 2008, PASP 120, 135

\bibitem[Conley et al.(2008)]{conley08} Conley, A., et al. 2008, \apj\ 681, 482

\bibitem[Cushing, Vacca, \& Rayner(2004)]{Cushing04} Cushing, M. C., Vacca, W. D. and Rayner, J. T. 2004, PASP 116, 362

\bibitem[Dom\'{\i}nguez, H\"oflich \& Straniero(2001)]{dominguez01} Dom\'{\i}nguez, I.,  H\"oflich, P., and Straniero, O. 2001, \apj\ 557, 126

\bibitem[Gamezo et al.(2003)]{gam03} Gamezo, V. N., Khokhlov, A. M., Oran, E. S., Chtchelkanova, A. Y., and Rosenberg, R. O. 2003 Sci 299, 77

\bibitem[Garnavich(1998)]{gar98} Garnavich, P., et al. 1998 \apj\ 493, 53

\bibitem[Guy et al.(2007)]{guy07} Guy, J. et al. 2007, A\&A, 466, 11

\bibitem[Hamuy et al.(1996)] {hamuy96} Hamuy, M., Phillips, M.M., Suntzeff, N.B., Schommer, R.A., Maza, J., Aviles, A. 1996, \aj\ 112, 2398

\bibitem[Hamuy, et al.(2002)]{Hamuy02} Hamuy, M., et al. 2002, \aj\ 124, 417

\bibitem[H\"oflich, Mueller \& Khokhlov(1993)]{hmk93} H\"oflich P., Mueller, E., and Khokhlov, A.  1993, A\&A, 268, 570

\bibitem[H\"oflich(1995)]{pah95} H\"oflich, P. 1995, \apj\ 443, 89

\bibitem[H\"oflich \& Khokhlov(1996)]{hk96} H\"oflich, P. and Khokhlov, A  1996, \apj\ 457, 500

\bibitem[H\"oflich et al.(1998)]{hwt98} H\"oflich, P., Wheeler J.C., Thielemann F.K 1998, \apj\ 495, 617

\bibitem[H\"oflich et al.(2002)]{pah02} H\"oflich P., Gerardy C., Fesen R.,  Sakai S. 2002, \apj\ 568, 791

\bibitem[Hoyle \& Fowler(1960)]{hoyle60} Hoyle, P., \& Fowler, W.A. 1960, \apj\ I132, 565

\bibitem[Jha, Riess, \& Kirshner (2007)]{jha07} Jha, S., Riess, A., \& Kirshner, R. 2007, \apj\ 659, 122

\bibitem[Khokhlov(1991)]{kho91} Khokhlov, A. M. 1991 A\&A 245, 114

\bibitem[Khokhlov, Mueller \& H\"oflich(1993)]{kmh93} Khokhlov, A., Mueller, E., and H\"oflich P.  1993, A\&A 270, 223
	
\bibitem[Khokhlov, Oran \& Wheeler(1997)]{kho97} Khokhlov, A. M., Oran, E. S., and Wheeler, J. C. 1997, \apj\ 478, 678
\bibitem[Kowalski et al.(2008)]{kowalski08} Kowalski, M., et al. 2008 \apj\ 686, 749

\bibitem[Marion et al.(2003)]{m03} Marion, G. H., H\"oflich, P. , Vacca, W. D. and Wheeler, J. C. 2003, \apj\ 591, 316

\bibitem[Marion et al.(2006)]{m06} Marion, G. H., H\"oflich, P. , Gerardy, C. L, Vacca, W. D. Wheeler, J. C., and Robinson, E. L.,  2006, \apj\ 645, 1392

\bibitem[Meikle et al.(1996)]{Meikle_96} Meikle, W. P. S., et al. 1996, MNRAS, 281, 263

\bibitem[Perlmutter et al.(1999)]{Perl_99} Perlmutter, S. et al. 1999, \apj\ 517, 565

\bibitem[Phillips et al.(1993)]{phillips93}  Phillips, M.M. 1993, \apj\ 413, 105

\bibitem[Phillips et al.(1999)]{phillips99}  Phillips, M.M.,  Lira P., Sunzeff N.B., Schommer R.A., Hamuy M., Maza J. 1999, \aj ~  118, 1766

\bibitem[Rayner et al.(1998)]{Rayner98} Rayner. J. T.,  Toomey, D. W., Onaka,   P. M., Denault, A. J.,  Stahlberger, W. E.,  Watanabe, D. Y. and Wang S.-I. 1998, Infrared Astronomical Instrumentation, ed. A. M. Fowler, Proc. SPIE, 3354, 468-479

\bibitem[Rayner et al.(2003)]{Rayner03} Rayner. J. T., Toomey, D. W., Onaka, P. M., Denault, A. J.,  Stahlberger, W. E.,  Vacca, W. D., Cushing, M. C. and Wang, S., 2003, PASP 115,362


\bibitem[Riess et al.(1995)]{Riess95}  Riess A.G., Press W.H., Kirshner R.P. 1995, \apj\  438, L17

\bibitem[Riess et al.(1998a)]{Riess98a} Riess A.G., et al. 1998a, \aj\ 116, 1009

\bibitem[Riess, et al.(1999)]{riess99} Riess, A. G., et al. 1999, \aj\ 117, 707


\bibitem[Rudy \& Puetter()]{Rudy} Rudy, R. J. The Aerospace Corporation and Puetter, R. UCSD, private communication

\bibitem[Schmidt, et al.(1998)]{schmidt98} Schmidt, B., et al. 1998, \apj\ 507, 46 

\bibitem[SUSPECT (2008)]{suspect} SUSPECT, Online Supernovae Spectrum Archive, http://bruford.nhn.ou.edu/~suspect/index.html 

\bibitem[Tanaka, et al.(2008)]{tanaka} Tanaka, M., et al. 2008, \apj\  677, 448

\bibitem[Vacca, Cushing \& Rayner(2003)]{Vacca03} Vacca, W. D., Cushing, M. C. and Rayner, J. T. 2003, PASP 115, 389

\bibitem[Wang, H\"oflich \& Wheeler(1998)]{wang98} Wang, L., H\"oflich, P., Wheeler, J. C. 1998, \apjl\  487, 29

\bibitem[Wang, et al.(2003)]{wang03} Wang, L., et al. 2003, \apjl\  591, 1110

\bibitem[Wang \& Wheeler(2008)]{wang08} Wang, L. \& Wheeler, J. C. 2008, \araa\ 46, 433

\bibitem[Weller \& Albrecht(2001)]{WA2001}Weller J., Albrecht A. 2001, {\sl Opportunities for future supernova studies of cosmic acceleration},  astro-ph/0008314

\bibitem[Wheeler et al.(1998)]{Wheeler98} Wheeler, J. C.,  H\"oflich, P., Harkness, R. P., Spyromilio, J. 1998, \apj\ 496, 908

\bibitem[Yoon \& Langer(2004a)]{yoon04a} Yoon, S.-C., \& Langer, N. 2004, A\&A 419, 623

\bibitem[Yoon \& Langer(2004b)]{yoon04b} Yoon, S.-C., \& Langer, N. 2004, A\&A 419, 645


\end{thebibliography}
\end{document}